\documentclass[a4paper]{article}
\usepackage{amsmath}
\usepackage{graphicx}
\usepackage{amsbsy}
\usepackage{epsfig}
\usepackage{cite}
\usepackage{amsfonts}
\usepackage{amssymb}
\long\def\symbolfootnote[#1]#2{\begingroup\def\thefootnote{\fnsymbol{footnote}}\footnote[#1]{#2}\endgroup}
\newcommand{\tpair}[2]{$\begin{array}[c]{r} #1 \\ #2 \end{array}$}
\newcommand{\tpaic}[2]{$\begin{array}[c]{c} #1 \\ #2 \end{array}$}

\begin{document}
\thispagestyle{empty}  \setcounter{page}{0}  
\begin{flushright}
November 2011\\
\end{flushright}

\vskip 4 true cm

\begin{center}{\huge FCNC portals to the dark sector}\\[1.9cm]

\textsc{Jernej F. Kamenik}$^{1,2,}$\symbolfootnote[1]{jernej.kamenik@ijs.si}
\textsc{ and Christopher Smith}$^{3,}$\symbolfootnote[2]{c.smith@ipnl.in2p3.fr}
\\[12pt]$^{1}$\textsl{J. Stefan Institute, Jamova 39, P. O. Box 3000, 1001 Ljubljana, Slovenia}\\[6pt]$^{2}$\textsl{Department of Physics, University of Ljubljana, Jadranska 19, 1000 Ljubljana, Slovenia}
\\[6pt]$^{3}$\textsl{Universit\'{e} Lyon 1 \& CNRS/IN2P3, UMR5822 IPNL,}
\textsl{Rue Enrico Fermi 4, 69622 Villeurbanne Cedex, France}\\[1.9cm]

\textbf{Abstract}
\begin{quote}
\noindent
The most general basis of operators parametrizing a low-scale departure from the SM particle content is constructed. The SM gauge invariance is enforced, and operators of lowest dimensions are retained separately for a new light neutral particle of spin 0, 1/2, 1, and 3/2. The basis is further decomposed into couplings to the SM Higgs/gauge fields, to pairs of quark/lepton fields, and to baryon/lepton number violating combinations of fermion fields. This basis is then used to systematically investigate the discovery potential of the rare FCNC decays of the $K$ and $B$ mesons with missing energy in the final state. The most sensitive decay modes in the $s\to d$, $b\to d$, and $b\to s$ sectors are identified and compared for each type of couplings to the new invisible state.
\newpage

\setcounter{tocdepth}{2}
\rule{\linewidth}{0.3mm}
\tableofcontents
\rule{\linewidth}{0.3mm}
\end{quote}
\end{center}

\section{Introduction}

Though extremely successful, the Standard Model (SM) is not expected to be valid up to arbitrary high energies. It certainly needs to be amended at the Planck scale, with the advent of quantum gravity. But then, some new dynamics should show up already at a much lower scale, though still above the electroweak scale, to avoid hierarchy problems. With this picture in mind, it seems natural to assume that all the New Physics (NP) degrees of freedom are heavier than the known SM particles, and decouple at low energy. In a fully model-independent way, the impact on the SM of any NP model can then be embedded into non-renormalizable effective operators, under the provision that these involve only the SM fields and satisfy the SM gauge symmetries. The full classification of these operators has been achieved many years ago, by Buchmuller and Wyler~\cite{BuchmullerW86} for the baryon- and lepton-number conserving operators, and by Weinberg~\cite{Dim6} for those violating these global symmetries.

Still, whether the SM particles are the only dynamical degrees of freedom within the electroweak energy range is far from established. Not only is the existence of new light particles not excluded, since they would evade direct detection when sufficiently weakly interacting, but their presence could even be welcome. Indeed, many NP models are built upon some spontaneously broken symmetries, and do often have remnants at low-energy in the form of massless or very light Goldstone bosons. A well-known example is the axion~\cite{Axion}, introduced to cure the strong CP problem of the SM. More crucially, there are now very strong indications that the universe is filled with dark matter, so there should be at least one new electrically neutral colorless particle, possibly lighter than the electroweak scale. Once opening that door, it is not such a drastic step to imagine a whole dark sector, i.e. a full-fledged set of darkly interacting dark particles only loosely connected to our own visible sector. Further, it should be stressed that adjoining a dark sector to the SM is always possible, does not need to be directly related to dark matter (so one would rather speak of a hidden sector), and is actually quite generic in supersymmetric models. For a recent review, including further physical motivations from string theory or extra dimensional settings, see e.g. Ref.~\cite{Hidden}.

In the present work, our main goal is to construct the lowest-dimensional effective interactions parametrizing a low-scale departure from the SM particle content. Specifically, we assume that there is a new particle of spin $0$, $1/2$, $1$, or $3/2$, neutral under the SM gauge group $SU(3)_{C}\otimes SU(2)_{L}\otimes U(1)_{Y}$, and write down the gauge invariant operators coupling this particle to SM fields. These effective interactions are not yet included and thus complement the NP operator basis of Refs.~\cite{BuchmullerW86,Dim6}. To our knowledge, such a complete basis has never been presented, though parts of it already appeared in the literature. In particular, those effective couplings between SM and dark particles which are renormalizable, sometimes called portals, have already been investigated~\cite{Portals}.

Our second goal is to constrain the effective operators. Since the new state is assumed neutral under the SM gauge group, it looks like a natural dark matter candidate. However, the viability of this hypothesis would require constraining its mass, couplings, and lifetime, and this in general requires more inputs about the dark sector dynamics. Here, we refrain from doing so and rather concentrate exclusively on the quark FCNC transitions $s\to dX$, $b\to dX$, and $b\to sX$, where $X$ collectively denotes any final state made of dark particles. Our focus on these modes, instead of for example leptonic observables or collider signals, is motivated on one hand by their extreme sensitivities to NP (as detailed in the next section), and on the other, by the next generation of experiments currently under construction. Indeed, rare $K$ decays are the main targets of the NA62 (CERN) and K0TO (J-Parc) experiments, while rare $B$ decays could be accessed at the Super-B (Italy) and Belle II (KEK) facilities. So, in the present work, we further assume that the dark particle is light enough to be directly produced in $K$ and/or $B$ meson decays, and sufficiently long-lived to escape detection in flavor factories. For all practical matters, this new particle is invisible, and would show up as missing energy in FCNC-induced rare $K$ and $B$ decays (for recent works along this line, see e.g. Refs.~\cite{Badin:2010uh,Previous}). 

Our analysis is organized according to the dark particle spin. To keep the discussion focused on the operator basis, we rely on extensive appendices to cover the issues of hadronic matrix elements and differential rates. So, before entering the discussion, the next section summarizes the main features of the observables considered here, i.e. the rare $K$ and $B$ decays. Then, given our focus on FCNC processes, a flavor-based classification of the dark operators is described in the following section, on which we rely throughout the paper.

\subsubsection*{Rare FCNC decays}

\begin{figure}[t]
\centering 
\includegraphics[width=4.5cm]{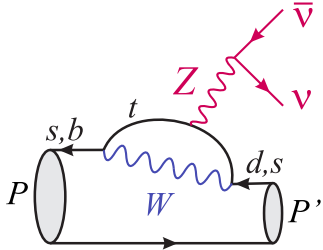}
\caption{The rare decays with missing energy in the SM, as induced by the Z penguin (W boxes are understood).}
\label{FigZpeng}
\end{figure}

The FCNC-induced decay modes are very suppressed in the SM, where the missing energy is carried away by a $\nu\bar{\nu}$ pair (see Fig.~\ref{FigZpeng}). So, even relatively small NP contributions could be evidenced. Specifically, to set the stage and get an idea of the sensitivity of the rare $K$ and $B$ decays, imagine that a NP operator of dimension $n$ contributes to $d^{I}\to d^{J}X$, with $I=2,3$, $J=1,2$ the quark generation indices. If its Wilson coefficient is set to one, then there is a scale $\Lambda$ such that the NP contribution equates the SM prediction for $d^{I}\to d^{J}\nu\bar{\nu}$,%
\begin{equation}
\frac{m_{I}^{n-6}}{\Lambda^{n-4}}\approx\frac{g^{2}}{M_{W}^{2}}\frac{g^{2}}{16\pi^{2}}|V_{tI}^{\ast}V_{tJ}|\;, \label{SMreach}
\end{equation}
with $m_{2,3}=m_{K,B}$, $g$ the $SU(2)_{L}$ coupling constant, and $V$ the CKM matrix. As shown in Table~\ref{Reach} as a function of the dimension, the SM loop factor combined with the CKM suppression pushes the scales $\Lambda$ well above the electroweak scale for $n\lesssim7$. On the contrary, the rare decay constraints cease to make sense for $SU(2)_{L}\otimes U(1)_{Y}$ invariant operators of dimension $n\gtrsim9$, since powers of $(H^{\dagger}H)/\Lambda^{2}\to v^{2}/\Lambda^{2}$ grow unchecked when $\Lambda\lesssim v$, where $v\approx246$ GeV is the SM Higgs vacuum expectation value.

\begin{table}[t] \centering
\begin{tabular}[c]{lccccc}\hline
& $n=5$ & $n=6$ & $n=7$ & $n=8$ & $n=9$\\\hline
$s\to d$ & $3.3\cdot10^{7}$ TeV & $130$ TeV & $2.0$ TeV & $0.25$ TeV & $0.07$ TeV\\
$b\to d$ & $1.3\cdot10^{5}$ TeV & $26$ TeV  & $1.5$ TeV & $0.37$ TeV & $0.16$ TeV\\
$b\to s$ & $2.7\cdot10^{4}$ TeV & $12$ TeV  & $0.9$ TeV & $0.25$ TeV & $0.11$ TeV\\\hline
\end{tabular}
\caption{Naive reach, in terms of scales $\Lambda$ and as a function of the effective operator dimension $n$, of the rare FCNC-induced $K$ and $B$ decays, as estimated from Eq.~(\ref{SMreach}) with the CKM values of Eq.~(\ref{MFVscaling}).}
\label{Reach}
\end{table}

Clearly, these scales are only indicative. The true sensitivity to a given dark operator depends essentially on two additional factors. First, the quark transitions $d^{I}\to d^{J}X$ have to be probed through hadronic processes. Hence, depending on the modes, hadronic matrix elements as well as phase-space factors can alter significantly the estimates of Table~\ref{Reach}. In the following, we compare the sensitivities of all the leading modes. Specifically, in the $K$ sector, we include the modes with the least number of pions and photons in the final states, i.e. $K\to X$, $K\to\pi X$, $K\to\gamma X$, and $K\to\pi\pi X$, and leave out the $K\to\pi\pi\pi X$ modes. Similarly, in the $B$ sector, the considered modes are the $B\to X$, $B\to(K,K^{\ast})X$, and $B\to(\pi,\rho)X$ decay channels. The $\gamma X$ channel, driven in the $K$ sector by the QED anomaly, is suppressed and difficult to reconstruct experimentally in the $B$ sector, and will thus not be included~\cite{Badin:2010uh}.

The second factor determining the true sensitivity of a given mode is related to the experimental strategies deployed to measure it. Since invisible states are not seen, the kinematical reconstruction is limited. In addition, these modes are so rare (in the SM) that they require very aggressive background suppressions. To this end, the central tool is the differential rate in terms of the kinematical parameters of the visible products. But this differential rate depends on the nature of the dark particle. Currently, most experimental analyses implicitly impose the SM differential rate (for $X=\nu\bar{\nu}$). This means that the current bounds cannot be directly translated to other types of final state particles. This motivates another goal of the present paper, which is to provide the full dictionary of the differential rates for all the leading effective interactions involving invisible final states. These spectra should be used by the experimentalists to derive bounds for each type of new invisible state.

For more details on these issues, including kinematics, matrix elements, current measurements or bounds, and experimental prospects for the various modes, we refer to appendix~\ref{AppKexp}~(\ref{AppBexp}) for $K$ ($B$) decays.

\subsubsection*{Flavor-based classification of the dark operators\label{Classification}}

At the electroweak scale, once the whole NP particle spectrum but the $X$ has been integrated out, the lowest dimensional operators can be split into three types according to their quark and lepton field contents:
\begin{equation}
\mathcal{H}_{eff}=\mathcal{H}_{mat}+\mathcal{H}_{int}+\mathcal{H}_{\Delta\mathcal{B},\Delta\mathcal{L}}\;.
\end{equation}
By definition, $\mathcal{H}_{int}$ contains only gauge and Higgs fields, while $\mathcal{H}_{mat}$ and $\mathcal{H}_{\Delta\mathcal{B},\Delta\mathcal{L}}$ contain at least one SM fermionic field. The operators of $\mathcal{H}_{\Delta\mathcal{B},\Delta\mathcal{L}}$ have a non-zero charge under the baryon ($\mathcal{B}$) or lepton ($\mathcal{L}$) number $U(1)$s. As in Ref.~\cite{BuchmullerW86,Dim6}, all the operators are to be written as manifestly invariant under $SU(3)_{C}\otimes SU(2)_{L}\otimes U(1)_{Y}$, i.e. in terms of the quark (lepton) doublets $Q$ $(L)$ and singlets $U,D$ $(E)$ of each flavor, of the $SU(3)_{C},SU(2)_{L},U(1)_{Y}$ field strengths $G_{\mu\nu}^{a},W_{\mu\nu}^{i},B_{\mu\nu}$, of the Higgs doublet $H$, as well as of covariant derivatives acting on these fields, insofar as these cannot be reduced using the SM equations of motion (EOM).

Due to their different field contents, these three types of operators do not contribute equally to the quark transitions $s\to dX$, $b\to dX$, and $b\to sX$, so let us organize them differently, in terms of four classes of scenarios, as shown in Table~\ref{TableClasses}.

Consider first the operators of $\mathcal{H}_{mat}$ involving down-type quarks (those with leptons or up-type quarks are obtained by substituting $D,Q\to E,L$ or $D,H\to U,H^{\ast}$). Up to possible partial derivatives acting on the quark or invisible fields, and omitting the
Dirac structures, the quark currents are%
\begin{equation}
\mathcal{H}_{mat}=\dfrac{c_{RL}^{IJ}}{\Lambda^{n}}H^{\dagger}\bar{D}^{I}Q^{J}\times X+\frac{c_{LR}^{IJ}}{\Lambda^{n}}H\bar{Q}^{I}D^{J}\times X+\dfrac{c_{LL}^{IJ}}{\Lambda^{n}}\bar{Q}^{I}Q^{J}\times X+\dfrac{c_{RR}^{IJ}}{\Lambda^{n}}\bar{D}^{I}D^{J}\times X\;.
\end{equation}
Those operators have a flavor structure, and thus can in principle induce $d^{I}\to d^{J}X$. Clearly, when analyzing the physics reach of rare $K$ and $B$ decays in terms of the scale $\Lambda$, the assumptions made on the $c^{I\neq J}$ are crucial. There are two main scenarios:

\begin{itemize}
\item[I.] The constraints derived from rare FCNC decays are the tightest when the NP flavor structure is generic,
\begin{equation}
c^{I\neq J}\sim\mathcal{O}(1)\;.
\end{equation}
As shown in Table~\ref{Reach}, the bounds on the NP scale $\Lambda$ are then often far above the electroweak scale. 

\item[II.] Since $\mathcal{H}_{mat}$ results from integrating out the whole NP particle spectrum, the flavor-breaking character of its operators could originate from dynamical effects not related to the dark sector. In that case, the NP dynamics would also quite naturally correct the visible FCNC operators, on which there are many tight experimental constraints from $K$ and $B$ physics~\cite{GeneralQuarkFlavors}. This is typically the case in supersymmetric settings, where the flavored soft-breaking terms cannot be fully generic. Phenomenologically, a simple way to account for such a non-generic NP flavor structure is to impose the Minimal Flavor Violation (MFV) ansatz~\cite{MFV}, i.e. force the quark currents to have the forms%
\begin{equation}
\bar{D}^{I}(\mathbf{Y}_{d}\mathbf{Y}_{u}^{\dagger}\mathbf{Y}_{u})^{IJ}Q^{J}\;,\;\;\bar{Q}^{I}(\mathbf{Y}_{u}^{\dagger}\mathbf{Y}_{u})^{IJ}Q^{J}\;,\;\bar{D}^{I}(\mathbf{Y}_{d}\mathbf{Y}_{u}^{\dagger}\mathbf{Y}_{u}\mathbf{Y}_{d}^{\dagger})^{IJ}D^{J}\;.
\end{equation}
In the down-quark mass-eigenstate basis, the diagonal $v\mathbf{Y}_{d}=\sqrt{2}\mathbf{m}_{d}$ tunes the chirality flips, while $v\mathbf{Y}_{u}=\sqrt{2}\mathbf{m}_{u}V$ parametrizes the flavor change ($\mathbf{m}_{u,d}$ denotes the diagonal quark mass matrices, $V$ the CKM matrix, and $v\approx246$ GeV the Higgs vacuum expectation value). So, MFV rescales the Wilson coefficients according to $c_{RL}^{IJ}=m_{d}^{I}c_{LL}^{IJ}/v$, $c_{LR}^{IJ}=c_{LL}^{IJ}m_{d}^{J}/v$, $c_{RR}^{IJ}=m_{d}^{I}m_{d}^{J}c_{LL}^{IJ}/v^{2}$, and%
\begin{equation}
c_{LL}^{I\neq J}\sim\lambda^{IJ}=\mathbf{Y}_{u}^{\dagger}\mathbf{Y}_{u}\approx V_{tI}^{\ast}V_{tJ}\to\left\{
\begin{array}[c]{l}
\lambda^{sd}\approx(-3.1+i1.3)\times10^{-4}\;,\\
\lambda^{bd}\approx(7.8-i3.1)\times10^{-3}\;,\\
\lambda^{bs}\approx(-4.0-i0.07)\times10^{-2}\;.
\end{array}
\right.
\label{MFVscaling}
\end{equation}
Upon these rescalings, the accessible scales $\Lambda$ are then much lower, especially for operators of low dimensions, and for $s\to d$ operators involving light right-handed quarks.
\end{itemize}

If the whole NP dynamics is flavor blind, then $X$ couples only to the flavor-diagonal quark currents and $c^{I\neq J}=0$. A flavor transition is of course still possible but it must proceed through the SM weak interactions, i.e. the flavor-blind quark currents must be dressed by a flavor-changing $W$ interaction, either between the quark lines or as a self-energy on these quark lines.

In particular, all the operators of $\mathcal{H}_{int}$ are of this type. Indeed, to contribute to FCNC processes, gauge fields have to be coupled to quarks, while Higgs fields are either coupled to quarks or left as external tadpoles (to be replaced by the vacuum expectation value after the electroweak symmetry breaking). The resulting couplings between quarks and $X$ can then be matched onto $\mathcal{H}_{mat}$, and satisfy\footnote{The contributions to $c^{I\neq J}$ are tiny because, as long as the $SU(2)_{L}\otimes U(1)_{Y}$ symmetry is exact, the gauge/Higgs fields of the $\mathcal{H}_{int}$ operators can couple to a flavor-changing quark current only at the cost of at least two Yukawa insertions, i.e. two Higgs tadpoles, on the quark line. At the high scale, these tadpoles cost a factor $\Lambda^{-2}$ compared to the flavor-blind version of the same operator. When $\Lambda$ is large, flavor-changing effects thus dominantly originate from electroweak scale GIM breaking, even though this necessitates an extra $W$ loop (see Table~\ref{TableClasses}).} $c^{I\neq J}\approx0$. Thus, the only difference with respect to the flavor-blind $\mathcal{H}_{mat}$ operators is that the $c^{II}$ coefficients are initially suppressed by some power of the SM coupling constants (at the scale $\Lambda$), by some loop factors, and by quark Yukawa couplings when a Higgs field is coupled to the quark line (so that $c^{33}\gg c^{11,22}$).

For reasons entirely pertaining to the SM dynamics, it is different to probe the couplings to heavy quarks, $c^{33}$, and to light quarks, $c^{11,22}$, so these constitute our third and fourth classes of scenarios.

\begin{itemize}
\item[III.] Let us assume we have an operator coupling $X$ to the top quark current. Dressing it with a $W$ exchange (see Table~\ref{TableClasses}), the necessary GIM breaking arises at the electroweak scale. From the rare decay perspective, this electroweak physics is local, so it can be matched onto the flavor-changing operators of Class II. The Wilson coefficients end up suppressed by the MFV scalings~(\ref{MFVscaling}) and by a loop and gauge coupling, i.e.
\begin{equation}
c^{I\neq J}\sim c^{33}k^{IJ}\;,\;\;k^{IJ}=\frac{g^{2}}{16\pi^{2}}\lambda^{IJ}\to\left\{
\begin{array}[c]{l}
k^{sd}\approx(-0.8+i0.4)\times10^{-6}\;,\\
k^{bd}\approx(2.1-i0.8)\times10^{-5}\;,\\
k^{bs}\approx(-1.1-i0.02)\times10^{-4}\;.
\end{array}
\right.  
\label{CKMscale}
\end{equation}
Typically, the bounds on the scale $\Lambda$ are brought down very significantly, often at around the electroweak scale. Still, the rare $K$ and $B$ decays remain ideal probes for such kind of effective couplings since a direct collider signal in the $t\bar{t}$ channel is presumably hidden by the large flavor-blind SM backgrounds.

\item[IV.] With $X$ coupled to the light quarks ($u$, $d$, or $s$), it is much trickier to derive bounds on the scale $\Lambda$ for three reasons. Firstly, $B$ physics is unable to provide competitive constraints, given the small CKM factors $V_{ub}^{\ast}V_{us}$ or $V_{ub}^{\ast}V_{ud}$. Secondly, though in the $K$ sector the weak transition is favored by the large $V_{us}^{\ast}V_{ud}$ for CP-conserving observables\footnote{CP-violating observables, for which $\operatorname{Im}(V_{us}^{\ast}V_{ud})=0$ (this standard CKM convention is used throughout the paper, see e.g. Ref.~\cite{GrossmanNir}), are induced by heavy quarks and fall into Class III. Due to the scaling~(\ref{CKMscale}), they are very suppressed.}, the light quarks are never integrated out and remain dynamical. This renders the theoretical control difficult since the $K$ physics scale is too low to allow for a perturbative QCD treatment. This is true both for the effective operators describing the weak transition and for the effective operators describing the production of the new invisible states. Thirdly, there are already many constraints on the flavor blind production of invisible particles, in particular from $\pi^{0}$ or quarkonium decays. Compared to the other classes, it is a priori not clear whether rare decays offer privileged windows. In the following, we will consider only specific scenarios where competitive bounds can be derived.
\end{itemize}%

\begin{table}[t] \centering
\begin{tabular}[c]{p{5.2cm}p{5.2cm}p{5.2cm}}\hline
\multicolumn{3}{c}{\rule{0in}{0.24in}$\mathcal{H}_{eff}(q^I\to q^J X)=\dfrac{c^{IJ}}{\Lambda^{n}_{\;}}\bar{q}^{I}q^{J}\times X$}\\\hline
\multicolumn{1}{c}{Flavor-violating\ ($c^{I\neq J}\neq 0$)} &
\multicolumn{2}{c}{Flavor-conserving ($c^{I\neq J} = 0$) \rule{0in}{0.19in}\smallskip}\\
\multicolumn{1}{c}{} & \multicolumn{1}{c}{Heavy quark: $q=(c),t\rule{0in}{0.19in}\smallskip$} & \multicolumn{1}{c}{Light quarks: $q=u,d,s,(c)$}\\
\multicolumn{1}{c}{\includegraphics[width=3.7cm]{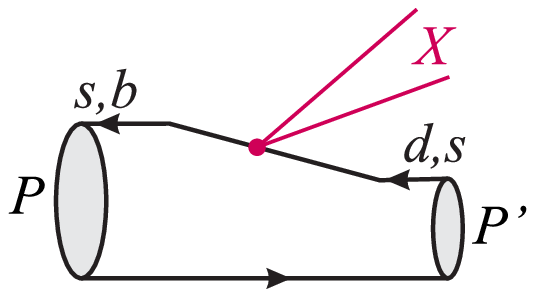}} &
\multicolumn{1}{c}{\includegraphics[width=3.7cm]{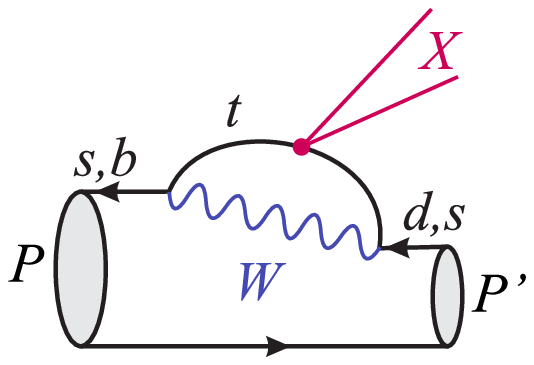}} &
\multicolumn{1}{c}{\includegraphics[width=3.7cm]{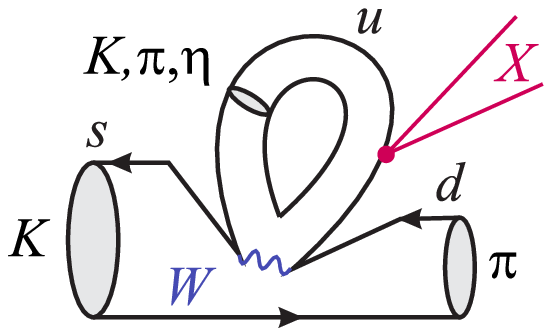}}\\
Bounds on $c^{IJ}/\Lambda^{n}$ directly derived from the $d^I\to d^J X$ processes. When MFV holds, $c^{IJ}\sim V_{tI}^{\ast}V_{tJ}$ times the appropriate chirality flip factors $m_{d^{I,J}}/v$, see Eq.~(\ref{MFVscaling}). & Same local operator basis, but with the coefficients rescaled as $c^{IJ}\to (g/(4\pi))^{2}V_{tI}^{\ast}V_{tJ}\times c^{33}$ times the appropriate chirality flip factors $m_{d^{I,J}}/v$, see Eq.~(\ref{CKMscale}). & Due to the small $V_{ub}^{\ast}$, $B$ decays are not competitive. For $K$ decays, the $q=u$ contributions are dominant but non local, and require controlling long-distance hadronic effects.\\\hline
\multicolumn{1}{c}{Class I, II} & \multicolumn{1}{c}{Class III} &
\multicolumn{1}{c}{Class IV}\\\hline
\end{tabular}
\caption
{Flavor-based classification of the operators involving dark particles, collectively denoted $X$. After the electroweak symmetry breaking, the $\mathcal{H}_{mat}$ operators are directly matched onto $\mathcal{H}_{eff}(q^{I}\to q^{J}X)$ and split into the four classes. The $\mathcal{H}_{int}$ operators collapse onto the Class III and/or IV flavor-blind operators once their gauge/Higgs fields are attached to quarks. The $\mathcal{H}_{\Delta\mathcal{B},\Delta\mathcal{L}}$ operators have different signatures, and do not fit in this classification. Note that the charm quark is considered heavy (light) for $K$ ($B$) decays.}
\label{TableClasses}
\end{table}

The final type of operators is a bit different and does not immediately fit in the above classification. First, in general, $\Delta\mathcal{B}$ and $\Delta\mathcal{L}$ dark operators cannot be present simultaneously, since an $X$ exchange would induce proton decay. A way out would be to impose MFV, which forces $\Delta\mathcal{L}$ operators to be proportional to the tiny neutrino masses~\cite{BLMFV}, rendering them irrelevant. The $\Delta\mathcal{B}$ operators are not particularly suppressed under MFV, but they never contribute to the FCNC-induced rare decays considered here. Indeed, these modes trivially conserve $\mathcal{B}$ since an odd number of baryons would be required in the final state. This is not possible for $K$ decays, while the signatures for $B$ decays should be experimentally clear, but are beyond our scope.

If MFV is not imposed and $\Delta\mathcal{B}$ operators are somehow disposed of, then $\Delta\mathcal{L}$ contributions to the $d^{I}\to d^{J}X$ processes are possible. But, the only $\Delta\mathcal{L}\neq0$ invisible final states either involve an odd number of neutrinos, or some $\Delta\mathcal{L}=2$ neutrino pairs $\nu_{L}\nu_{L}$. Since a neutrino field in an effective operator costs $\Lambda^{-3/2}$, these are in general significantly suppressed compared to the operators of $\mathcal{H}_{mat}$ and $\mathcal{H}_{int}$. The only exceptions are those contributing to $d^{I}\to d^{J}\nu_{L}\psi$ or $d^{I}\to d^{J}\nu_{L}\Psi$. As will be discussed in the relevant sections, because $\nu_{L}$ is part of the lepton doublet, these operators are always accompanied by the charge-current transitions $d^{I}\to u^{J}\ell^{-}\psi$ or $d^{I}\to u^{J}\ell^{-}\Psi$, which may offer better windows than the rare FCNC transitions.

\section{Invisible spin-1/2 fermion}

When the new invisible fermion is neutral under the SM gauge group and is produced in pairs, imposing the $SU(3)_C \otimes SU(2)_{L}\otimes U(1)_{Y}$ gauge invariance reduces the basis to the usual ten chiral currents:%
\begin{align}
\mathcal{H}_{mat}^{\mathrm{\bar{\psi}\psi}}  &  =\frac{c_{LL}^{V}}{\Lambda
^{2}}\bar{Q}\gamma_{\mu}Q\times\bar{\psi}_{L}\gamma^{\mu}\psi_{L}%
+\frac{c_{LR}^{V}}{\Lambda^{2}}\bar{Q}\gamma_{\mu}Q\times\bar{\psi}_{R}%
\gamma^{\mu}\psi_{R}+\frac{c_{RL}^{V}}{\Lambda^{2}}\bar{D}\gamma_{\mu}%
D\times\bar{\psi}_{L}\gamma^{\mu}\psi_{L}+\frac{c_{RR}^{V}}{\Lambda^{2}}%
\bar{D}\gamma_{\mu}D\times\bar{\psi}_{R}\gamma^{\mu}\psi_{R}\nonumber\\
&  \;\;\;\;+\frac{c_{LR}^{S}}{\Lambda^{3}}H^{\dagger}\bar{D}Q\times\bar{\psi
}_{L}\psi_{R}+\frac{c_{LL}^{S}}{\Lambda^{3}}H^{\dagger}\bar{D}Q\times\bar
{\psi}_{R}\psi_{L}+\frac{c_{RR}^{S}}{\Lambda^{3}}H\bar{Q}D\times\bar{\psi}%
_{L}\psi_{R}+\frac{c_{RL}^{S}}{\Lambda^{3}}H\bar{Q}D\times\bar{\psi}_{R}%
\psi_{L}\nonumber\\
&  \;\;\;\;+\frac{c_{LL}^{T}}{\Lambda^{3}}H^{\dagger}\bar{D}\sigma_{\mu\nu
}Q\times\bar{\psi}_{R}\sigma^{\mu\nu}\psi_{L}+\frac{c_{RR}^{T}}{\Lambda^{3}%
}H\bar{Q}\sigma_{\mu\nu}D\times\bar{\psi}_{L}\sigma^{\mu\nu}\psi_{R}\;,
\label{Hspin12}%
\end{align}
with the definition $\sigma^{\mu\nu}\equiv i(\gamma^{\mu}\gamma^{\nu}+g^{\mu\nu})$, and where $Q$ ($D$) stands for the left-handed quark doublet (right-handed down-quark singlet). Similar operators can be written down for the up-quark right-handed singlet or for leptonic transitions, and the generalization to a two Higgs-doublet model is straightforward.

The coefficients are not assumed real, and their flavor indices are understood. For example, written in full for the $s\to d$, $b\to s$, and $b\to d$ sectors which concern us here:%
\begin{equation}
\frac{c_{LL}^{V}}{\Lambda^{2}}\bar{Q}\gamma_{\mu}Q\times\bar{\psi}_{L}\gamma^{\mu}\psi_{L}\equiv\frac{c_{LL}^{V,sd}}{\Lambda^{2}}\bar{s}\gamma_{\mu}P_{L}d\times\bar{\psi}_{L}\gamma^{\mu}\psi_{L}+\frac{c_{LL}^{V,bs}}{\Lambda^{2}}\bar{b}\gamma_{\mu}P_{L}s\times\bar{\psi}_{L}\gamma^{\mu}\psi
_{L}+\frac{c_{LL}^{V,bd}}{\Lambda^{2}}\bar{b}\gamma_{\mu}P_{L}d\times\bar{\psi}_{L}\gamma^{\mu}\psi_{L}+h.c.\;.
\label{FlavInd}
\end{equation}
There are only two tensor operators because the identity $2\sigma^{\mu\nu}\gamma_{5}=i\varepsilon^{\mu\nu\alpha\beta}\sigma_{\alpha\beta}$ forbids $\bar{Q}\sigma_{\mu\nu}D\times\bar{\psi}_{R}\sigma^{\mu\nu}\psi_{L}$ and $\bar{D}\sigma_{\mu\nu}Q\times\bar{\psi}_{L}\sigma^{\mu\nu}\psi_{R}$. The dimension-seven operators involving the SM Higgs field are retained because their suppression is only by $v/\Lambda$. By contrast, true dimension-seven operators involving an additional derivative are more severely suppressed by $m_{K,B,\psi}/\Lambda$, and are thus not included.

\begin{table}[t!] \centering
\begin{tabular}[c]{cc|p{0.3cm}|c|cc|c|ccc|c}\hline
\thinspace &  & $d$ & $K_{L}\to XX$ & \multicolumn{2}{|c|}{$K^{i}\to\pi^{j}XX$} & $K_{L}\to\gamma XX$ &
\multicolumn{3}{|c|}{$K^{i}\to\pi^{j}\pi^{k}XX$} & Behavior\\
\multicolumn{2}{c|}{$i,j,...$} &  &  & $L,0$ & $+,+$ &  & $+,+,0$ & $L,+,-$ & $L,0,0$ & in $m_{X}$ for\\\cline{1-10}%
\multicolumn{2}{c|}{$m_{X}^{\max}$} &  & $m_{K}/2$ & \multicolumn{2}{|c|}{$m_{\pi}\;($exp. cut$)$} & $m_{K}/2$ &
\multicolumn{3}{|c|}{$(m_{K}-2m_{\pi})/2$} & $K\to\pi XX$\\\hline
$\dfrac{1}{2}$ & \multicolumn{1}{|c|}{\tpaic{f_{XS}}{f_{XP}}} & $\bar{7}$ & $74$ & $24$ & $19$ & $-$ & $3$ & $6$ & $5$ &
\raisebox{-0.4\height}{\includegraphics[totalheight=1cm]{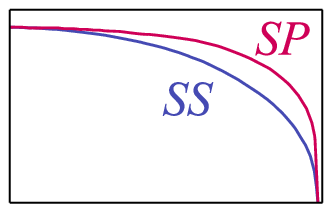}}\\
& \multicolumn{1}{|c|}{\tpaic{f_{VV}}{f_{VA}}} & $6$ & $-$ & $140$ & $98$ & $28$ & $3$ & $4$ & $-$ &
\raisebox{-0.4\height}{\includegraphics[totalheight=1cm]{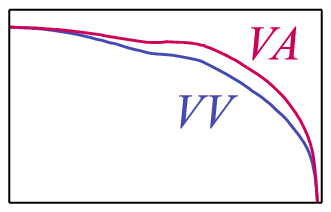}}\\
& \multicolumn{1}{|c|}{\tpaic{f_{AV}}{f_{AA}}} & $6$ & \tpaic{-}{372^{\ast}} & $-$ & $-$ & $-$ & $10$ & $23$ & $18$ & $-$\\
& \multicolumn{1}{|c|}{\tpaic{f_{TT}}{f_{\tilde{T}T}}} & $\bar{7}$ & $-$ & $11$ & $9$ & $11$ & $4$ & $5$ & $2$ &
\raisebox{-0.4\height}{\includegraphics[totalheight=1cm]{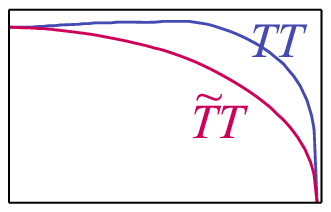}}\\\hline
$0$ & \multicolumn{1}{|c|}{$g_{XS}$} & $\bar{6}$ & $24000$ & $7200$ & $5000$ & $-$ & $380$ & $1100$ & $910$ & \\
& \multicolumn{1}{|c|}{\tpaic{g_{VV}}{g_{AV}}} & 6 & $-$ & \tpaic{99}{-} & \tpaic{70}{-} & \tpaic{20}{-} & \tpaic{2}{7} & \tpaic{3}{16} & \tpair{-}{13} & \raisebox{-0.4\height}{\includegraphics[totalheight=1cm]{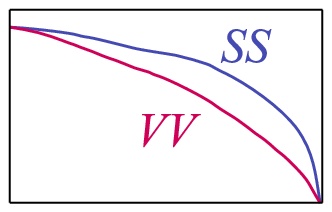}}\\\hline
$1$ & \multicolumn{1}{|c|}{\tpaic{h_{XS}}{h_{XP}}} & $\bar{8}$ & $4.5$ & $1.6$ & $1.3$ & $-$ & $0.3$ & $0.5$ & $0.4$ &
\raisebox{-0.4\height}{\includegraphics[totalheight=1cm]{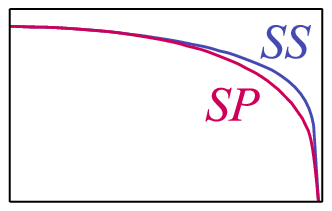}}\\\hline
$\dfrac{3}{2}$ & \multicolumn{1}{|c|}{\tpaic{f_{XS}}{f_{XP}}} & $\bar{9}$ & $0.51$ & $0.19$ & $0.20$ & $-$ & $0.040$ & $0.067$ & $0.063$
& \raisebox{-0.4\height}{\includegraphics[totalheight=1cm]{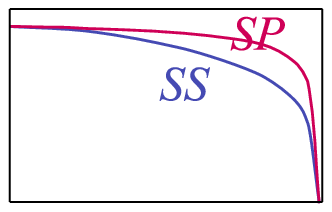}}\\
& \multicolumn{1}{|c|}{\tpaic{f_{VV}}{f_{VA}}} & $8$ & $-$ & $0.12$ & $0.11$ & $0.07$ & $0.014$ & $0.017$ & $-$ &
\raisebox{-0.4\height}{\includegraphics[totalheight=1cm]{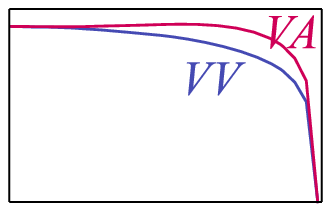}}\\
& \multicolumn{1}{|c|}{\tpaic{f_{AV}}{f_{AA}}} & $8$ & \tpaic{-}{0.32^{\ast}} & $-$ & $-$ & $-$ & $0.027$ & $0.038$ & $0.034$ & $-$\\
& \multicolumn{1}{|c|}{\tpaic{f_{TS}}{f_{TP}}} & $\bar{9}$ & $-$ & $0.10$ & $0.09$ & $0.10$ & $0.034$ & $0.039$ & \multicolumn{1}{l|}{\tpaic{.016^{\ast}}{.021^{\ast}}} & \raisebox{-0.4\height}{\includegraphics[totalheight=1cm]{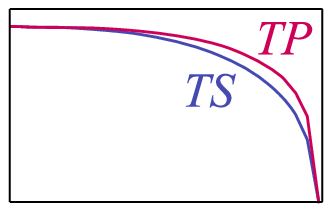}}\\
& \multicolumn{1}{|c|}{\tpaic{f_{TT}}{f_{\tilde{T}T}}} & $\bar{9}$ & $-$ & $0.12$ & $0.10$ & $0.13$ & $0.046$ & $0.053$ & $0.038$
& \raisebox{-0.4\height}{\includegraphics[totalheight=1cm]{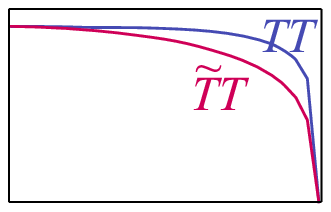}}\\
& \multicolumn{1}{|c|}{\tpaic{f_{\tilde{T}S}}{f_{\tilde{T}P}}} & $\bar{9}$ & $-$ & \tpaic{.08^{\ast}}{.10^{\ast}} & \tpaic{.07^{\ast}}{.08^{\ast}} & $0.10$ & $0.039$ & $0.045$ & $0.033$ & \raisebox{-0.5\height}{\includegraphics[totalheight=1cm]{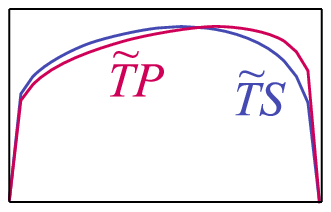}}\\\hline
\end{tabular}%
\caption{Pair production of invisible states: Scales $\Lambda$ (in TeV) accessible for each operator, assuming $10^{-10}$ bounds on all the branching ratios. The operator dimensions $d$ are written with a bar for those involving a Higgs field, and thus a $v/\Lambda$ factor. For the $K\to\pi XX$ modes, the differential rates are integrated over the pion momentum window $[140,195]\cup[211,229]$ MeV, see App. \ref{AppKexp}. The short-hands XS (XP) stand for SS or PS (SP or PP), as appropriate. The quoted $m_{X}^{\max}$ are indicative; the bounds are derived for $m_{X}=0$ or, if the rate vanishes (signaled by (*)), for $m_{X}=100$ MeV. For those which do not vanish, most $V-A$, $S-P$, or $T-\tilde{T}$ degeneracies are lifted when $m_{X}\neq 0$. Still, the dependences of the $\Lambda$s on $m_X$ are rather weak. The bounds stay within an order of magnitude of their values at $m_{X}=0$, except for $m_{X}$ close to the kinematical threshold, where they sharply drop towards zero. This is shown in the last column for $K^+\to\pi^+\nu\bar{\nu}$, where $\Lambda$ (normalized to the value quoted in the table) is plotted against $m_{X}$ over the range $[0,m_{\pi}]$.}
\label{TableBND2}
\end{table}

\begin{table}[t!] \centering
\begin{tabular}
[c]{cl|p{0.3cm}|lll|ll}\hline
&  & $d$ & $B^{0}\to XX$ & $B\to\pi XX$ & $B\to\rho XX$ & $B\to KXX$ & $B\to K^{\ast}XX$\\\hline
\multicolumn{2}{c|}{$m_{X}^{\max}$} &  & $m_{B^{0}}/2$ & $(m_{B}-m_{\pi})/2$ & $(m_{B}-m_{\rho})/2$ & $(m_{B}-m_{K})/2$ & $(m_{B}-m_{K^{\ast}})/2$\\\hline
$1/2$ & \multicolumn{1}{|l|}{$f_{SS,SP}$} & $\bar{7}$ & $-$ & $3.2$ & $-$ & $3.0\,(4.6)$ & $-$\\
& \multicolumn{1}{|l|}{$f_{PS,PP}$} & $\bar{7}$ & $2.7$ & $-$ & $2.1$ & $-$ & $1.7(4.0)$\\
& \multicolumn{1}{|l|}{$f_{VV,VA}$} & $6$ & $-$ & $10$ & $5.3$ & $9.7\,(19)$ & $4.3(15)$\\
& \multicolumn{1}{|l|}{$f_{AV}$} & $6$ & $-$ & $-$ & $7.6$ & $-$ & $6.2(21)$\\
& \multicolumn{1}{|l|}{$f_{AA}$} & $6$ & $6.2^{\ast}$ & $-$ & $7.6$ & $-$ & $6.2(21)$\\
& \multicolumn{1}{|l|}{$f_{TT}$} & $\bar{7}$ & $-$ & $2.6$ & $3.1$ & $2.4\,(3.7)$ & $2.7(6.2)$\\
& \multicolumn{1}{|l|}{$f_{\tilde{T}T}$} & $\bar{7}$ & $-$ & $2.6$ & $3.0$ & $2.4\,(3.7)$ & $2.6(6.0)$\\\hline
0 & \multicolumn{1}{|l|}{$g_{SS}$} & $\bar{6}$ & $-$ & $82$ & $-$ & $75(140)$ & $-$\\
& \multicolumn{1}{|l|}{$g_{PS}$} & $\bar{6}$ & $50$ & $-$ & $46$ & $-$ & $37(130)$\\
& \multicolumn{1}{|l|}{$g_{VV}$} & $6$ & $-$ & $7.2$ & $-$ & $6.8(13)$ & $-$\\
& \multicolumn{1}{|l|}{$g_{AV}$} & $6$ & $-$ & $-$ & $4.7$ & $-$ & $3.9(13)$\\\hline
1 & \multicolumn{1}{|l|}{$h_{SS,SP}$} & $\bar{8}$ & $-$ & $0.61$ & $-$ & $0.56(0.79)$ & $-$\\
& \multicolumn{1}{|l|}{$h_{PS,PP}$} & $\bar{8}$ & $0.56$ & $-$ & $0.42$ & $-$ & $0.36(0.67)$\\\hline
3/2 & \multicolumn{1}{|l|}{$f_{SS,SP}$} & $\bar{9}$ & $-$ & $0.18$ & $-$ & $0.17\,(0.22)$ & $-$\\
& \multicolumn{1}{|l|}{$f_{PS,PP}$} & $\bar{9}$ & $0.18$ & $-$ & $0.13$ & $-$ & $0.10(0.19)$\\
& \multicolumn{1}{|l|}{$f_{VV,VA}$} & $8$ & $-$ & $0.15$ & $0.10$ & $0.14\,(0.19)$ & $0.09(0.17)$\\
& \multicolumn{1}{|l|}{$f_{AV}$} & $8$ & $-$ & $-$ & $0.12$ & $-$ & $0.11(0.20)$\\
& \multicolumn{1}{|l|}{$f_{AA}$} & $8$ & $0.14^{\ast}$ & $-$ & $0.12$ & $-$ & $0.11(0.20)$\\
& \multicolumn{1}{|l|}{$f_{TT,\tilde{T}T}$} & $\bar{9}$ & $-$ & $0.16$ & $0.16$ & $0.15\,(0.19)$ & $0.15(0.24)$\\
& \multicolumn{1}{|l|}{$f_{TS,TP}$} & $\bar{9}$ & $-$ & $0.14$ & $0.12$ & $0.13\,(0.17)$ & $0.11(0.18)$\\
& \multicolumn{1}{|l|}{$f_{\tilde{T}S}$} & $\bar{9}$ & $-$ & $0.11^{\ast}$ & $0.14$ & $0.09^{\ast}\,(0.12^{\ast})$ & $0.12(0.20)$\\
& \multicolumn{1}{|l|}{$f_{\tilde{T}P}$} & $\bar{9}$ & $-$ & $0.13^{\ast}$ & $0.14$ & $0.11^{\ast}\,(0.15^{\ast})$ & $0.12(0.20)$\\\hline
\end{tabular}
\caption{Pair production of invisible states: Scales $\Lambda$ (in TeV) accessible for the various operators with present (future) measurements (see App.~\ref{AppBexp}). The operator dimensions $d$ are written with a bar for those involving a Higgs field, and thus a $v/\Lambda$ factor. We assume $m_{X}=0$ everywhere, except when the rate vanishes in this limit (indicated by *). Specifically, the $B \to XX$ bound on $f_{AA}$, the $B\to\pi(K) XX$ bounds on $g_{T}$, $f_{TS,TP}$, and the $B\to K^{*}XX$ bounds on $g_{T,V,A}$ are derived for $m_{X}\simeq 2$ GeV. The behaviors of the scales $\Lambda$ as functions of $m_{X}$ are shown for $X=\psi$ in Fig. \ref{FigBehaveB}.}%
\label{TableBNDB2}
\end{table}

After the electroweak symmetry breaking, the gauge-invariant operators are rewritten in terms of the (pseudo)scalar, (axial)vector, and (pseudo)tensor currents. This minimizes the interferences between the currents in physical observables, since these operators dominantly produce the invisible fermions in different states. The change of basis is%
\begin{align}
f_{VV,VA}  &  =\frac{c_{LR}^{V}\pm c_{LL}^{V}\pm c_{RL}^{V}+c_{RR}^{V}}{4},\;\;f_{SS,SP}=\frac{v}{\Lambda}\frac{c_{LR}^{S}\pm c_{LL}^{S}\pm c_{RL}^{S}+c_{RR}^{S}}{4}\;,\;f_{TT,\tilde{T}T}=\frac{v}{\Lambda}\frac{c_{RR}^{T}\pm c_{LL}^{T}}{2}\;,\nonumber\\
f_{AV,AA}  &  =\frac{c_{RR}^{V}\pm c_{RL}^{V}\mp c_{LL}^{V}-c_{LR}^{V}}{4},\;f_{PS,PP}=\frac{v}{\Lambda}\frac{c_{RR}^{S}\pm c_{RL}^{S}\mp c_{LL}^{S}-c_{LR}^{S}}{4}\;, \label{Fspin12}%
\end{align}
where $f_{XY}$ tunes $\bar{q}^{I}\Gamma_{X}q^{J}\times\bar{\psi}\Gamma_{Y}\psi$ with $\Gamma_{V,A,S,P,T,\tilde{T}}=\gamma^{\mu},\gamma^{\mu}\gamma_{5},1,\gamma_{5},\sigma_{\mu\nu},\sigma_{\mu\nu}\gamma_{5}$, and flavor indices are understood. The corresponding differential rates are listed in Appendix~\ref{AppKspin12}~(\ref{AppBspin12}) for $K$ ($B$) decays, and the physics reach is summarized in Tables~\ref{TableBND2}~(\ref{TableBNDB2}). Specifically, the entries of these tables are obtained by turning on each $f_{i}$ in turn, while keeping the others to zero. We do not do this at the level of the $c_{i}$ in order to minimize the interferences among the NP contributions. The value of each $|f_{i}|$ is set to one for the vector and axial vector currents, and to $v/\Lambda$ for the others, in order to keep track of the $SU(2)_{L}\otimes U(1)_{Y}$ structure of the underlying operators. Since $K_{L,S}$ are approximate CP eigenstates, the CP phase of the $f_{i}$ must be kept arbitrary. Effectively, we turn on $\operatorname{Im}f_{i}$ and $\operatorname{Re}f_{i}$ to one (or $v/\Lambda$) separately, and the tightest bound (largest $\Lambda$) is indicated in Table~\ref{TableBND2}.

To derive the entries of Tables~\ref{TableBND2} and~\ref{TableBNDB2}, specific experimental bounds on the rare decay branching ratios are used, as detailed in  App.~\ref{AppKexp}~(\ref{AppBexp}) for $K$ ($B$) decays. Note that the scales $\Lambda$ corresponding to tighter or looser experimental bounds are immediately obtained by a simple rescaling of the numbers in Tables~\ref{TableBND2} and~\ref{TableBNDB2}, given the dimensions of the operators~(\ref{Hspin12}) and the definitions~(\ref{Fspin12}).

\begin{figure}[t]
\centering \includegraphics[width=15cm]{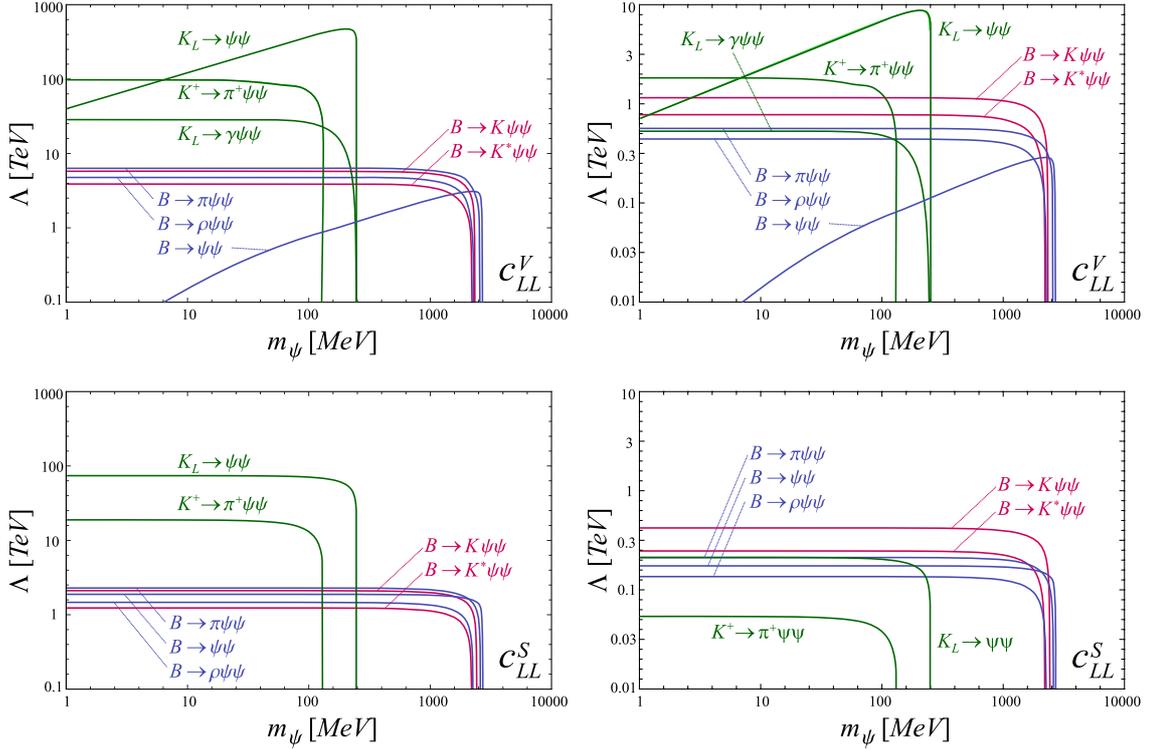}
\caption{Scales $\Lambda$ (in TeV) accessible using the rare $K$ and $B$ decays, for the vector current $c_{LL}^{V}$ and the scalar current $c_{LL}^{S}$. The plots on the right are obtained when MFV is imposed on the Wilson
coefficients.}%
\label{FigKandB}%
\end{figure}

Finally, in Fig.~\ref{FigKandB}, we compare the sensitivity of the various $K$ and $B$ decay modes for two illustrative examples, $c_{LL}^{V}$ and $c_{LL}^{S}$, with and without imposing MFV. In the former case, the coefficients are set to $(c_{LL}^{V})^{IJ}=\lambda^{IJ}$ and $(c_{LL}^{S})^{IJ}=\lambda^{IJ}m_{d^{I}}/v$, see Eq.~(\ref{MFVscaling}). Actually, to avoid dragging a relative factor $m_{b}/v$, the scales $\Lambda$ are plotted taking a two Higgs doublet model of type II at large $\tan\beta = v_u/v_d$ (where $v_u$ and $v_d$ are the two Higgs vacuum expectation values), so that $m_{b}/v_{d}\approx1$ and $(c_{LL}^{S})^{sd}$ is simply suppressed by $m_{s}/m_{b}$. As can be seen in Fig.~\ref{FigKandB}, this chirality flip is expensive in the $K$ sector, and pulls the $K\to\pi\psi\bar{\psi}$ constraint an order of magnitude below those from rare $B$ decays. By contrast, for the vector current, similar constraints are drawn from rare $K$ and $B$ decays when MFV is imposed.%

\begin{figure}[t]
\centering \includegraphics[width=0.95\textwidth]{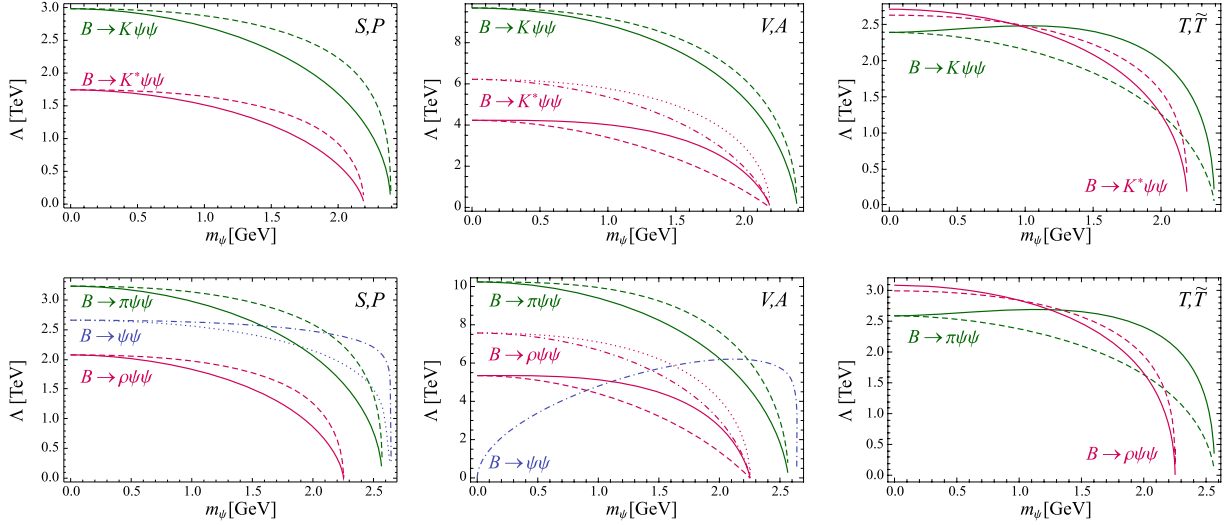}
\caption{Evolution of the scales $\Lambda$ associated with the pair production of invisible fermions as a function of the fermion mass. The values at $m_{\psi}=0$ correspond to those quoted in Table~\ref{TableBNDB2}, except for $B\to\psi\bar{\psi}$, given at $m_{\psi}=2$ GeV. Note that these $B\to(K^{(\ast)},\rho,\pi)\psi\bar{\psi}$ bounds assume flat experimental acceptances and full phase-space coverage, which is not true in the present experimental analyses, see App.~\ref{AppBexp}.}%
\label{FigBehaveB}%
\end{figure}

Let us now turn to the operators not involving the SM fermions, but rather the SM gauge and Higgs fields. Using these fields EOM, as well as partial integration and identities like $2\sigma^{\mu\nu}\gamma_{5}=i\varepsilon^{\mu\nu\alpha\beta}\sigma_{\alpha\beta}$, the leading operators reduce to%
\begin{align}
\mathcal{H}_{int}^{\mathrm{\bar{\psi}\psi}} &  =\frac{c_{RL}^{B}}{\Lambda
}B_{\mu\nu}\times\bar{\psi}_{R}\sigma^{\mu\nu}\psi_{L}+\frac{c_{LR}^{B}}{\Lambda}B_{\mu\nu}\times\bar{\psi}_{L}\sigma^{\mu\nu}\psi_{R}+\frac{c_{RL}^{H}}{\Lambda}H^{\dagger}H\times\bar{\psi}_{R}\psi_{L}+\frac{c_{LR}^{H}}{\Lambda}H^{\dagger}H\times\bar{\psi}_{L}\psi_{R}\nonumber\\
&  \;\;\;\;+\frac{c_{LL}^{H}}{\Lambda^{2}}iH^{\dagger}\overleftrightarrow{\mathcal{D}}_{\mu}H\times\bar{\psi}_{L}\gamma^{\mu}\psi_{L}+\frac{c_{RR}^{H}}{\Lambda^{2}}iH^{\dagger}\overleftrightarrow{\mathcal{D}}_{\mu}H\times\bar{\psi}_{R}\gamma^{\mu}\psi_{R}\nonumber\\
&  \;\;\;\;+\frac{c_{RL,i}^{G}}{\Lambda^{3}}Q_{i}^{G}\times\bar{\psi}_{R}\psi_{L}+\frac{c_{LR,i}^{G}}{\Lambda^{3}}Q_{i}^{G}\times\bar{\psi}_{L}\psi_{R}\nonumber\\
&  \;\;\;\;+\frac{c_{RL}^{HB}}{\Lambda^{3}}(H^{\dagger}H)B_{\mu\nu}\times\bar{\psi}_{R}\sigma^{\mu\nu}\psi_{L}+\frac{c_{LR}^{HB}}{\Lambda^{3}}(H^{\dagger}H)B_{\mu\nu}\times\bar{\psi}_{L}\sigma^{\mu\nu}\psi_{R}\nonumber\\
&  \;\;\;\;+\frac{c_{RL}^{HW}}{\Lambda^{3}}(H^{\dagger}\sigma^{i}H)W_{\mu\nu}^{i}\times\bar{\psi}_{R}\sigma^{\mu\nu}\psi_{L}+\frac{c_{LR}^{HW}}{\Lambda^{3}}(H^{\dagger}\sigma^{i}H)W_{\mu\nu}^{i}\times\bar{\psi}_{L}\sigma^{\mu\nu}\psi_{R}\;,\label{Spin12Gauge}%
\end{align}
where $\overleftrightarrow{\mathcal{D}}_{\mu}=\overleftarrow{\mathcal{D}}_{\mu}-\overrightarrow{\mathcal{D}}_{\mu}$ and $Q_{i}^{G}$ stand for each of the following gauge-invariant combinations of SM fields%
\begin{equation}
Q_{i}^{G}=\mathcal{D}^{\mu}H^{\dagger}\mathcal{D}_{\mu}H,\;(H^{\dagger}H)^{2},\;B_{\mu\nu}B^{\mu\nu},\;W_{\mu\nu}^{i}W_{i}^{\mu\nu},\;G_{\mu\nu}^{a}G_{a}^{\mu\nu},\;B_{\mu\nu}\tilde{B}^{\mu\nu},\;W_{\mu\nu}^{i}\tilde{W}_{i}^{\mu\nu},\;G_{\mu\nu}^{a}\tilde{G}_{a}^{\mu\nu}\;.\label{Qgauge}%
\end{equation}
But for the last three CP-odd combinations, the $Q_{i}^{G}$ monomials are precisely the dimension-four gauge and Higgs couplings of the SM Lagrangian. Operators with partial derivatives acting on $\psi$ are systematically discarded.

The leading dimension-five operators involve either $B_{\mu\nu}$ or $H^{\dagger}H$. The $B_{\mu\nu}$ operators effectively assign a (mass-dependent) hypercharge to the invisible fermions, since under partial integration
\begin{equation}
B_{\mu\nu}\times\bar{\psi}_{R}\sigma^{\mu\nu}\psi_{L}=2iB^{\mu}\times\bar{\psi}_{R}\overleftrightarrow{\mathcal{D}}_{\mu}\psi_{L}+2m_{\psi}B_{\mu}\times (\bar{\psi}_{L}\gamma^{\mu}\psi_{L}+\bar{\psi}_{R}\gamma^{\mu}\psi_{R})\;.
\label{milicharge}%
\end{equation}
With $m_{\psi}/\Lambda\ll1$ but otherwise arbitrary, the second term describes millicharged fermions, a scenario to be discussed later on. The $H^{\dagger}H$ operators effectively correct the $\psi$ mass after the electroweak symmetry breaking, with (setting $c_{LR,RL}^H$ to one)
\begin{equation}
\delta m_{\psi}=v^{2}/\Lambda\;.
\end{equation}
Taken at face value, naturality would thus prefer $\Lambda>24\,(240)$ TeV if $m_{\psi}\approx\delta m_{\psi}<m_{B}/2$ $(m_{K}/2)$, as required to produce these states in rare $B$ ($K$) decays. As seen in Tables~\ref{TableBND2} and~\ref{TableBNDB2}, this would push the NP effects from the FCNC operators~(\ref{Hspin12}) just beyond accessibility. Of course, these numbers are only indicative, a specific model need not produce these dimension-five operators, and even if generated, a sizeable deviation from naturality cannot be ruled out.

All the other $\mathcal{H}_{int}^{\mathrm{\bar{\psi}\psi}}$ operators are, for our purpose, only marginally relevant. As explained in Sec.~\ref{Classification}, they reduce to the four-fermion operators of Eq.~(\ref{Hspin12}) once Higgs and gauge fields are coupled to quarks, with the four-fermion Wilson coefficients obeying the SM scalings~(\ref{CKMscale}). For example, the dimension-six operators with $H^{\dagger}\overleftrightarrow{\mathcal{D}}_{\mu}H$ reduce, after the electroweak symmetry breaking, to an effective $Z$ coupling to $\psi$ which can be treated in perfect analogy to the SM $Z$ coupling to $\nu\bar{\nu}$ (see Fig.~\ref{FigZpeng}). The rare decays proceed through the flavor-changing hadronic $Z$ penguin, and the four-fermion operators scale like $c_{LL(RR)}^{IJ}\sim gc_{LL(RR)}^{H}k^{IJ}$ with $k^{IJ}$ given in Eq.~(\ref{CKMscale}).

The final class of operators involves only one dark fermion field, and thus, from Lorentz invariance, violates either lepton ($\mathcal{L}$) or baryon ($\mathcal{B}$) number:%
\begin{align}
\mathcal{H}_{\Delta\mathcal{B},\Delta\mathcal{L}}^{\mathrm{\psi}}  &
=c_{0}^{\Delta\mathcal{L}}H\times\bar{\psi}_{R}L+\frac{c_{1}^{\Delta\mathcal{L}}}{\Lambda^{2}}(H^{\dagger}H)H\times\bar{\psi}_{R}L+\frac{c_{B}^{\Delta\mathcal{L}}}{\Lambda^{2}}B_{\mu\nu}H\times\bar{\psi}_{R}\sigma^{\mu\nu}L+\frac{c_{W}^{\Delta\mathcal{L}}}{\Lambda^{2}}W_{\mu\nu}^{i}H\sigma_{i}\times\bar{\psi}_{R}\sigma^{\mu\nu}L\nonumber\\
&  \;\;\;\;+\frac{c_{2}^{\Delta\mathcal{L}}}{\Lambda^{2}}\bar{E}L\times\bar{\psi}_{R}L+\frac{c_{3}^{\Delta\mathcal{L}}}{\Lambda^{2}}\bar{D}Q\times\bar{\psi}_{R}L+\frac{c^{\Delta\mathcal{B}}}{\Lambda^{2}}\bar{D}D^{C}\times\bar{\psi}_{R}U^{C}+h.c.\;. \label{Spin12BL}%
\end{align}
For simplicity, the possible Dirac, $SU(2)_{L}$, and flavor structures are understood, as well as the operators with $\psi_{R}\to\psi_{L}^{C}$. Operators with three dark fermion fields are at least of dimension seven, hence not included.

As explained in Sec.~\ref{Classification}, the phenomenology of the $\Delta\mathcal{B}$ or $\Delta\mathcal{L}$ operators is different from that induced by those of $\mathcal{H}_{mat}^{\mathrm{\bar{\psi}\psi}}$ and $\mathcal{H}_{int}^{\mathrm{\bar{\psi}\psi}}$. Since $c^{\Delta\mathcal{B}}$ and $c_{i}^{\Delta\mathcal{L}}$ cannot be simultaneously large as a tree-level $\psi$ exchange would induce proton decay, and since $c^{\Delta\mathcal{B}}$ cannot be probed with the rare FCNC-induced decays considered here, let us concentrate on the $\Delta\mathcal{L}$ operators. The renormalizable $c_{0}^{\Delta\mathcal{L}}$ interaction, as well as $c_{1}^{\Delta\mathcal{L}}$ after the electroweak symmetry breaking, mix $\psi$ with $\nu_{L}$. As a result, these operators are bounded by neutrino masses, $\psi_{R}$ behaves effectively as a right-handed neutrino, and the $c_{2,3}^{\Delta\mathcal{L}}$ effective interactions include both FCNC and charged-current interactions. Since the renormalizable operator (the so-called neutrino portal) is not suppressed by the NP scale, its coefficient $c_{0}^{\Delta\mathcal{L}}$ must be tiny. So, the effective interactions tuned by $c_{2,3}^{\Delta\mathcal{L}}$ can be sizeable only if the NP dynamics responsible for such a suppression does not apply equally to all the $\Delta\mathcal{L}$ operators. 

In this respect, the semileptonic operator tuned by $c_{3}^{\Delta\mathcal{L}}$ may be the most likely to escape such a suppression. It is the only one directly accessible with the quark FCNC transitions. However, the processes induced by the charged current component of the $c_{3}^{\Delta\mathcal{L}}$ operator, $\bar{d}^{I}(1-\gamma_{5})u^{J}\times\bar{\psi}(1-\gamma_{5})\ell^{K}$, may give tighter constraints. Indeed, since there is no possible interference with the SM contributions, this operator enhances all the kinematically-allowed (semi-)leptonic decays of the $K^{+}$, $B^{+}$, and $D^{+}$, as well as hadronic $\tau$ decays. In particular, the most interesting channels are the $K^{+},B^{+},D^{+}\to\ell^{+}\psi$ transitions, whose rates are%
\begin{equation}
\Gamma^{NP}(P\to\ell^{K}\psi)=\frac{m_{P}\lambda(1,r_{\ell}%
^{2},r_{\psi}^{2})^{1/2}(1-r_{\ell}^{2}-r_{\psi}^{2})}{64\pi}\left(
\frac{f_{P}}{m_{d^{I}}+m_{u^{J}}}\right)  ^{2}\left(  \frac{m_{P}^{2}}%
{\Lambda^{2}}(c_{3}^{\Delta\mathcal{L}})^{IJK}\right)  ^{2}\;,
\label{semiLept}
\end{equation}
where $P=B,K,D$ for $I,J=(3,1),(2,1),(1,2)$, $r_{i}=m_{i}/m_{P}$, $f_{P}$ is the $P^{\pm}$ decay constant (defined as in Ref.~\cite{PDG}), and the kinematical function $\lambda$ is defined in Eq.~(\ref{LKin}). Indeed, the SM charged-current Fermi operators are vectorial, hence these transitions are helicity-suppressed in the SM (proportional to $m_{\ell^K}$). On the contrary, the relative strengths of the $P^{+}\to \ell^{+}\psi$ transitions for $\ell = e,\mu,\tau$ depend only on that between $(c_{3}^{\Delta\mathcal{L}})^{IJK}$, $K=1,2,3$. For example, if $(c_{3}^{\Delta\mathcal{L}})^{IJK}$ is independent of $K$, the NP effects would be most easily seen for $e\psi$ final states.

Such an enhancement may be welcome in the $B$ sector. The persistent tension between the Belle and BaBar measurements of $\mathcal{B}(B\to \tau \nu) = (1.68 \pm 0.31)\cdot10^{-4}$ ~\cite{Ikado:2006un, :2008gx, Hara:2010dk, :2010rt} and the predictions of the global CKM fits within the SM (see Ref.~\cite{Lenz:2010gu} for a recent review) can be addressed provided $m_\psi<3.5$~GeV. Assuming $(c_3^{\Delta \cal L})^{313}\approx 1$, the reconciliation of the discrepancy of the order $\Delta \mathcal{B}(B\to\tau X) \simeq 10^{-4}$ would point towards $\Lambda\lesssim 5$~TeV, where the equality is reached for  $m_\psi=0$. Note, though, that a non-universal lepton flavor structure in $(c_{3}^{\Delta\mathcal{L}})^{IJK}$ is necessary, in order to circumvent the current bounds from $B\to e\nu$ and $B\to \mu\nu$ at the level of $10^{-6}$~\cite{PDG}.
 
Actually, if the $c_{3}^{\Delta\mathcal{L}}$ couplings are fully universal in their quark and lepton indices (i.e. $(c_{3}^{\Delta\mathcal{L}})^{IJK}\sim\mathcal{O}(1)$ for all $I,J,K$), and if $m_{\psi}\ll m_K$, the best constraints currently come from the $K_{\ell2}$ universality test\footnote{The $\pi_{\ell2}$ universality test is also constraining when $m_{\psi} < m_{\pi}$ (we thank D. Bryman for pointing this out). Up to trivial substitutions, $R_{\pi}$ is given by Eq.~(\ref{univlept}) even when $\pi\to\mu\psi$ is kinematically forbidden. A future measurement of $R_{\pi}$ at the $10^{-3}$ level~\cite{Pil2} would give $\Lambda\gtrsim67$ TeV. Though lower than from $R_{K}$, the two are equally sensitive if $(c_{3}^{\Delta\mathcal{L}})^{suK}/(c_{3}^{\Delta\mathcal{L}})^{duK}\sim V_{us}/V_{ud}$.}, where the NP effect gets enhanced by the small electron mass:%
\begin{align}
R_{K}^{\exp}  &  =\frac{\Gamma^{SM}(K\to e\nu_{e})+\Gamma^{NP}(K\to e\psi)}{\Gamma^{SM}(K\to\mu\nu_{\mu})+\Gamma^{NP}(K\to\mu\psi)}\nonumber\\ &  \approx R_{K}^{SM}\left(  1+\frac{2(1-r_{\psi}^{2})^{2}}{r_{e}^{2}}\left( \frac{m_{K}}{m_{s}}\frac{c_{3}^{\Delta\mathcal{L}}}{g^{2}|V_{us}|}\frac{M_{W}^{2}}{\Lambda^{2}}\right) ^{2}\right)  \approx R_{K}^{SM}\left( 1+(1-r_{\psi}^{2})^{2}\left(  \frac{22\;\text{TeV}}{\Lambda}\right)^{4}\right)  \;. \label{univlept}
\end{align}
From $R_{K}^{SM}=2.477(1)\cdot10^{-5}$~\cite{RKSM} and $R_{K}^{\exp}=2.487(13)\cdot10^{-5}$~\cite{UTNA62}, we require $R_{K}^{\exp}-R_{K}^{SM}\lesssim0.013\cdot10^{-5}$ which translates as $\Lambda\gtrsim82$ TeV for $m_{\psi}\ll m_{K}$. This is comparable to the scales probed with the FCNC modes.

\subsubsection*{Some models with light weakly coupled fermions}

The operator basis~(\ref{Hspin12}) can be used to describe the SM \textbf{transitions to neutrino pairs}, $\psi_{L}=\nu_{e},\nu_{\mu},\nu_{\tau}$, $\psi_{R}=0$, by setting all the $c_{i}$ to zero but for $c_{LL}^{V}$. The SM rates for $K$ decays are given in App.~\ref{AppSMrates}, and those for $B$ decays in App.~\ref{AppBSMrates}. With $\psi_{R}=0$, all but the vector FCNC operators $c_{LL}^{V}$ and $c_{RL}^{V}$ drop out. Since most NP models do not modify the particle spectrum below the electroweak scale, their effect simply enter as corrections to these two coefficients~\cite{SomeNP,Altmannshofer:2009ma}, and lead to the same vector-like spectrum as the SM contribution. Note that contrary to true dark fermions, the neutrinos are not neutral under $SU(2)_{L}\otimes U(1)_{Y}$, and thus the basis~(\ref{Hspin12}) is not complete since it is lacking the charged current semileptonic operators. Said differently, if neutrinos had not yet been discovered, charge-current interactions would offer better windows.

The presence of a light \textbf{right-handed (sterile) neutrino} corresponds to $\psi_{R}=\nu_{R}$, $\psi_{L}=0$ in Eqs.~(\ref{Hspin12}) and~(\ref{Spin12BL}). Of course, once $\psi_{R}$ is identified with $\nu_{R}$, it can be given a charge under $U(1)_{\mathcal{L}}$ so that all these operators become $\Delta\mathcal{L}=0$. Some NP dynamics is nevertheless needed to couple $\nu_{R}$, a gauge-singlet, to the quark currents. This can arise in the $\nu$MSM model~\cite{nMSM}, based on the $c_{0}^{\Delta\mathcal{L}}$ coupling and with the quark flavor transition induced by the SM weak interaction. Alternatively, in some left-right symmetric models~\cite{KiyoMT98}, the $c_{4}^{\Delta\mathcal{L}}$ operator can arise from combined $W_{L}$ and $W_{R}$ boxes. Note that these two examples are matched onto the operators of Eq.~(\ref{Spin12BL}), for which lepton universality tests are competitive with FCNC decays, as our analysis of the previous section shows.

In the MSSM, the\textbf{ lightest neutralino} $\chi_{1}$ is another electrically neutral weakly interacting particle which could be pair produced in rare decays. Though this particle is not neutral under the SM gauge group, it must be produced in pairs when R parity is conserved. Because the neutralino is a Majorana fermion obeying%
\begin{equation}
\bar{\psi}^{M}\gamma^{\mu}\psi^{M}=\bar{\psi}^{M}\sigma^{\mu\nu}\psi^{M}=0\;,
\end{equation}
the corresponding operators disappear. In the general MSSM, there are then tree-level processes contributing to the combinations $c_{LL}^{V}+c_{LR}^{V}$, $c_{RL}^{V}+c_{RR}^{V}$, and to the scalar and pseudoscalar currents $c_{XY}^{S}$, $X,Y=L,R$. The flavor transitions are tuned by the off-diagonal entries of the down squark $LL$, $RR$, and $LR$ mass terms, and the overall scale $\Lambda$ is set by the exchanged down squark mass. This (Class II) scenario was analyzed in detail in Ref.~\cite{Neutralinos}, to which we refer for more information.

Another possible new fermion is the \textbf{axino} $\tilde{a}$, the fermionic superpartner of the axion~\cite{Axinos}. Depending on the model, it could be light enough to be produced in $K$ and $B$ decays. Note, though, that its Lagrangian couplings are flavor-blind, and further, involve the superpartners of the SM particles when $R$ parity is conserved. So, the effective interactions are not only suppressed by the large scale $\Lambda=f_{a}$, but also by the sparticle mass scale, loop factors, and the already tightly constrained flavor-violation occurring in the squark sector. We will not consider this scenario further because, with scales $f_{a}$ above $10^{6}$ TeV~\cite{Axinos} (or even much higher in some models), the fermionic operators are far too suppressed, and signals should be more readily accessible using the single scalar axion production discussed in the next section.

A final example is the dark sector \textbf{millicharged fermion} $\psi_{\varepsilon}$~\cite{Millicharged}. Typically, these fermions arise when there is a new dark $U(1)$ field $V^{\mu}$ coupled to the SM $U(1)_Y$ through the kinetic mixing $\varepsilon B^{\mu\nu}V_{\mu\nu}$, as well as to some new fermion states initially neutral under the SM gauge group~\cite{Holdom}. After diagonalizing the two $U(1)$ fields, the dark sector fermions end up coupled to the SM photon, but with an arbitrary electric charge $\varepsilon e$. Alternatively, this scenario could follow from the $B_{\mu\nu}$ operator in Eq.~(\ref{milicharge}).

When $m_{\psi_{\varepsilon}}\lesssim m_{e}$, various very tight cosmological and astrophysical bounds hold on $\varepsilon$~\cite{Millicharged}, ruling out any signal in meson decays. On the contrary, for $m_{\psi_{\varepsilon}}\sim\mathcal{O}(100-1000$ MeV), the bounds are much less tight, with $\varepsilon$ as large as $10^{-2}$ still possible. Since $\psi_{\varepsilon}$ couples to quarks through the photon field, its coupling is flavor-blind (Class III or IV of Table~\ref{TableClasses}). The bounds in Table~\ref{TableBND2} are adapted to this scenario by setting $\Lambda=M_{W}$ and rescaling the Wilson coefficients as $c^{IJ}\to\varepsilon e^{2}k^{IJ}$ with $k^{IJ}$ given in Eq.~(\ref{CKMscale}). Alternatively, the physics reach can be more accurately estimated by looking at the $K$ and $B$ radiative modes with a Dalitz pair, and rescaling their branching ratio by $\varepsilon^{2}$,%
\begin{equation}
\mathcal{B}(P\to P^{\prime}\bar{\psi}_{\varepsilon}\psi_{\varepsilon
})\approx\varepsilon^{2}\times\mathcal{B}(P\to P^{\prime}\ell^{+}%
\ell^{-})\;.
\end{equation}
Up to simple phase-space corrections accounting for $m_{\ell}\neq m_{\psi_{\varepsilon}}$, this is valid as long as the $Z$ boson does not dominate. From this, no $B$ decay rate appears large enough to reach the interesting $\varepsilon\lesssim10^{-3}$ region. For example, with $\mathcal{B}(B\to (K,K^{\ast})\ell^{+}\ell^{-})$ in the $10^{-7}$ range, a bound on $\mathcal{B}(B\to(K,K^{\ast})\bar{\psi}_{\varepsilon}\psi_{\varepsilon})$ at the $10^{-11}$ level would be needed, far below the experimental prospects. Similarly, in the $K$ sector, assuming about $10^{13}$ kaon decays, the only mode for which one could theoretically reach down to $\varepsilon\approx10^{-3}$ is $K_{L}\to\gamma\bar{\psi}_{\varepsilon}\psi_{\varepsilon}$, since $\mathcal{B}(K_{L}\to\gamma e^{+}e^{-})=(9.4\pm0.4)\cdot10^{-6}$~\cite{PDG}.

So, the rare FCNC decays do not appear competitive when $\psi_{\varepsilon}$ couples to ordinary matter exclusively through the photon. But, let us stress that if the couplings to matter of the SM and dark $U(1)$s are not perfectly aligned, the CKM suppression of Eq.~(\ref{CKMscale}) may be evaded, and rare decays would become prime sources of information. Also, as we will see in Sec.~\ref{Vectors}, in case the dark photon has a non-zero mass, it may be directly produced and competitive constraints can be derived.

\section{Invisible spin-0 boson}

If there is a light scalar particle neutral under the SM gauge group and under any dark gauge symmetry, it can be produced alone. The simplest effective operators are then of dimension five%
\begin{equation}
\mathcal{H}_{mat}^{\mathrm{\phi}}=\frac{c_{VL}^{\phi}}{\Lambda}\bar{Q}%
\gamma_{\mu}Q\times\partial^{\mu}\phi+\frac{c_{VR}^{\phi}}{\Lambda}\bar
{D}\gamma_{\mu}D\times\partial^{\mu}\phi+\frac{c_{SL}^{\phi}}{\Lambda
}H^{\dagger}\bar{D}Q\times\phi+\frac{c_{SR}^{\phi}}{\Lambda}H\bar{Q}%
D\times\phi\;. \label{Axion}%
\end{equation}
No operator involving the tensor current or the SM field strengths arises, as these would require more derivatives and thus would be suppressed by additional factors of $\mathcal{O}(m_{K,B}/\Lambda)$. Actually, the first two operators (as well as $(\bar{Q}\!\not\!\mathcal{D}Q)\phi$, not explicitly included) collapse to the third and fourth upon using the tree-level quark EOM~\cite{BuchmullerW86}%
\begin{equation}
i\!\not\!\mathcal{D}Q=\mathbf{Y}_{u}^{\dagger}UH^{\ast}+\mathbf{Y}_{d}^{\dagger}DH\;,\;i\!\not\!\mathcal{D}U=\mathbf{Y}_{u}QH^{\ast\dagger}\;,\;i\!\not\!\mathcal{D}D=\mathbf{Y}_{d}QH^{\dagger}\;, \label{EOM}%
\end{equation}
i.e., at the cost of the chirality flips
\begin{equation}
i(c_{SL}^{\phi})^{IJ}=(\mathbf{Y}_{d})^{II}(c_{VL}^{\phi})^{IJ}-(c_{VR}^{\phi})^{IJ}(\mathbf{Y}_{d})^{JJ},\;\;i(c_{SR}^{\phi})^{IJ}=(c_{VR}^{\phi})^{IJ}(\mathbf{Y}_{d}^{\dagger})^{II}-(c_{VL}^{\phi})^{IJ}(\mathbf{Y}_{d}^{\dagger})^{JJ}\;.
\label{SubsAxion}
\end{equation}
Though in the rest of the paper, the EOM are always enforced, we prefer to keep all three operators here because they correspond to well-defined scenarios. On one hand, derivative couplings are characteristic of non-linearly realized symmetries, while on the other hand, the $H^{\dagger}\bar{D}Q\times\phi$ operator is effectively a dimension-four Yukawa-like coupling after the electroweak symmetry breaking.

If the scalar field is charged under some dark sector symmetry, or if the $\mathbb{Z}_{2}$ symmetry under $\phi\to-\phi$ is imposed, it must be produced in pairs. The simplest effective operators are then of dimension six:%
\begin{equation}
\mathcal{H}_{mat}^{\mathrm{\phi\phi}}=\frac{c_{VL}^{\phi\phi}}{\Lambda^{2}}%
\bar{Q}\gamma^{\mu}Q\times i\phi^{\dagger}\overleftrightarrow{\partial}_{\mu
}\phi+\frac{c_{VR}^{\phi\phi}}{\Lambda^{2}}\bar{D}\gamma^{\mu}D\times
i\phi^{\dagger}\overleftrightarrow{\partial}_{\mu}\phi+\frac{c_{SL}^{\phi\phi}%
}{\Lambda^{2}}H^{\dagger}\bar{D}Q\times\phi^{\dagger}\phi+\frac{c_{SR}^{\phi\phi}%
}{\Lambda^{2}}H\bar{Q}D\times\phi^{\dagger}\phi\;.
\label{H2scalars}%
\end{equation}
The minus sign of $\overleftrightarrow{\partial}$ in the first two operators is required as the plus combinations reduce to the third and fourth operators by partial integration and use of the quark EOM~(\ref{EOM}). Tensor currents are not included since they are suppressed by a factor of $\mathcal{O}(m_{K,B}/\Lambda)$.

Bounds are derived on the operators involving quark currents of definite $C$ and $P$, tuned by the couplings%
\begin{equation}
g_{S,P}=\frac{v}{\Lambda}\frac{c_{SR}^{\phi}\pm c_{SL}^{\phi}}{2}\;,\;
g_{V,A}=\frac{c_{VR}^{\phi}\pm c_{VL}^{\phi}}{2}\;,\;
g_{SS,PS}=\frac{v}{\Lambda}\frac{c_{SR}^{\phi\phi}\pm c_{SL}^{\phi\phi}}{2}\;,\;
g_{VV,AV}=\frac{c_{VR}^{\phi\phi}\pm c_{VL}^{\phi\phi}}{2}\;,
\end{equation}
where flavor indices are understood\footnote{Note that the $c_{VR,VL}^{\phi(\phi)}$ couplings are hermitian matrices in flavor space (see Eq.~(\ref{FlavInd})), while $(c_{SL}^{\phi(\phi)\ast})^{JI} =(c_{SR}^{\phi(\phi)})^{IJ}$.}. The corresponding differential rates are listed in Appendix~\ref{AppKspin0}~(\ref{AppBspin0}) for $K$ ($B$) decays. The physics reach is summarized in Tables~\ref{TableBND1} and~\ref{TableBNDB1} for one scalar in the final state, and in Tables~\ref{TableBND2} and~\ref{TableBNDB2} for two scalars. As expected from Table~\ref{Reach}, the lower dimensionality of the operators~(\ref{Axion}) and~(\ref{H2scalars}) translates as much higher accessible scales compared to spin 1/2 final states.

The simplest effective operators involving Higgs and gauge fields are
\begin{align}
\mathcal{H}_{int}^{\mathrm{\phi,3\phi}}  &  =\mu^{\prime}\;H^{\dagger}%
H\times\phi+\frac{c_{i}^{\phi,G}}{\Lambda}Q_{i}^{G}\times\phi+\frac{c_{i}%
^{3\phi}}{\Lambda}H^{\dagger}H\times\phi^{3}\;, \label{HScalarGauge1} \\
\mathcal{H}_{int}^{\mathrm{\phi\phi,4\phi}}  &  =\lambda^{\prime}\;H^{\dagger
}H\times\phi^{\dagger}\phi+\frac{c_{i}^{\phi\phi,H}}{\Lambda^{2}}H^{\dagger
}\overleftrightarrow{\mathcal{D}}_{\mu}H\times \phi^{\dagger}\overleftrightarrow
{\partial}\hspace{0in}^{\mu}\phi+\frac{c_{i}^{\phi\phi,G}}{\Lambda^{2}}%
Q_{i}^{G}\times\phi^{\dagger}\phi+\frac{c_{i}^{4\phi}}{\Lambda^{2}}H^{\dagger
}H\times(\phi^{\dagger}\phi)^{2}\;,\label{HScalarGauge2}
\end{align}
where $Q_{1-8}^{G}$ are defined in Eq.~(\ref{Qgauge}). Only operators of dimension less or equal to those in Eqs.~(\ref{Axion}) and~(\ref{H2scalars}) are considered. The possibility to write renormalizable couplings between $\phi$ and $H$ embodies the so-called Higgs portal. One consequence is a mixing between $\phi$ and $H$, which has already been investigated in details, see e.g. Ref.~\cite{Portals}, and will not be considered further here (but for a comment on $\lambda^{\prime}$ in the next section). The other operators are a priori subleading compared to the Higgs portal, simply because of their higher dimensions. Note, though, that when $\phi$ is pseudoscalar, most of the $\mathcal{H}_{int}^{\mathrm{\phi,3\phi}}$ couplings drop out in the CP limit, leaving only those to $F_{\mu\nu}\tilde{F}^{\mu\nu}$ with $F_{\mu\nu}$ any one of the SM field strengths.

\begin{table}[t] \centering
\begin{tabular}[c]{cc|p{0.3cm}|cc|c|ccc}\hline
&  & $d$ & \multicolumn{2}{|c|}{$K^{i}\to\pi^{j}X$} & $K_{L}\to\gamma X$ & \multicolumn{3}{|c}{$K^{i}\to\pi^{j}\pi^{k}X$}\\
\multicolumn{2}{c|}{$i,j,...$} &  & $L,0$ & $+,+$ &  & $+,+,0$ & $L,+,-$ & $L,0,0$\\\hline
\multicolumn{2}{c|}{$m_{X}^{\max}$} &  & \multicolumn{2}{|c|}{$2m_{\pi}\;($exp. cut$)$} & $m_{K}$ & \multicolumn{3}{|c}{$m_{K}-2m_{\pi}$}\\\hline
0 & \multicolumn{1}{|c|}{$g_{S}$} & $\bar{5}$ & $3.0\cdot10^{12}$ & $1.5\cdot10^{12}$ & $-$ & $-$ & $-$ & $-$\\
  & \multicolumn{1}{|c|}{$g_{P}$} & $\bar{5}$ & $-$ & $-$ & $-$ & $0.8\cdot10^{11}$ & $1.2\cdot10^{11}$ & $0.3\cdot10^{11}$\\
  & \multicolumn{1}{|c|}{$g_{V}$} & $5$ & $1.2\cdot10^9$ & $0.6\cdot10^9$ & $-$ & $-$ & $-$ & $-$\\
  & \multicolumn{1}{|c|}{$g_{A}$} & $5$ & $-$ & $-$ & $-$ & $3.6\cdot10^7$ & $5.4\cdot10^7$ & $1.2\cdot10^7$\\\hline
1 & \multicolumn{1}{|c|}{$h_{T}$} & $\bar{6}$ & $8.2\cdot10^3\,^{\ast}$ & 
   $5.8\cdot10^3\,^{\ast}$ & $7.6\cdot10^3$ & $2.3\cdot10^3$ & $3.2\cdot10^3$ & $-$\\
  & \multicolumn{1}{|c|}{$h_{\tilde{T}}$} & $\bar{6}$ & -- & -- & 
   $7.6\cdot10^3$ & $2.3\cdot10^3$ & $3.2\cdot10^3$ & $1.2\cdot10^3\,^{\ast}$\\\hline
\end{tabular}
\caption{Production of a single invisible particle: Scales $\Lambda$ (in TeV) accessible for the
various operators, assuming bounds on the branching ratios of $10^{-10}$ for each mode. As for Table \ref{TableBND2}, the ranges of accessible invisible masses indicated in the first line are indicative, as the bounds are derived setting $m_{X}=0$ (except for those channels which vanish, denoted by (*), for which $m_{X}=100$ MeV). These scales naively decrease when $m_{X}$ increases due to the phase-space suppressions, though the experimental acceptances need to be taken into account (see App.~\ref{AppKexp}). For the production of an invisible vector, see also Fig.~\ref{FigVectorK}.}
\label{TableBND1}
\end{table}

\begin{table}[t] \centering
\begin{tabular}[c]{cl|p{0.3cm}|ll|ll}\hline
&  & $d$ & $B\to\pi X$ & $B\to\rho X$ & $B\to KX$ & $B\to K^{\ast}X$\\\hline
\multicolumn{2}{c|}{$m_{X}^{\max}$} &  & $m_{B}-m_{\pi}$ & $m_{B}-m_{\rho}$ & $m_{B}-m_{K}$ & $m_{B}-m_{K^{\ast}}$\\\hline
0 & \multicolumn{1}{|l|}{$g_{S}$} & $\bar{5}$ & $2\cdot10^{7}$ & $-$ & $1\cdot10^{7}(5\cdot10^{7})$ & $-$\\
  & \multicolumn{1}{|l|}{$g_{P}$} & $\bar{5}$ & $-$ & $7\cdot10^{6}$ & $-$ & $5\cdot10^{6}(6\cdot10^{7})$\\
  & \multicolumn{1}{|l|}{$g_{V}$} & $5$ & $3\cdot10^{5}$ & $-$ & $2\cdot10^{5}(9\cdot10^{5})$ & $-$\\
  & \multicolumn{1}{|l|}{$g_{A}$} & $5$ & $-$ & $1\cdot10^{5}$ & $-$ & $8\cdot10^{4}(1\cdot10^{6})$\\\hline
1 & \multicolumn{1}{|l|}{$h_{T}$} & $\bar{6}$ & $210^{\ast}$ & $260$ & $210^{\ast}(400^{\ast})$ & $220(770)$\\
  & \multicolumn{1}{|l|}{$h_{\tilde{T}}$} & $\bar{6}$ & $-$ & $260$ & $-$ & $220(770)$\\\hline
\end{tabular}
\caption
{Production of a single invisible particle: Scales $\Lambda$ (in TeV) accessible for the various operators with present (future) measurements (see App.~\ref{AppBexp}). We assume $m_{X}=0$ everywhere, except for those channels which vanish, denoted by (*), for which $m_{X}=2$ GeV. For the production of vector states, see also Fig.~\ref{FigVectorB}.}
\label{TableBNDB1}
\end{table}

Finally, the presence of a neutral scalar field does not open new possibilities compared to the SM to construct $\Delta\mathcal{B}$ and $\Delta\mathcal{L}$ couplings. The simplest operators are obtained by multiplying by $\phi$ (or by $\phi^{\dagger}\phi$ if $\phi$ is not neutral) those of Ref.~\cite{Dim6}, and are either of dimension seven and $\Delta\mathcal{B}=\Delta\mathcal{L}=1$, or dimension six and $\Delta\mathcal{B}=0,\Delta\mathcal{L}=2$:%
\begin{align}
\mathcal{H}_{\Delta\mathcal{B},\Delta\mathcal{L}}^{\mathrm{\phi}} &
=\frac{c^{\Delta\mathcal{L}=2}}{\Lambda^{2}}H\bar{L}^{C}LH\times
\phi+\frac{c_{1}^{\Delta\mathcal{B}=\Delta\mathcal{L}}}{\Lambda^{3}}\bar
{Q}^{C}Q\times\bar{Q}^{C}L\times\phi+\frac{c_{2}^{\Delta\mathcal{B}%
=\Delta\mathcal{L}}}{\Lambda^{3}}\bar{Q}^{C}\sigma_{i}Q\times\bar{Q}^{C}%
\sigma^{i}L\times\phi\nonumber\\
&  +\frac{c_{3}^{\Delta\mathcal{B}=\Delta\mathcal{L}}}{\Lambda^{3}}\bar{Q}%
^{C}Q\times\bar{U}^{C}E\times\phi+\frac{c_{4}^{\Delta\mathcal{B}%
=\Delta\mathcal{L}}}{\Lambda^{3}}\bar{D}^{C}U\times\bar{U}^{C}E\times
\phi+\frac{c_{5}^{\Delta\mathcal{B}=\Delta\mathcal{L}}}{\Lambda^{3}}\bar
{Q}^{C}L\times\bar{D}^{C}U\times\phi+h.c.\;.\label{HSBL}%
\end{align}
The $\Delta\mathcal{L}=2$ operator can produce an invisible $\nu_{L}\nu_{L}\phi$ final state. However, not only is this operator of higher dimension compared to those of Eq.~(\ref{Axion}), but its contribution to the rare decays proceeds through a neutral Higgs penguin (Class III in Table~\ref{TableClasses}), and is thus extremely suppressed. The $\Delta\mathcal{B}=\Delta\mathcal{L}=1$ operators have no impact on the $\Delta\mathcal{B}=0$ rare FCNC decays considered here. In addition, if $m_{\phi}<m_{p,n}$, those involving light flavors can induce $\Delta\mathcal{B}=1$ proton or neutron decays. In that case, without highly non-generic flavor structures for the $c_{i}^{\Delta\mathcal{B}=\Delta\mathcal{L}}$, the scale $\Lambda$ must be close to the Planck scale. On the other hand, if $m_{\phi}>m_{p,n}$, these operators could still induce exotic $B$ decays into an odd number of baryons plus missing energy.%

\subsubsection*{Some models with light weakly coupled (pseudo)scalars}

Typical non-linear symmetry realizations lead to derivative couplings of the type $c_{VL,VR}^{\phi}$ to the dark scalar field, with $\Lambda$ given by the symmetry breaking scale $F$. Well-known examples of such light scalar states are the \textbf{axion}, resulting from the breaking of the PQ symmetry~\cite{Axion}, or the \textbf{familon}, originating from the breaking of some family symmetry~\cite{Familons}. Only the latter naturally lead to FCNC couplings, since by design the family symmetry relates the three generations. For the axion, the dominant effect comes from its flavor-blind coupling to light quarks (Class IV in Table~\ref{TableClasses}). This is dominated by long-distance effects, specifically by the mixing of the axion with the light neutral mesons, as analyzed e.g. in Ref.~\cite{AxionKpi} to which we refer for more details. Let us mention also that in some axion models, there is no direct coupling to light quarks, but rather couplings to the QED or QCD field strengths of the form $\phi F_{\mu\nu}\tilde{F}^{\mu\nu}$ or $\phi G_{\mu\nu}^{a}\tilde{G}_{\alpha}^{\mu\nu}$, as included in Eq.~(\ref{HScalarGauge1}).

Similar derivative couplings arise from models of \textbf{meta-stable supersymmetry breaking}~\cite{ISS}, where a light pseudo Nambu-Goldstone boson $\phi=P$ may be present. As discussed e.g. in Ref.~\cite{PNGB}, through the exchange of three gauge bosons, axion-like effective couplings to quarks are generated with $c_{VL,VR}^{\phi}\sim\alpha_{i}^{3}(\Lambda)$ and $\alpha_{i}$ either the weak, strong, or hypercharge coupling. The scale $\Lambda$ is the supersymmetry breaking scale, which can be around $10-100$ TeV in gauge-mediated supersymmetry breaking scenarios. These couplings are dominantly flavor blind, so according to our general rule~(\ref{CKMscale}), the flavor-changing vertices scale\footnote{Note that this estimate differs from Ref.~\cite{PNGB}, where the flavor-violation is required to arise at the scale $\Lambda$ from effective dimension seven operators. As explained in Sec.~\ref{Classification}, dressing the effective dimension-five flavor blind $\bar{q}\gamma^{\mu}q\partial_{\mu}P$ coupling with a $W$ exchange allows for an electroweak-scale GIM breaking, so that the operator retains its $1/\Lambda$ suppression.} as $c_{VL,VR}^{\phi,IJ}\sim\alpha_{i}^{3} (\Lambda)k^{IJ}$. Thus, rare $K$ and $B$ decays seem unable to reach the interesting range $\Lambda>10$ TeV. If supersymmetric particles are allowed to propagate in the loop(s), the operator may be enhanced, though the tight constraints on the flavor violation in the squark sector probably limit the accessible scales to $\Lambda\lesssim10$ TeV range.

Another scenario involving dark light scalar particles is the \textbf{Next to Minimal Supersymmetric Standard Model} (NMSSM). But, in this case, the scalar mass is often larger than $m_{B}$~\cite{NMSSM}, the scalar may decay too fast (e.g. to $\ell^{+}\ell^{-}$) to be considered as an asymptotic state in $K$ and $B$ decays, and the flavor transitions need to be induced by the (supersymmetrized) weak interaction. When MFV holds, the effective operators are then very suppressed (Class II scenario, see Eq.~(\ref{MFVscaling})). For all these reasons, the NMSSM does not appear as a likely scenario where our effective operators could play a role. Still, our formulas for the differential rates can be directly adapted to that case, and improve on the analysis of Ref.~\cite{NMSSMbounds} by including more observables, and by a better treatment of the hadronic matrix elements.

A more generic example is the \textbf{singlet scalar model} for dark matter, denoted $S$. In its simplest form, it enforces the $\mathbb{Z}_{2}$ symmetry and includes only the renormalizable coupling $\lambda^{\prime}$ of Eq.~(\ref{HScalarGauge2}), with $\phi=S$ a real field. The capabilities of rare decays in probing this scenario were discussed in Ref.~\cite{WIMP}. As described there, being flavor blind, the weak interactions are needed to induce the $c_{SR,SL}^{\phi\phi}$ coupling from the $\lambda^{\prime}$ one (which couples $SS$ to $t\bar{t}$ through a tree-level Higgs exchange). So, in this case, the key to interpret the numbers in Table~\ref{TableBND2} is the identification $c_{SR,SL}^{\phi\phi,IJ}/\Lambda^{2}\to\lambda^{\prime}k^{IJ}/m_{h}^{2}$, with $k^{IJ}$ given in Eq.~(\ref{CKMscale}). See Ref.~\cite{WIMP} for more details.

As a final example, the new invisible scalars could be the \textbf{sgoldstinos} $S$ and $P$, the scalar superpartners of the goldstino~\cite{sgoldstinos}, for which the scale $\Lambda$ corresponds to the fundamental scale of supersymmetry breaking. If light enough, these scalars could be produced alone or in pairs in $K$ and/or $B$ decays. In the former case, the coupling is of the form of $c_{SR,SL}^{\phi}$, and is tuned by the $LR$ squark mass insertions. In the latter case, the derivative interactions are of the form of $c_{VL}^{\phi\phi}$ and $c_{VR}^{\phi\phi}$ with $\phi=S$ and $\phi^{\dagger}=P$, and are tuned by $LL$ and/or $RR$ squark mass insertions. Note that since these squark mass insertions are rather tightly constrained by visible $K$ and $B$ observables, the accessible scales are limited by the MFV rescalings~(\ref{MFVscaling}).

\section{Invisible spin-1 boson}

Compared to scalar and fermionic invisible particles, the presence of an invisible vector particle (denoted $V$) is significantly more difficult to parametrize. A consistent description of a neutral massive vector boson coupled to SM fermions is notoriously delicate. At the same time, adding a $U(1)$ gauge group to the SM is one of the simplest possible extensions~\cite{THDM,Holdom,U1DarkH}. Furthermore, various theoretical models, whose motivations originate for example from strings~\cite{U1String}, extra dimensions~\cite{U1DBrane}, or dark matter theories~\cite{DMU1}, predict such new long or medium range forces, i.e. with masses in the MeV to a few GeV range. Such a dark vector boson would have many phenomenological implications, and has been intensely investigated recently~\cite{Searches}. It could also be produced in rare $B$ and $K$ decays, where it would show up as missing energy if sufficiently long-lived. In this respect, most models do also induce a coupling to leptons, but as long as the $V\to\ell^{+}\ell^{-}$ vertex is not significantly larger than $\bar{q}q\to V$, our bounds should hold since producing a Dalitz pair through a virtual $V$ exchange would push the rates well beyond the experimental reach.

There are several ways to deal with light vector states, which we organize into three scenarios. For the first, the simplest FCNC operators involving the dark vector field are constructed. Those lead to decay rates diverging in the $m_{V}\to0$ limit. This singularity is then treated assuming some kind of Higgs mechanism takes place in the dark sector (in a way similar to Ref.~\cite{U1DarkH}). For the second scenario, the vector field is supposed to couple to SM matter fields only through its field strength, in a (dark) gauge-invariant way. This automatically ensures a safe $m_{V}\to0$ limit, but significantly increases the dimensionality of the effective FCNC couplings. Finally, the third scenario considers only low-energy effective couplings of $V$ to conserved flavor-blind quark currents~\cite{THDM}, as can arise for example from the couplings of $V$ to SM gauge fields through the kinetic mixing~\cite{Holdom}.

\subsection{Simplest FCNC operators\label{Simplest}}

For the first scenario, we consider the lowest-dimensional flavor-changing operators involving a vector field neutral under the SM gauge group, which are simply
\begin{equation}
\mathcal{H}_{mat}^{\mathrm{V}}[\text{I}]=\varepsilon_{L}^{V}\bar{Q}\gamma_{\mu}Q\times V^{\mu}+\varepsilon_{R}^{V}\bar{D}\gamma_{\mu}D\times V^{\mu}.
\label{HvI}%
\end{equation}
Thanks to the Lorentz condition $\partial_{\mu}V^{\mu}=0$, the leading correction starts at $\mathcal{O}(1/\Lambda^{2})$.

The flavor-changing quark currents $\bar{Q}^{I}\gamma_{\mu}Q^{J}$ and $\bar{D}^{I}\gamma_{\mu}D^{J}$ are not conserved when $I\neq J$ (see Eq.~(\ref{EOM})). As a result, a naive rate computation with the polarization sum%
\begin{equation}
\mathcal{P}_{\mu\nu}(k)=\sum_{pol}\varepsilon_{\mu}^{\ast}\varepsilon_{\nu}=-g^{\mu\nu}+\frac{k^{\mu}k^{\nu}}{m_{V}^{2}}\;, \label{PolVec}%
\end{equation}
diverges as $m_{V}\to0$. This well known phenomenon is related to the impossibility of defining consistently the massless limit without an active gauge symmetry. It is interesting to compare this to the behavior of the dimension-four operators originating from the $Z$ penguin in the SM (see Fig.~\ref{FigZpeng}). There, the Ward identity is violated only once the electroweak symmetry breaking takes place, and%
\begin{equation}
\frac{g^{2}V_{tI}^{\ast}V_{tJ}}{(4\pi\Lambda_{SM})^{2}}\bar{Q}^{I}\gamma_{\mu}Q^{J}\times H^{\dagger}\mathcal{D}^{\mu}H\to\frac{g^{2}V_{tI}^{\ast}V_{tJ}}{(4\pi\Lambda_{SM})^{2}}gv^{2}\bar{Q}^{I}\gamma_{\mu}Q^{J}\times Z^{\mu}\;.
\label{Zpeng}%
\end{equation}
With $\Lambda_{SM}\sim v$, this effective interaction is simply proportional to $g^{3}$, up to loop factors. However, the mass of the $Z$ is also of $\mathcal{O}(v)$, and thus it can never be on-shell once using the effective operator formalism. Instead, dimension-six four-fermion operators proportional to $g^{4}/M_{Z}^{2}\sim g^{2}/v^{2}$ are relevant at low energy, see Eq.~(\ref{SMreach}).

In our case, $V$ is light enough to be produced in meson decays. If it also gets its mass through some spontaneous symmetry breaking, then\footnote{Actually, we should write $m_{V}\sim \varepsilon v_{dark}$ and $\varepsilon_{L,R}^{V}\sim \varepsilon ^n$ for some $\varepsilon$ and $n>0$. The simplest situation corresponds to $n=1$.} $m_{V}\sim\varepsilon_{L,R}^{V}v_{dark}$ for some $v_{dark}$ presumably similar or larger than $v$. But as long as $v_{dark}>0$, the limit $m_{V}\to0$ requires $\varepsilon_{L,R}^{V}\to0$, which never diverges for physical processes. Actually, enforcing $\varepsilon_{L,R}^{V}\sim m_{V}/v_{dark}$ in the $m_{V}\to0$ limit, any decay rate behaves as%
\begin{equation}
\Gamma(P\to P^{\prime}V)=\frac{1}{2m_{P}}\int d\Phi_{P^{\prime}V}(\varepsilon_{L,R}^{V}\mathcal{M}^{\mu})\mathcal{P}_{\mu\nu}(k)(\varepsilon_{L,R}^{V}\mathcal{M}^{\nu})^{\ast}\overset{m_{V}\to0}{=}\frac{1}{2m_{P}}\int d\Phi_{P^{\prime}V}\mathcal{M}^{\mu}\frac{k_{\mu}k_{\nu}}{v_{dark}^{2}}\mathcal{M}^{\nu\ast}\;, \label{VectorScalar}%
\end{equation}
with $\mathcal{M}^{\mu}$ the hadronic matrix element $\langle P^{\prime}|\bar{Q}^{I}\gamma_{\mu}Q^{J}|P\rangle$ or $\langle P^{\prime}|\bar{D}^{I}\gamma_{\mu}D^{J}|P\rangle$ and $d\Phi_{P^{\prime}V}$ the phase-space integrations. As expected from the equivalence theorem, this is precisely the rate one would derive from the axionic operators $\bar{Q}^{I}\gamma_{\mu}Q^{J}\times\partial^{\mu}\phi$ and $\bar{D}^{I}\gamma_{\mu}D^{J}\times\partial^{\mu}\phi$ considered in the previous section, Eq.~(\ref{Axion}) with $\Lambda=v_{dark}$.

For the pair-production of dark vectors, the simplest operators are of dimension six:
\begin{align}
\mathcal{H}_{mat}^{\mathrm{VV}}[\text{I}] &  =\frac{c_{DL}^{VV}}{\Lambda^{2}}\bar{Q}\gamma_{\mu}\overleftrightarrow{\mathcal{D}}_{\nu}Q\times V^{\mu}V^{\nu}+\frac{c_{L}^{VV}}{\Lambda^{2}}\bar{Q}\gamma_{\mu}Q \times V_{\nu}V^{\mu\nu}+\frac{c_{L}^{V\tilde{V}}}{\Lambda^{2}}\bar{Q}\gamma_{\mu}Q\times V_{\nu}\tilde{V}^{\mu\nu}+\frac{c_{SL}^{VV}}{\Lambda^{2}}H^{\dagger}\bar{D}Q\times V_{\mu}V^{\mu}\nonumber\\
&  \;+\frac{c_{DR}^{VV}}{\Lambda^{2}}\bar{D}\gamma_{\mu}\overleftrightarrow{\mathcal{D}}_{\nu}D\times V^{\mu}V^{\nu}+\frac{c_{R}^{VV}}{\Lambda^{2}}\bar{D}\gamma_{\mu}D\times V_{\nu}V^{\mu\nu}+\frac{c_{R}^{V\tilde{V}}}{\Lambda^{2}}\bar{D}\gamma_{\mu}D\times V_{\nu}\tilde{V}^{\mu\nu}+\frac{c_{SR}^{VV}}{\Lambda^{2}}H\bar{Q}D\times V_{\mu}V^{\mu}\;,\label{HVVI}%
\end{align}
with the field strength $V_{\mu\nu}\equiv\partial_{\mu}V_{\nu}-\partial_{\nu}V_{\mu}$ and its dual $\tilde{V}^{\mu\nu}\equiv\frac{1}{2}\varepsilon^{\mu\nu\rho\sigma}V_{\rho\sigma}$. Note that those involving $V^{\mu\nu}$ can be reduced to $\partial_{\nu}(\bar{Q}\gamma_{\mu}Q)\times V^{\mu}V^{\nu}$ and $\partial_{\nu}(\bar{D}\gamma_{\mu}D)\times V^{\mu}V^{\nu}$, which are orthogonal to those tuned by $c_{DL,DR}^{VV}$ (which are missing in Ref.~\cite{Badin:2010uh}). These operators are given for completeness, but will not be considered further for two reasons. First, in the present minimal theoretical setting, there is no reason for the renormalizable operators of Eq.~(\ref{HvI}) to be absent, and those would clearly offer better windows. Second, the leading operators of $\mathcal{H}_{mat}^{\mathrm{VV}}[$I$]$ are those tuned by $c_{SL,SR}^{VV}$ since the others are comparatively suppressed by $m_{K,B}/v$. But these operators also arise from the $H^{\dagger}\bar{D}Q\times V_{\mu\nu}V^{\mu\nu}$ and $H \bar{Q}D\times V_{\mu\nu}V^{\mu\nu}$  operators considered in the next scenario (albeit in a rescaled form, see Eq.~(\ref{DirectVV})). As explained there, these operators could become leading in the presence of a non-abelian dark gauge invariance, which would forbid single vector production.

\subsubsection*{Phenomenology}

Let us consider separately the vector $\varepsilon_{V}(\bar{d}^{I}\gamma_{\mu}d^{J})\times V^{\mu}$ and axial-vector $\varepsilon_{A}(\bar{d}^{I}\gamma_{\mu}\gamma_{5}d^{J})\times V^{\mu}$ couplings. When $m_{V}\neq0$, both terms of the polarization sum~(\ref{PolVec}) contribute, so experimental bounds translate as constraints in the $(m_{V},\varepsilon_{V})$ or $(m_{V}$, $\varepsilon_{A})$ planes. As explained above, we identify the regions where $m_{V}/\varepsilon_{V,A}>v$ as physical (for a similar reasoning, see e.g. Ref.~\cite{U1DarkH}).

\begin{figure}[t!]
\centering      \includegraphics[width=15cm]{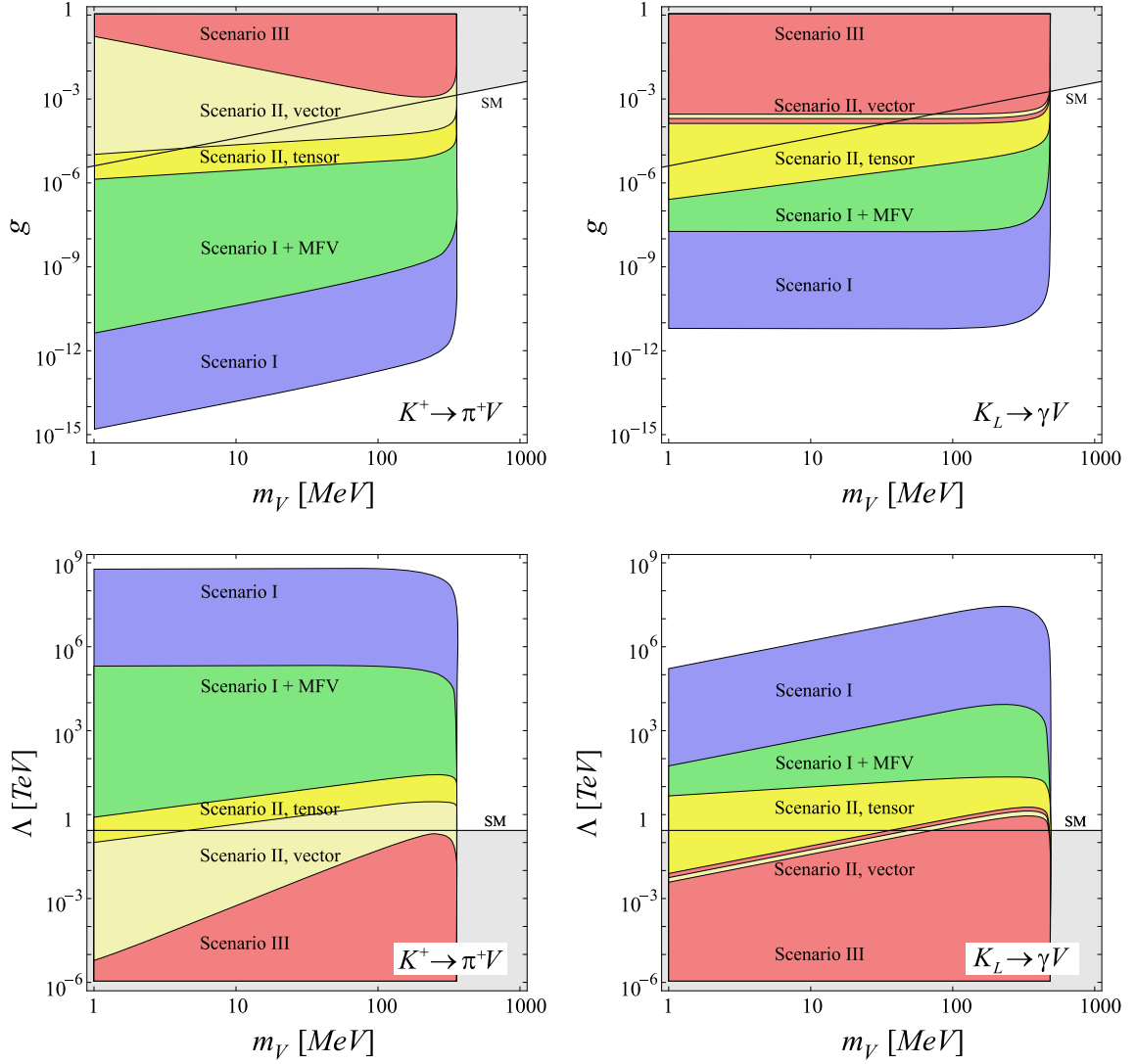}  \caption{Above: Exclusion regions drawn from a bound on the $K^{+}\to\pi^{+}V$ (left) and $K_{L}\to\gamma V$ (right) branching ratios at the $10^{-10}$ level. The plots on the first line show the coupling $g$ as a function of $m_{V}$ (in MeV) for the scenario I (blue, $g=\varepsilon_{V}$), scenario I with MFV (green, $g=\varepsilon_{V}|V_{ts}^{\ast}V_{td}|$), scenario II from the tensor operators $\bar{s}\sigma^{\mu\nu}d\times V_{\mu\nu}$ (yellow) and II from the vector operators $\bar{s}\gamma_{\mu}d\times\partial_{\nu}V^{\mu\nu}$ (light yellow), and scenario III ($g=\varepsilon e$). The grey area represents the region where $m_{V}/g=v_{dark}<v\simeq246$~GeV. The plots in the second line show the same, but replace $g$ with $\Lambda=m_{V}/g$ (in TeV).}
\label{FigVectorK}
\end{figure}

\begin{figure}[t!]
\centering      \includegraphics[width=15cm]{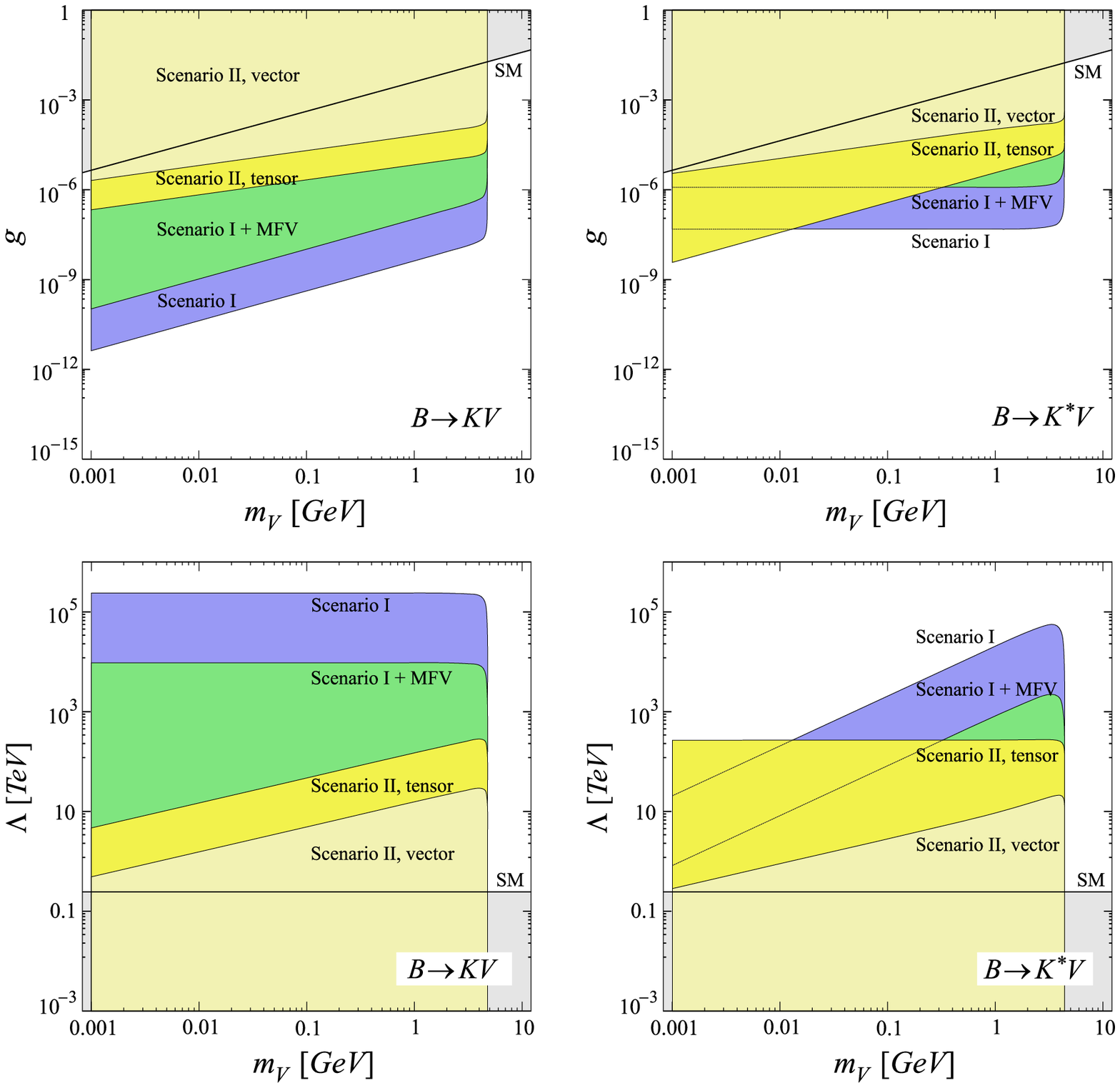}
\caption{Constraints on the coupling $g$ against $m_{V}$ from $B\to KV$ (left) and $B\to K^{\ast}V$ (right) for the scenario I (blue, $g=\epsilon_{V}$), scenario I with MFV (green, $g=\epsilon|V_{tb}^{\ast}V_{ts}|$), scenario II from the tensor operators $\bar{b}\sigma_{\mu\nu}s\times V^{\mu\nu}$ (yellow) and II from the vector operators $\bar{b}\gamma_{\mu}s\times\partial_{\nu}V^{\mu\nu}$ (light yellow). The gray areas represent the regions where $m_{V}/g=v_{dark}<v\simeq246$~GeV. Lower plots: Same as above, but in terms of $\Lambda=m_{V}/g$ (in TeV).}
\label{FigVectorB}
\end{figure}

As shown in Figs.~\ref{FigVectorK} and~\ref{FigVectorB}, the pair $(m_{V},\varepsilon_{V})$ is very constrained in both the $K$ and $B$ sectors, with typically $m_{V}/\varepsilon_{V}\gg v$. Indeed, because the second term of Eq.~(\ref{PolVec}) is growing as $1/m_{V}^{2}$ in the $m_{V}\to0$ limit, $\varepsilon_{V}$ has to be correspondingly tiny to pass the bounds on the branching ratio. Said differently, the dark spontaneous symmetry breaking scale $v_{dark}$ in Eq.~(\ref{VectorScalar}) has to be much larger than the electroweak scale. Interestingly, this remains true even when the flavor-violating part of $\varepsilon_{V}$ satisfies MFV (green regions in Figs.~\ref{FigVectorK} and~\ref{FigVectorB}), i.e. is rescaled as $\varepsilon_{V}\to|V_{tI}^{\ast}V_{tJ}|\varepsilon_{V}$, see Eq.~(\ref{MFVscaling}). So, even if there is no theoretical requirement for the hidden symmetry breaking scale to be very different from that of the visible sector, current FCNC constraints nonetheless require $v_{dark}\gg v$.

For comparison, the right panel in Fig.~\ref{FigVectorK} shows the constraints one would get from a similar bound at the $10^{-10}$ level on the $K\to\gamma V$ channel. As this process is induced by the anomaly, the coupling is of the form%
\begin{equation}
\mathcal{L}_{anom}=\frac{eN_{c}}{12\pi^{2}F_{\pi}}(\operatorname{Re}(\varepsilon_{V})K_{L}+i\operatorname{Im}(\varepsilon_{V})K_{S})F_{\mu\nu}\tilde{V}^{\mu\nu}\;,
\end{equation}
with $F^{\mu\nu}$ the QED field strength. So, the current is effectively conserved, the $1/m_{V}^{2}$ term in Eq.~(\ref{PolVec}) drops out, and the $m_{V}\to0$ limit becomes smooth.

\subsection{Gauge invariant FCNC operators}

For the second scenario, we impose that only the dark field strength $V_{\mu\nu}\equiv\partial_{\mu}V_{\nu}-\partial_{\nu}V_{\mu}$ and its dual $\tilde{V}^{\mu\nu}\equiv\frac{1}{2}\varepsilon^{\mu\nu\rho\sigma}V_{\rho\sigma}$ occur in the flavor-changing couplings. This restores gauge invariance and ensures a smooth $m_{V}\to0$ limit, but increases the dimension of the simplest operators. Depending on whether the vector is assumed abelian or non-abelian, the lowest-dimensional operators are either dimension six,%
\begin{align}
\mathcal{H}_{mat}^{\mathrm{V}}[\text{II}]  &  =+\frac{c_{L}^{V}}{\Lambda^{2}}\bar{Q}\gamma_{\mu}\overleftrightarrow{\mathcal{D}}_{\nu}Q\times V^{\mu\nu}+\frac{c_{L}^{\prime V}}{\Lambda^{2}}\bar{Q}\gamma_{\mu}Q\times\partial_{\nu}V^{\mu\nu}+\frac{c_{TL}^{V}}{\Lambda^{2}}H^{\dagger}\bar{D}\mathcal{\sigma}_{\mu\nu}Q\times V^{\mu\nu}\nonumber\\
&  \;\;\;\;+\frac{c_{R}^{V}}{\Lambda^{2}}\bar{D}\gamma_{\mu}\overleftrightarrow{\mathcal{D}}_{\nu}D\times V^{\mu\nu}+\frac{c_{R}^{\prime V}}{\Lambda^{2}}\bar{D}\gamma_{\mu}D\times\partial_{\nu}V^{\mu\nu}+\frac{c_{TR}^{V}}{\Lambda^{2}}H\bar{Q}\mathcal{\sigma}_{\mu\nu}D\times V^{\mu\nu}\;, \label{HvII1}%
\end{align}
or dimension eight,%
\begin{align}
\mathcal{H}_{mat}^{\mathrm{VV}}[\text{II}]  &  =\frac{c_{SL}^{VV}}{\Lambda^{4}}H^{\dagger}\bar{D}Q\times V_{\mu\nu}V^{\mu\nu}+\frac{c_{SL}^{V\tilde{V}}}{\Lambda^{4}}H^{\dagger}\bar{D}Q\times V_{\mu\nu}\tilde{V}^{\mu\nu}+\frac{c_{SR}^{VV}}{\Lambda^{4}}H\bar{Q}D\times V_{\mu\nu}V^{\mu\nu}+\frac{c_{SR}^{V\tilde{V}}}{\Lambda^{4}}H\bar{Q}D\times V_{\mu\nu}\tilde{V}^{\mu\nu}\nonumber\\
&  \;+\frac{c_{L}^{VV}}{\Lambda^{4}}\bar{Q}\gamma^{\mu}\overleftrightarrow{\mathcal{D}}_{\nu}Q\times V_{\mu\rho}V^{\rho\nu}+\frac{c_{L}^{V\tilde{V}}}{\Lambda^{4}}\bar{Q}\gamma^{\mu}\overleftrightarrow{\mathcal{D}}_{\nu}Q\times V_{\mu\rho}\tilde{V}^{\rho\nu}+\frac{c_{L}^{\tilde{V}V}}{\Lambda^{4}}\bar{Q}\gamma^{\mu}\overleftrightarrow{\mathcal{D}}_{\nu}Q\times \tilde{V}_{\mu\rho}V^{\rho\nu}\nonumber\\
&  \;+\frac{c_{R}^{VV}}{\Lambda^{4}}\bar{D}\gamma^{\mu}\overleftrightarrow{\mathcal{D}}_{\nu}D\times V_{\mu\rho}V^{\rho\nu}+\frac{c_{R}^{V\tilde{V}}}{\Lambda^{4}}\bar{D}\gamma^{\mu}\overleftrightarrow{\mathcal{D}}_{\nu}D\times V_{\mu\rho}\tilde{V}^{\rho\nu}+\frac{c_{R}^{\tilde{V}V}}{\Lambda^{4}}\bar{D}\gamma^{\mu}\overleftrightarrow{\mathcal{D}}_{\nu}D\times \tilde{V}_{\mu\rho}V^{\rho\nu}\;. \label{HvII2}%
\end{align}
To reach this minimal basis, a number of identities and approximations were used. First, for the operators with a single vector field, the dual field strength $\tilde{V}^{\mu\nu}$ is absent from $\mathcal{H}_{mat}^{\mathrm{V}}[$II$]$ since it can always be reduced to operators involving $V^{\mu\nu}$ using the quark EOM, integration by part, and the Chisholm identity~\cite{BuchmullerW86}. 

For the operators with two field strengths, we first note that summation over the generators of the adjoint representation of the dark gauge group may be needed to enforce gauge invariance. However, to keep things simple, we consider that after the dark sector symmetry breaking, only one (possibly complex) vector field is light enough to be produced at low energy, and simply discard the adjoint index. In addition, the non-abelian terms of the field strengths are systematically removed since the signatures we are after involve only two vectors. These restrictions, together with $\mathcal{D}_{\mu}V^{\mu\nu}=-m_{V}^{2}V^{\nu}$ and $\mathcal{D}_{\mu}\tilde{V}^{\mu\nu}=0$, permit to reduce all the operators with derivatives acting on the field strengths (as well as those involving $\bar{Q}\gamma^{\mu}(\overleftarrow{\mathcal{D}}_{\nu}+\overrightarrow{\mathcal{D}}_{\nu})Q$). Finally, tensor operators involving two field strengths cannot be constructed since $V_{\mu\rho}V_{\,\,\nu}^{\rho}$ is symmetric under $\mu\leftrightarrow\nu$, while $H^{\dagger}\bar{D}\sigma^{\mu\nu}Q\times \tilde{V}_{\mu\rho}V_{\,\,\nu}^{\rho}$ reduces to $H^{\dagger}\bar{D}\sigma^{\mu\nu}Q\times V_{\mu\rho}V_{\,\,\nu}^{\rho}$ upon using $2\sigma^{\mu\nu}\gamma_{5}=i\varepsilon^{\mu\nu\alpha\beta}\sigma_{\alpha\beta}$.

\subsubsection*{Reduction and phenomenology}

Not all the operators are relevant phenomenologically. First, we can discard all those involving a derivative acting on the quark fields, since they are comparatively suppressed by $\mathcal{O}(m_{K,B}/v)$. Then, consider the operators involving $\partial_{\mu}V^{\mu\nu}$ in $\mathcal{H}_{mat}^{\mathrm{V}}[$II$]$. Upon enforcing the non-gauge invariant free EOM $\partial_{\mu}V^{\mu\nu}=-m_{V}^{2}V^{\nu}$, they collapse to the dimension-four couplings of $\mathcal{H}_{mat}^{\mathrm{V}}[$I$]$, Eq.~(\ref{HvI}), with the identification%
\begin{equation}
\frac{c_{L,R}^{\prime V}}{\Lambda^{2}}(\bar{Q}\gamma_{\mu}Q,\bar{D}\gamma_{\mu}D) \times \partial_{\nu}V^{\mu\nu}\to\frac{m_{V}^{2}c_{L,R}^{\prime V}}{\Lambda^{2}}( \bar{Q}\gamma_{\mu}Q,\bar{D}\gamma_{\mu}D)\times V^{\mu}\;\;\Rightarrow\;\frac{m_{V}^{2}}{\Lambda^{2}%
}c_{L,R}^{\prime V}\to\varepsilon_{L,R}^{V}\;. \label{ReducV1}%
\end{equation}
This is thus an explicit realization of the $\mathcal{H}_{mat}^{\mathrm{V}}[$I$]$ operators, though with a major difference compared to the previous section. In the $m_{V}\to0$ limit, the coupling $\varepsilon_{L,R}^{V}$ here effectively scales at least like $m_{V}^{2}$ instead of $m_{V}$, and the $P\to P^{\prime}V$ rates vanish when $m_{V}\to0$ (compare with Eq.~(\ref{VectorScalar})). The precise scaling is not fixed though, because that between $c_{L,R}^{\prime V}$ and $m_{V}$ is not known. If the vector gains its mass at the scale $\Lambda$ from some spontaneous symmetry breaking, all we can say is that $m_{V}\sim g\Lambda$ for some $g$, and $c_{L,R}^{\prime V}\sim g^{n}$ for some $n\geq0$, so that $\varepsilon_{L,R}^{V}\sim(m_{V}/\Lambda)^{2+n}$. This scenario (with $n=1$, i.e. $g=\varepsilon_{V}=(m_{V}/\Lambda)^{3}$) is shown by the light yellow regions in Figs.~\ref{FigVectorK} and~\ref{FigVectorB}. Comparing these regions with those corresponding to the first scenario (in blue), the rescaling~(\ref{ReducV1}) is very expensive in terms of accessible scales.

It is interesting to compare the reduction~(\ref{ReducV1}) to that of the magnetic operator of $\mathcal{H}_{mat}^{\mathrm{V}}[$II$]$. After integrating by part and using the quark EOM~(\ref{EOM}),
\begin{align}
\frac{c_{T}^{V}}{\Lambda^{2}}H^{\dagger}\bar{D}^{I}\mathcal{\sigma}_{\mu\nu}Q^{J}\times V^{\mu\nu}  &  =2\frac{c_{T}^{V}}{\Lambda^{2}}H^{\dagger}H((\bar{Q}\mathbf{Y}_{d}^{\dagger})^{I}\gamma_{\mu}Q^{J}+\bar{D}^{I}\gamma_{\mu}(\mathbf{Y}_{d}^{\dagger}D)^{J})\times V^{\mu}\nonumber\\
&  \;\;\;\;-2i\frac{c_{T}^{V}}{\Lambda^{2}}H^{\dagger}\bar{D}^{I}\overleftrightarrow{\mathcal{D}}_{\mu}Q^{J}\times V^{\mu}-2\frac{c_{T}^{V}}{\Lambda^{2}}\mathcal{D}^{\nu}H^{\dagger}\bar{D}^{I}\mathcal{\sigma}_{\mu\nu}Q^{J}\times V^{\mu}\;. \label{ReducV2}%
\end{align}
The first two terms match those of $\mathcal{H}_{mat}^{\mathrm{V}}[$I$]$, Eq.~(\ref{HvI}) after electroweak symmetry breaking. Since we started from a gauge-invariant operator, gauge invariance is now hidden in the quark-mass dependent relationships among the couplings $\varepsilon_{L,R}^{V}$ of $\mathcal{H}_{mat}^{\mathrm{V}}[$I$]$, as well as of those of the higher-dimensional operators in Eq.~(\ref{ReducV2}). It is only upon imposing these relationships that all the terms originating from the $k^{\mu}k^{\nu}/m_{V}^{2}$ piece of the polarization sum~(\ref{PolVec}) cancel out. Though this is another physically sound interpretation of the $\mathcal{H}_{mat}^{\mathrm{V}}[$I$]$ operators, it is much easier phenomenologically to consider directly the operator $H^{\dagger}\bar{D}^{I}\mathcal{\sigma}_{\mu\nu}Q^{J}\times V^{\mu\nu}$ of $\mathcal{H}_{mat}^{\mathrm{V}}[$II$]$ from which they derive.

Note, finally, that a similar reduction starting with the last two operators of $\mathcal{H}_{mat}^{\mathrm{VV}}[$II$]$ can also be done, leading to the leading dimension-six non-gauge invariant operator of Eq.~(\ref{HVVI})
\begin{equation}
\frac{c_{SL}^{VV}}{\Lambda^{4}}H^{\dagger}\bar{D}Q\times V_{\mu\nu}V^{\mu\nu}\ni\frac{m_{V}^{2}c_{SL}^{VV}}{\Lambda^{4}}H^{\dagger}\bar{D}Q\times V_{\nu}V^{\nu}\;.
\label{DirectVV}
\end{equation}
As explained before, in the present work, we consider this operator exclusively in the gauge invariant form, since the dimension-four couplings of Eq.~(\ref{HvI}) dominate if no dark gauge invariance is enforced.

After the reductions described above, the only operators relevant for rare FCNC decays are those involving the Higgs field, which we write after the electroweak symmetry breaking as ($I>J$)%
\begin{align}
\mathcal{H}_{mat}^{\mathrm{V},\mathrm{VV}}[\text{II}]  &  =+\frac{h_{SS}^{IJ}}{\Lambda^{3}}\bar{d}^{I}d^{J}\times V_{\mu\nu}V^{\mu\nu}+\frac{h_{PS}^{IJ}}{\Lambda^{3}}\bar{d}^{I}\gamma_{5}d^{J}\times V_{\mu\nu}V^{\mu\nu}+\frac{h_{T}^{IJ}}{\Lambda}\bar{d}^{I}\mathcal{\sigma}_{\mu\nu}d^{J}\times V^{\mu\nu}\nonumber\\
&  \;\;\;\;+\frac{ih_{SP}^{IJ}}{\Lambda^{3}}\bar{d}^{I}d^{J}\times V_{\mu\nu}\tilde{V}^{\mu\nu}+\frac{ih_{PP}}{\Lambda^{3}}\bar{d}^{I}\gamma_{5}d^{J}\times V_{\mu\nu}\tilde{V}^{\mu\nu}+\frac{h_{\tilde{T}}^{IJ}}{\Lambda}\bar{d}^{I}\mathcal{\sigma}_{\mu\nu}\gamma_{5}d^{J}\times V^{\mu\nu}+h.c.\;\;,
\end{align}
with, omitting flavor indices for simplicity:%
\begin{equation}
h_{T,\tilde{T}}=\frac{v}{\Lambda}\frac{c_{TR}^{V}\pm c_{TL}^{V}}{2},\;
h_{SS,PS}=\frac{v}{\Lambda}\frac{c_{SR}^{VV}\pm c_{SL}^{VV}}{2},\;
ih_{SP,PP}=\frac{v}{\Lambda}\frac{c_{SR}^{V\tilde{V}}\pm c_{SL}^{V\tilde{V}}}{2}\;.
\end{equation}
The rates and differential rates for $K$ ($B$) decays are in Appendix~\ref{AppKspin1}~(\ref{AppBspin1}), while the physics reach are in Tables~\ref{TableBND2} to~\ref{TableBNDB1}. Also, the constraints on the tensor operators are shown by the dark yellow regions in Figs.~\ref{FigVectorK} and~\ref{FigVectorB}, setting $h_{T,\tilde{T}}=(v/\Lambda)g$ and $\Lambda=m_{V}/g$.

\subsection{Flavor-blind operators}

For the last scenario, the invisible vector boson is allowed to couple to conserved quark currents only, so that the singular term in the polarization sum~(\ref{PolVec}) automatically cancels out in decay rate computations. To implement this, we have to drop the $SU(2)_{L}\otimes U(1)_{Y}$ gauge
invariance requirement (see below), and couple $V$ to quarks as%
\begin{equation}
\mathcal{H}_{eff}^{\mathrm{V}}[\text{III}]=J_{\mu}[c_{q}]\times V^{\mu}\;,\;\;\;\;
J_{\mu}[c_{q}]\equiv\sum_{q=u,d,s,c,b,t}c_{q}\bar{q}\gamma_{\mu}q\;. \label{ConsCurr}%
\end{equation}
The $c_{q}$ cannot be completely arbitrary but must reflect the flavor structures present in the SM. If that was not the case, then the FCNC couplings~(\ref{HvI}) of the first scenario are in general present~\cite{LangackerP00}, and those would completely dominate in $B$ and $K$ decays.

Using the MFV language, there are two possible flavor structures. Either the $c_{q}$ are universal or they are proportional to the Yukawa couplings. In the latter case, the top current would completely dominate, corresponding to Class III in the nomenclature of Table~\ref{TableClasses}. In the former case, the remaining freedom corresponds to
\begin{equation}
c_{u}=c_{c}=c_{t}\equiv c_{U}\;,\;\;c_{d}=c_{s}=c_{b}\equiv c_{D}\;,
\label{universal}%
\end{equation}
with $c_{U}$ and $c_{D}$ a priori arbitrary. Note that this two-parameter freedom means that it is always possible to write $J_{\mu}[c_{q}]$ in terms of the electromagnetic and the baryon number currents only, as was pointed out in Ref.~\cite{Fayet}. When the invisible vector is aligned with the photon, rather tight constraints are set by flavor-blind observables, e.g. the muon $g-2$, quarkonium decays, beam dump experiments (see e.g. Ref.~\cite{Hidden} for a recent analysis). While we will compare our constraints with those, let us stress that consistency does not require $V_{\mu}$ to couple to leptons\footnote{Anomaly cancellation may require the leptons to be charged under the dark gauge symmetry, depending on the detailed particle content and dynamics of the dark sector. We do not consider such constraints.}. If it is leptophobic, many of these limits can be evaded. Hence, only the purely hadronic production of $V$ in $K$ or $B$ decays will be considered here, and should be compared e.g. with those from $\pi^{0}$ or quarkonium decays with missing energy.

As explained in Sec.~\ref{Classification}, it is very different to probe the couplings to heavy (Class III) or to light quarks (Class IV), so we will analyze each situation in turn in the next two sections. But before that, let us return to the issue of an $SU(2)_{L}\otimes U(1)_{Y}$ gauge invariant origin for the conserved currents in Eq.~(\ref{ConsCurr}).

\subsubsection*{Flavor-blind $SU(2)_{L}\otimes U(1)_{Y}$ invariant couplings}

In Sec.~\ref{Simplest}, the $1/m_{V}^{2}$ singularity of the decay rates arising from the FCNC couplings of Eq.~(\ref{HvI}) was interpreted in terms of a dark symmetry breaking scale $v_{dark}\gtrsim v$. Specifically, we enforced that the coupling constant $\varepsilon$ of $V$ to quarks and its mass satisfy $m_{V}\sim v_{dark}\varepsilon$, so that it is always the finite combination $\varepsilon^{2}/m_{V}^{2}\to1/v_{dark}^{2}$ which occurs in decay rates. But, generic vector or axial-vector quark currents are conserved for massless quarks, which is the case when $SU(2)_{L}\otimes U(1)_{Y}$ is exact (for simplicity, we disregard the chiral current anomalies, which do not concern us here). So, the presence of the apparent $1/m_{V}^{2}$ singularity could be due to the electroweak symmetry breaking instead of to the dark symmetry breaking.

To make this statement more concrete (see also the discussion in Ref.~\cite{HFCNC}), let us write down the flavor-blind renormalizable couplings between $V$ and SM gauge, Higgs, and quark fields (the flavor-blind summation over $I=1,2,3$ is understood)%
\begin{subequations}
\label{HvGauge}%
\begin{align}
\mathcal{H}_{int}^{\mathrm{V}}[\text{III}] &  =\varepsilon_{1}B_{\mu\nu}\times V^{\mu\nu}+\varepsilon_{2}iH^{\dagger}\overleftrightarrow{\mathcal{D}}_{\mu}H\times V^{\mu}+\varepsilon_{\theta}B_{\mu\nu}\times\tilde{V}^{\mu\nu}+\varepsilon_{H}(H^{\dagger}H)\times V_{\mu}V^{\mu}\;,\\
\mathcal{H}_{mat}^{\mathrm{V}}[\text{III}] &  =\varepsilon_{B}(\bar{Q}^{I}\gamma_{\mu}Q^{I}+\bar{D}^{I}\gamma_{\mu}D^{I}+\bar{U}^{I}\gamma_{\mu}U^{I})\times V^{\mu}+\varepsilon_{D}(\bar{D}^{I}\gamma_{\mu}D^{I})\times V^{\mu}+\varepsilon_{U}(\bar{U}^{I}\gamma_{\mu}U^{I})\times V^{\mu}\;.
\label{BaryonC}%
\end{align}
\end{subequations}
These dimension-four couplings presumably dominate over the higher dimensional flavor-blind operators, which we do not list here (See Ref.~\cite{Zprime}).

Only the $\varepsilon_{1}$, $\varepsilon_{2}$, $\varepsilon_{\theta}$, and $\varepsilon_{B}$ couplings are compatible with a dark sector gauge invariance associated to $V$, since they vanish under $V^{\mu}\to\partial^{\mu}\phi$ (see Eq.~(\ref{VectorScalar})) upon partial integration, and using the free Higgs boson and quark EOM. The $\varepsilon_{B}$ coupling involves the (conserved) baryon number current, and is thus directly matched onto Eq.~(\ref{ConsCurr}). The other couplings, $\varepsilon_{H}$, $\varepsilon_{D}$, and $\varepsilon_{U}$, would break the dark gauge invariance and are thus discarded. Note that $\varepsilon_{H}$ must in any case be tiny if $V$ is to be light enough to be produced in rare decays. Indeed, the $(H^{\dagger}H)\times V_{\mu}V^{\mu}$ coupling gives a mass to $V$ after the electroweak symmetry breaking, so barring a large cancellation between the dark and visible Higgs sectors, $\varepsilon_{H}\lesssim m_{K(B)}^{2}/v^{2}\approx10^{-6}(10^{-4})$. We can further discard the $\varepsilon_{\theta}$ term, which is a total derivative and is relevant only for magnetic monopoles~\cite{U1Mono}.

The two remaining couplings are $\varepsilon_{1}$ and $\varepsilon_{2}$. The first one is the celebrated kinetic mixing~\cite{Holdom}. Since $B_{\mu}=\cos\theta_{W}A_{\mu}-\sin\theta_{W}Z_{\mu}$, we can rewrite it as%
\begin{equation}
B_{\mu\nu}\times V^{\mu\nu}=2\cos\theta_{W}J_{\nu}^{em}\times V^{\nu}+2\sin\theta_{W}Z_{\nu}\times \partial_{\mu}V^{\mu\nu}\;. \label{KinMix}
\end{equation}
The piece proportional to the (conserved) electromagnetic current $\partial^{\mu}F_{\mu\nu}=-J_{\nu}^{em}$ can be matched onto Eq.~(\ref{ConsCurr}), while the other one mixes $Z$ and $V$. The $\varepsilon_{2}$ coupling also generates a direct mixing $Z_{\mu}\times V^{\mu}$ after the electroweak symmetry breaking, since $H^{\dagger}\mathcal{D}_{\mu}H\to gv^{2}Z_{\mu}$.

The $Z-V$ mixings induced by $\varepsilon_{1}$ and $\varepsilon_{2}$ are very different. The kinetic mixing $Z_{\nu}\times\partial_{\mu}V^{\mu\nu}$ is safe in the $m_{V}\to0$ limit, because it is insensitive to the electroweak symmetry breaking. Actually, once the $Z$ is integrated out, this $Z_{\nu}\times \partial_{\mu}V^{\mu\nu}$ vertex together with the SM flavor-changing hadronic vertex of Eq.~(\ref{Zpeng}) generate the gauge-invariant operators $c_{L}^{\prime V}$ of Eq.~(\ref{HvII1}), with $c_{L}^{\prime V}\sim\varepsilon_{1}g^{3}V_{tI}^{\ast}V_{tJ}$ and $\Lambda\sim M_{Z}$. On the other hand, doing the same with the $\varepsilon_{2}$ coupling generates the dangerous operators of the first scenario, Eq.~(\ref{HvI}), with%
\begin{equation}
(\varepsilon_{L}^{V})^{IJ}\sim\varepsilon_{2}gv^{2}\times\frac{g^{2}}{M_{W}^{2}}\frac{1}{16\pi^{2}}V_{tI}^{\ast}V_{tJ}\sim\varepsilon_{2}\times\frac{g}{16\pi^{2}}V_{tI}^{\ast}V_{tJ}\;. \label{ModelISM}
\end{equation}
However, in parallel to the above FCNC operator, $\varepsilon_{2}$ also corrects the $V$ mass as $\delta m_{V}^{2}\sim\varepsilon_{2}^{2}v^{2}$. As a result, $m_{V}^{phys}\to0$ requires $\varepsilon_{2}\to0$, ensuring again the safety of all decay rates in the massless limit. This shows that indeed, the visible symmetry breaking scale can play the same role as a dark symmetry breaking scale. The only difference, besides $v\neq v_{dark}$ in general, is that the former relies on the SM dynamics to drive the FCNC, and thus brings in the loop and CKM suppression factors, see Eq.~(\ref{ModelISM}). The allowed range of $\varepsilon_{2}$ values can be derived from the green areas in Figs.~\ref{FigVectorK} and~\ref{FigVectorB}, up to the rescaling by $g/16\pi^{2}\sim10^{-3}$. Note that $\varepsilon_{2}$ values acceptable for rare decays ensure that $\delta m_{V}^{2}\sim\varepsilon_{2}^{2}v^{2}$ can be safely neglected.

The above electroweak mechanism ensuring a safe massless limit may render the extension to a two Higgs doublet model desirable. Indeed, there would then be two different $\varepsilon_{H}$ couplings, and an additional conserved current can be constructed~\cite{THDM}, whose charges are aligned with those of the Peccei-Quinn (PQ) symmetry~\cite{PecceiQ77}. Combined with the conserved lepton number current, this allows in principle to make $V$ completely leptophobic~\cite{leptophobic}. The cost being the presence of a flavor-blind axial-vector quark current, not conserved at low energy. The corresponding $1/m_{V}^{2}$ singularity is nevertheless under control thanks to the additional sources of $Z-V$ mixing present in the PQ current, i.e. tuned by the same coupling constant. We will not further elaborate on this construction, but just retain that a leptophobic setting is in principle possible, allowing to evade the many low-energy constraints based on the $\bar{e}\gamma_{\mu}e\times V^{\mu}$ and $\bar{\mu}\gamma_{\mu}\mu\times V^{\mu}$ couplings~\cite{Hidden,Searches}.

Apart from the $Z-V$ mixing effects, matched onto Eq.~(\ref{HvI}) or Eq.~(\ref{HvII1}), the $\varepsilon_{B}$ and $\varepsilon_{1}$ couplings are genuine $SU(2)_{L}\otimes U(1)_{Y}$ invariant realization of the two conserved quark currents of Eq.~(\ref{ConsCurr}). Let us now see how to constrain them from rare decays.

\subsubsection*{Phenomenological constraints on the couplings to heavy quarks}

Once the heavy quarks are integrated out along with the weak bosons, the presence of $V$ in $K$ and $B$ physics is felt through the operators of the second scenario, in particular $\mathcal{H}_{mat}^{\mathrm{V}}[$II$]$. For example, the last operator in Eq.~(\ref{HvII1}) is induced in complete analogy to the electromagnetic operators describing $b\to s\gamma$ and $s\to d\gamma$ in the SM. This situation is thus simple to account by adapting the coupling of $\mathcal{H}_{mat}^{\mathrm{V}}[$II$]$ according to Eq.~(\ref{CKMscale}), and setting the scale $\Lambda$ at $M_{W}$. Alternatively, a more precise estimate can be obtained when the new invisible vector boson is very light and aligned with the photon (in the quark sector). If we set $c_{U}=2/3\varepsilon e$ and $c_{D}=-\varepsilon e/3$, the branching ratios for $b\to sV$ and $s\to dV$ are obtained by rescaling by $\varepsilon^{2}$ the SM predictions for the $b\to s\gamma$ and $s\to d\gamma$ processes, up to simple phase-space corrections.

Specifically, in the $B$ sector, the branching ratio for $b\to sV$ is
\begin{equation}
\mathcal{B}(b\to sV)=|\varepsilon|^{2}\mathcal{B}(b\to s\gamma)^{\text{SM}}\;,\;\mathcal{B}(b\to s\gamma)^{\text{SM}}=(3.15\pm0.23)\cdot10^{-4}~\text{\cite{bsg}\ ,}
\end{equation}
when $m_{V}\ll m_{B}$ and $E_{\gamma,V}>1.6$ GeV. This cut on the photon energy is actually at the opposite end of phase-space compared to those set for $b\to sX$. But even without a definite prediction, it is clear that the expected sensitivity of about $10^{-5}$ in the $B\to(K,K^{\ast})X$ channels would at best probe $\varepsilon$ down to a few percent. For comparison, typical bounds on $\varepsilon$ derived from flavor-blind hadronic observables are currently down to the $10^{-3}$ range~\cite{Hidden,Searches}.

The situation is worse in the $K$ sector, where only CP-violating observables are sensitive to the short-distance ($c$ and $t$) magnetic operator. As analyzed in Ref.~\cite{RadPaper}, those are beyond experimental reach even in the SM case, and thus cannot be used to set constraints on $\varepsilon$. This is actually clear from Table~\ref{TableBND1}: rescaling by $k^{sd}\sim10^{-6}$, the scale $\Lambda$ ends up well below the electroweak scale.

So, rare $K$ and $B$ decays are rather ineffective at constraining the presence of a new flavor-blind vector coupled exclusively to heavy quarks. Fortunately, in many cases, as e.g. from Eq.~(\ref{KinMix}), universality holds and this vector must also couple to light quarks, where the situation is much better.

\subsubsection*{Phenomenological constraints on the couplings to light quarks\label{Vectors}}

In this case, the CKM factors strongly favor the $K$ sector to derive competitive bounds. At the $K$ mass scale, only the $u$, $d$, and $s$ quarks are active quark degrees of freedom. Adopting a matrix notation in the $q=(u,d,s)$ flavor space, $\mathcal{H}_{eff}^{\mathrm{V}}[$III$]$ takes the
form%
\begin{equation}
\mathcal{H}_{eff}^{\mathrm{V}}[\text{III}]=e\;\bar{q}\gamma_{\mu}\mathbf{Q}^{\prime}q\times V^{\mu},\;\;\mathbf{Q}^{\prime}=\varepsilon\mathbf{Q}+\varepsilon^{\prime}\mathbf{1}\;,\;\varepsilon\equiv\frac{c_{U}-c_{D}}{e},\;\varepsilon^{\prime}\equiv\frac{c_{U}+2c_{D}}{3e}\;,
\end{equation}
with $\mathbf{1}=\operatorname*{diag}(1,1,1)$, $\mathbf{Q}=\operatorname*{diag}(2/3,-1/3,-1/3)$, and $e$ the QED coupling constant. So, from the point of view of low energy physics, there are only two
possibilities: either $V_{\mu}$ is effectively aligned with the photon ($\varepsilon$ term) or its charges are proportional to baryon number ($\varepsilon^{\prime}$ term)~\cite{Fayet}. This $\mathcal{H}_{eff}^{\mathrm{V}}[$III$]$ coupling must be directly embedded within Chiral Perturbation Theory (ChPT)~\cite{ChPT}. At the leading $p^{2}$ order, the $V^{\mu}$ field enters only through the covariant derivative acting on the meson fields%
\begin{equation}
D_{\mu}U=\partial_{\mu}U-ieA_{\mu}\left[  \mathbf{Q},U\right]  -iV_{\mu}\left[  \mathbf{Q}^{\prime},U\right]  =\partial_{\mu}U-ie(A_{\mu}+\varepsilon V_{\mu})\left[  \mathbf{Q},U\right]  \;. \label{ChPTcd}
\end{equation}
The $\varepsilon^{\prime}$ term cancels out in the commutator, leaving $V_{\mu}$ coupled exactly like the photon $A_{\mu}$. This ensures the absence of a direct $K\to\pi V$ coupling at leading order, relegating them to $\mathcal{O}(p^{4})$. Such a direct leading order coupling only exists when the $d$ and $s$ charges are different. Indeed, in that case, the generator $\mathbf{Q}^{\prime}$ would no longer commute with that of the weak interaction. This is another way to see that when the universality~(\ref{universal}) fails, the dimension-four FCNC couplings of $\mathcal{H}_{mat}^{\mathrm{V}}[$I$]$ should be allowed.

To get bounds on $\varepsilon$ is rather immediate since the phenomenology is completely analogous to that of the radiative $K$ decays (see Ref.~\cite{RadPaper} for a recent review). It suffices to consider the dominant radiative modes and replace one photon by $V$. When it is massless, the rates are obtained from those in QED simply by rescaling by $\varepsilon^{2}$ (or by $\alpha^{\prime}/\alpha$, if one defines $\alpha^{\prime}\equiv\varepsilon^{2}\alpha$). When massive, the amplitudes are essentially the same, the polarization sum is identical (since the QED Ward identity holds), so the main impact is a reduced sensitivity due to the truncated phase-space. Assuming that about $10^{13}$ kaon decays will be analyzed in the next generation of experiments, and bounds in the $10^{-12}$ range are set, the reach in $\varepsilon$ is thus naively
\begin{equation}
\varepsilon^{2}\lesssim\frac{\mathcal{B}(K\to n\pi+m\gamma+V)}{\mathcal{B}(K\to n\pi+(m+1)\gamma)}\sim\frac{10^{-12}}{\mathcal{B}(K\to n\pi+(m+1)\gamma)}\;, \label{NaiveVB}
\end{equation}
for massless (or very light) vector boson $V$, and $n+m>0$, $n<4$. Since the current bounds on $\varepsilon$ are down to the $10^{-3}$ range, competitive bounds could be obtained from all the modes with $\mathcal{B}(K\to n\pi+(m+1)\gamma))\gtrsim10^{-7}$. Those are listed in Table~\ref{TableVB}.

\begin{table}[t]
\centering                                                        
\begin{tabular}
[c]{lll}\hline
& Experiment (for $V=\gamma$)~\cite{PDG} & Indicative reach\\\hline
$K_{L}\to\gamma V$ & $5.47(4)\cdot10^{-4}$ & 
$|\varepsilon,\varepsilon^{\prime}|\lesssim4\cdot10^{-5}\;[m_{V}\ll m_{K}]$\\
$K_{L}\to\pi^{0}\gamma V$ & $1.273(34)\cdot10^{-6}$ & 
$|\varepsilon|\lesssim1\cdot10^{-3}\;[m_{V}\ll m_{K}]$\\
$K_{S}\to\pi^{0}V$ & $-$ & 
$|\varepsilon|\lesssim3\cdot10^{-3}\;[m_{V}\approx m_{\pi}]$\\
$K_{L}\to\pi^{+}\pi^{-}V$ & $2.83(11)\cdot10^{-5}\;[DE]$ & 
$|\varepsilon|\lesssim2\cdot10^{-4}\;[m_{V}\ll m_{K}]$\\\hline
$K^{+}\to\pi^{+}\gamma V$ & $1.10(32)\cdot10^{-6}$ & 
$|\varepsilon|\lesssim1\cdot10^{-3}\;[m_{V}\ll m_{K}]$\\
$K^{+}\to\pi^{+}V$ & $-$ & 
$|\varepsilon|\lesssim5\cdot10^{-4}\;[m_{V}\approx m_{\pi}]$\\
$K^{+}\to\pi^{+}\pi^{0}V$ & $6.0(4)\cdot10^{-6}\;[DE]$ & 
$|\varepsilon|\lesssim4\cdot10^{-4}\;[m_{V}\ll m_{K}]$\\\hline
\end{tabular}
\caption{Indicative experimental reach of the radiative decays for a new light vector boson aligned with the photon ($\varepsilon$) or with baryon number ($\varepsilon^{\prime}$), assuming about $10^{13}$ kaon decays are observed and bounds in the $10^{-12}$ range are set. For the $K^{+}\to\pi^{+}V$ mode, the reach in $|\varepsilon|$ drops as $m_V$ decreases (with limits at the $10^{-3}$ level for $m_V\approx50$ MeV, $10^{-2}$ for $m_V\approx5$ MeV, and only $10^{-1}$ for $m_V\approx0.5$ MeV).}%
\label{TableVB}%
\end{table}

The $K\to\pi V$ channel is special because $K\to\pi\gamma$ is forbidden due to gauge invariance. Further, even when off-shell, $K\to\pi\gamma^{\ast}$ vanishes at leading order in ChPT, and so does $K\to\pi V$. The leading contribution thus starts at $\mathcal{O}(p^{4})$, from loops and local counterterms, and is approximately given by~\cite{DEIP98}%
\begin{equation}
\mathcal{A}(K^{+}(P)\to\pi^{+}V(q))=\varepsilon\frac{eG_{F}}{8\pi^{2}}a_{+}\left(  q^{2}P^{\mu}-q^{\mu}P\cdot q\right)  \varepsilon_{\mu}^{\ast }(q)\;, \label{KSAS}
\end{equation}
with $a_{+}$ an $\mathcal{O}(1)$ constant. The $K_{S}$ rate is expressed similarly in terms of the $\mathcal{O}(1)$ constant $a_{S}$, while that for $K_{2}\approx K_{L}$ is CP-violating and thus driven by heavy quarks (since $\operatorname{Im}(V_{us}^{\ast}V_{ud})=0$). From this amplitude, we get the rate%
\begin{equation}
\Gamma(K^{+}\to\pi^{+}V)=\varepsilon^{2}\alpha\frac{G_{F}^{2}m_{K}^{5}}{1024\pi^{4}}|a_{+}|^{2}\frac{\lambda^{3/2}(1,r_{\pi}^{2},r_{V}^{2})}{8\pi}r_{V}^{2}\;,
\end{equation}
with $\lambda(1,r_{\pi}^{2},r_{V}^{2})$ the kinematical function defined in Eq.~(\ref{LKin}), and $r_{i}=m_{i}/m_{K}$. Normalizing with the $K^{+}\to\pi^{+}e^{+}e^{-}$ process to get rid of $a_{+}$,%
\begin{equation}
\mathcal{B}(K^{+}\to\pi^{+}V)=\frac{\varepsilon^{2}}{\alpha}\frac{3\lambda^{3/2}(1,r_{\pi}^{2},r_{V}^{2})}{8\Phi_{e}}r_{V}^{2}\times\mathcal{B}\left(  K^{+}\to\pi^{+}e^{+}e^{-}\right)  \;,
\end{equation}
with $\Phi_{e}\approx0.145$ the $K^{+}\to\pi^{+}e^{+}e^{-}$ phase-space factor, and $\mathcal{B}\left(  K^{+}\to\pi^{+}e^{+}e^{-}\right)  =(3.00\pm0.09)\cdot10^{-7}$~\cite{PDG}. Reminiscent of $K\nrightarrow\pi\gamma$, the rate vanishes in the $m_{V}\to0$ limit. This seriously hampers the reach in $|\varepsilon|$, as indicated in Table~\ref{TableVB}. Even in the most favorable window $m_{\pi}\lesssim m_{V}\lesssim2m_{\pi}$, $K^{+}\to\pi^{+}V$ is less sensitive to $|\varepsilon|$ than $K_{L}\to\gamma V$ by about an order of magnitude, see the red regions in Fig.~\ref{FigVectorK}.

To get bounds on $\varepsilon^{\prime}$ is more difficult because it cancels out from the $\mathcal{O}(p^{2})$ Lagrangian, and thus also from the $\mathcal{O}(p^{4})$ meson loops (see Eq.~(\ref{ChPTcd})). It can thus occur only in some local counterterms involving the $V_{\mu\nu}$ field strength (of vanishing anomalous dimension since there are no divergent loop contributions), and in the odd-parity anomalous $\mathcal{O}(p^{4})$ Lagrangian. The former are very suppressed compared to the loop contributions induced by QED and by the $\varepsilon$ piece, and will be neglected~\cite{RadPaper}.

Concentrating on the odd-parity sector, only the $K_{L}\to\gamma V$ mode appears useful to constrain $\varepsilon^{\prime}$ (its anomalous amplitude, driven entirely by the up quark~\cite{GST}, is sensitive to both $\varepsilon$ and $\varepsilon^{\prime}$). The magnetic direct emission amplitudes in $K\to\pi\pi\gamma$ are significantly more suppressed and difficult to access experimentally. Rescaling $K_{L}\to\gamma\gamma$ according to Eq.~(\ref{NaiveVB}) shows that a bound on $K_{L}\to\gamma V$ at the $10^{-12}$ level would probe couplings down to at least $|\varepsilon,\varepsilon^{\prime}|\lesssim10^{-4}$. Interestingly, this is more than an order of magnitude better than using the flavor-blind transition $\pi^{0}\to\gamma V$, for which the best limit is $3.3\cdot10^{-5}$, i.e. $|\varepsilon,\varepsilon^{\prime}|\lesssim 6\cdot10^{-3}$~\cite{PDG}. Further, the range of accessible $V$ masses is evidently larger in $K$ decays.

\subsection{Baryon and lepton number violating operators}

As for the dark scalar, Lorentz invariance requires an even number of SM fermion fields. However, the vector field index allows for alternative chiral structures compared to the Weinberg operators~\cite{Dim6}. Specifically, keeping only operators of leading dimensions,
\begin{subequations}
\label{HVBL}
\begin{align}
\mathcal{H}_{\Delta\mathcal{L}=2}^{\mathrm{V,VV}} &  =\frac{c_{1}^{\text{I}}}{\Lambda^{3}}H\bar{L}^{C}\mathcal{D}_{\mu}LH\times V^{\mu}+\frac{c_{2}^{\text{I}}}{\Lambda^{3}}H\bar{L}^{C}L\mathcal{D}_{\mu}H\times V^{\mu}+\frac{c_{3}^{\text{I}}}{\Lambda^{3}}H\bar{L}^{C}LH\times V_{\mu}V^{\mu}\nonumber\\ &
\;\;\;\;+\frac{c_{1}^{\text{II}}}{\Lambda^{3}}H\bar{L}^{C}\sigma_{\mu\nu}LH\times V^{\mu\nu}+h.c.\;,\\
\mathcal{H}_{\Delta\mathcal{B}=-\Delta\mathcal{L}}^{\mathrm{V,VV}} &
=\frac{c_{1,b}^{\text{I}}}{\Lambda^{3}}\bar{D}^{C}b_{\mu\nu}D\times\bar{E}\gamma^{\nu}D\times V^{\mu}+\frac{c_{2,b}^{\text{I}}}{\Lambda^{3}}\bar{D}^{C}b_{\mu\nu}D\times\bar{L}\gamma^{\nu}Q\times V^{\mu}+h.c.\;,\\
\mathcal{H}_{\Delta\mathcal{B}=\Delta\mathcal{L}}^{\mathrm{V,VV}} &
=\frac{c_{1,a}^{\text{I}}}{\Lambda^{4}}J_{a}^{\mu}\times V_{\mu}+\frac{c_{2,a}^{\text{I}}}{\Lambda^{4}}J_{a}\times V_{\mu}V^{\mu}+\frac{c_{1,a}^{\text{II}}}{\Lambda^{4}}J_{a}^{\mu\nu}\times V_{\mu\nu}+h.c.\;,
\end{align}
\end{subequations}
where the tensor $b_{\mu\nu}$ stands for $g_{\mu\nu}$ or $\sigma_{\mu\nu}$, the set $a=QQQL,QQUE,DUUE,DUQL$ denotes the gauge-singlet combinations of fields of the Weinberg operators~\cite{Dim6}, the corresponding currents are defined as%
\begin{subequations}
\begin{align}
J_{ABCD} &  =\bar{A}^{C}B\times\bar{C}^{C}D\;,\\
J_{ABCD}^{\mu} &  =\bar{A}^{C}\mathcal{D}^{\mu}B\times\bar{C}^{C}D,\;\bar{A}^{C}B\times\bar{C}^{C}\mathcal{D}^{\mu}D,\;\bar{A}^{C}\overleftarrow{\mathcal{D}}\hspace{0in}^{\mu}B\times\bar{C}^{C}D\;,\\
J_{ABCD}^{\mu\nu} &  =\bar{A}^{C}\sigma^{\mu\nu}B\times\bar{C}^{C}D,\;\bar{A}^{C}B\times\bar{C}^{C}\sigma^{\mu\nu}D,\;\bar{A}^{C}\gamma^{\mu}B\times\bar{C}^{C}\gamma^{\nu}D\;,
\end{align}
\end{subequations}
and where the $SU(2)_{L}$ triplet contraction for the $QQQL$ current is understood. Note also that $c_{1}^{\text{II}}$ is antisymmetric in flavor space, $c_{3}^{\text{I}}$ is symmetric, while $c_{1,2}^{\text{I}}$ have no particular symmetry, though one of them is redundant when flavor-diagonal. For the tensor current, which of the three Dirac structures does exist depends on the chiralities of the fermions involved (and may require reordering the fields using Fierz identities). The only other possible vector current is $J_{D^{C}ULL}^{\mu}$, but it vanishes upon Fierzing due to the antisymmetric $SU(2)_{L}$ contraction of the two lepton doublets and thus requires two more Higgs fields. The superscripts I and II refer to the scenarios discussed previously, i.e. separates those operators which explicitly break a dark gauge invariance associated with $V$ from those which do not.

All these interactions have high dimensions, especially compared to the renormalizable couplings to SM particles of Eqs.~(\ref{HvI}) and~(\ref{HvGauge}). Phenomenologically, they may be relevant only when $\Lambda$ is not too large, which requires the absence of direct FCNC couplings with $V$, see Figs.~\ref{FigVectorK} and~\ref{FigVectorB}. At the same time, the $\Delta\mathcal{B}=\pm\Delta\mathcal{L}$ interactions can induce nucleon decay when $m_{V}<m_{n}$ (note that $\Delta\mathcal{B}=-\Delta\mathcal{L}$ interactions contain at least two $d$ quarks and do not contribute to $p^{+}$ decay at the leading electroweak order), and thus require either the scale $\Lambda$ to be extremely high, or the Wilson coefficients to have highly non-generic flavor structures~\cite{BLMFV}.

A scenario with $m_{B}>m_{V}>m_{n}$ is interesting since astrophysical, leptonic, and nucleon decay bounds are essentially circumvented. In that case, rare $B$ decays into an odd number of baryons (together with an invisible $V$) may offer the best windows for the $\Delta\mathcal{B}=\pm\Delta\mathcal{L}$ interactions. Note, though, that if these interactions occur concurrently to those of Eq.~(\ref{HvI}),~(\ref{HvII1}), or~(\ref{HvGauge}), the $V$ may occur as an intermediate state, bringing back the tight proton decay constraints. It remains to be seen whether in that case, signals in $B$ decays are nevertheless possible. Such virtual exchanges are beyond our scope, since in the present work, we require the dark particle to be sufficiently long-lived to escape as missing energy in rare decays.

Finally, as for the dark scalar scenario, the $\Delta\mathcal{L}=2$ interactions can produce three-body invisible $\nu_L\nu_L V$ final states, but are not particularly interesting for FCNC decays. Indeed, they are unable to induce quark flavor transitions, and the FCNC decays would thus proceed through an extremely suppressed hadronic Higgs penguin.

\section{Invisible spin-3/2 fermion}

Spin-$3/2$ particles are described by Rarita-Schwinger fields, denoted $\Psi_{\mu}$, which transform as spinors with a vector index under the Lorentz group. The corresponding Lagrangian kinetic term can be written as~\cite{Gravitino}
\begin{equation}
\mathcal{L}_{kin}=-\frac{1}{2}\epsilon^{\mu\nu\rho\sigma}\overline{\Psi}_{\mu}\gamma_{5}\gamma_{\nu}\partial_{\rho}\Psi_{\sigma}-\frac{1}{4}m_{\Psi}\overline{\Psi}_{\mu}[\gamma^{\mu},\gamma^{\nu}]\Psi_{\nu}\,. \label{RSLagr}
\end{equation}
In addition, these fields are also subject to the conditions $\!\not\!\Psi=0$ (spin-$3/2$ projection), $(i\slash \hspace{-0.19cm}\partial-m_{\Psi})\Psi^{\mu}=0$ (Dirac equation), and $\partial_{\mu}\Psi^{\mu}=0$ (Lorenz condition). For external states, the spin summation is
performed as~\cite{Gravitino}%
\begin{equation}
\Pi(p)_{\mu\nu}=\sum_{spin}u(p)_{\mu}^{s}\bar{u}(p)_{\nu}^{s}=-(\slash\hspace{-0.19cm}p+m_{\Psi})(P_{\mu\nu}-\frac{1}{3}P_{\mu\rho}P_{\nu\sigma}\gamma^{\rho}\gamma^{\sigma})\;,\;\;P_{\alpha\beta}\equiv g_{\alpha\beta}-\frac{p_{\alpha}p_{\beta}}{m_{\Psi}^{2}}. \label{RSPol}%
\end{equation}
The sum over spin for $v(p)_{\mu}^{s}$ spinors is given by $-\Pi(-p)_{\mu\nu}$.

We distinguish the possible operators for the pair-production of these fields by their Lorenz structures, which can be scalar, vector, or tensor-like (even though Eq.~(\ref{RSLagr}) is written down for a Majorana field, $\Psi$ will be taken as complex from now on). Taking into account the above-stated conditions reduces the possible leading operators to%
\begin{align}
\mathcal{H}_{mat}^{\overline{\Psi}\Psi}  &  =+\frac{c_{L}^{V}}{\Lambda^{2}}\bar{Q}\gamma_{\mu
}Q\times\overline{\Psi}^{\rho}\gamma^{\mu}\Psi_{\rho}+\frac{c_{R}^{V}}%
{\Lambda^{2}}\bar{D}\gamma_{\mu}D\times\overline{\Psi}^{\rho}\gamma^{\mu}%
\Psi_{\rho}+\frac{c_{L}^{A}}{\Lambda^{2}}\bar{Q}\gamma_{\mu}Q\times
\overline{\Psi}^{\rho}\gamma^{\mu}\gamma_{5}\Psi_{\rho}+\frac{c_{R}^{A}%
}{\Lambda^{2}}\bar{D}\gamma_{\mu}D\times\overline{\Psi}^{\rho}\gamma^{\mu
}\gamma_{5}\Psi_{\rho}\nonumber\\
&  \;\;\;\;+\frac{c_{L}^{S}}{\Lambda^{3}}H^{\dagger}\bar{D}Q\times
\overline{\Psi}^{\mu}\Psi_{\mu}+\frac{c_{L}^{P}}{\Lambda^{3}}H^{\dagger}%
\bar{D}Q\times\overline{\Psi}^{\mu}\gamma_{5}\Psi_{\mu}+\frac{c_{R}^{S}%
}{\Lambda^{3}}H\bar{Q}D\times\overline{\Psi}^{\mu}\Psi_{\mu}+\frac{c_{R}^{P}%
}{\Lambda^{3}}H\bar{Q}D\times\overline{\Psi}^{\mu}\gamma_{5}\Psi_{\mu
}\nonumber\\
&  \;\;\;\;+\frac{c_{LS}^{T}}{\Lambda^{3}}H^{\dagger}\bar{D}\sigma_{\mu\nu
}Q\times\overline{\Psi}\hspace{0in}^{[\mu}\Psi^{\nu]}+\frac{c_{LP}^{T}%
}{\Lambda^{3}}H^{\dagger}\bar{D}\sigma_{\mu\nu}Q\times\overline{\Psi}%
\hspace{0in}^{[\mu}\gamma_{5}\Psi^{\nu]}+\frac{c_{LT}^{T}}{\Lambda^{3}%
}H^{\dagger}\bar{D}\sigma_{\mu\nu}Q\times\overline{\Psi}_{\rho}\sigma^{\mu\nu
}\Psi^{\rho}\nonumber\\
&  \;\;\;\;+\frac{c_{RS}^{T}}{\Lambda^{3}}H\bar{Q}\sigma_{\mu\nu}%
D\times\overline{\Psi}\hspace{0in}^{[\mu}\Psi^{\nu]}+\frac{c_{RP}^{T}}%
{\Lambda^{3}}H\bar{Q}\sigma_{\mu\nu}D\times\overline{\Psi}\hspace{0in}^{[\mu
}\gamma_{5}\Psi^{\nu]}+\frac{c_{RT}^{T}}{\Lambda^{3}}H\bar{Q}\sigma_{\mu\nu
}D\times\overline{\Psi}_{\rho}\sigma^{\mu\nu}\Psi^{\rho}\;,
\label{Hspin3/2}
\end{align}
where $\overline{\Psi}\hspace{0in}^{[\mu}\Gamma\Psi^{\nu]}=i(\overline{\Psi}\hspace{0in}^{\mu}\Gamma\Psi^{\nu}-\overline{\Psi}\hspace{0in}^{\nu}\Gamma\Psi^{\mu})/2$. Only the leading operators of each kind are kept; operators with additional derivatives are systematically discarded. For the vectorial couplings, the operators involving $\varepsilon^{\mu\nu\rho\sigma}\overline{\Psi}_{\nu}\gamma_{\rho}\Psi_{\sigma}$ and $\varepsilon^{\mu\nu\rho\sigma}\overline{\Psi}_{\nu}\gamma_{5}\gamma_{\rho}\Psi_{\sigma}$ have been reduced to the others using the Chisholm identity. Finally, tensor structures are similarly reduced using the Chisholm identity together with $\sigma_{\mu\nu}\varepsilon^{\mu\nu\rho\sigma }=-2i\sigma^{\rho\sigma}\gamma_{5}$, which in particular permits to get rid of the $\epsilon^{\mu\nu\sigma\rho}\overline{\Psi}_{\sigma}\gamma_{5}\Psi_{\rho}$ and $\overline{\Psi}^{\rho}\sigma^{\mu\nu}\gamma_{5}\Psi_{\rho}$ structures.

The effective couplings of dimensions up to six involving gauge or Higgs fields are easy to construct from those in Eq.~(\ref{Spin12Gauge}),%
\begin{align}
\mathcal{H}_{int}^{\overline{\Psi}\Psi}  &  =+\frac{c_{H}^{S}}{\Lambda}H^{\dagger}%
H\times\overline{\Psi}^{\mu}\Psi_{\mu}+\frac{c_{H}^{P}}{\Lambda}H^{\dagger
}H\times\overline{\Psi}^{\mu}\gamma_{5}\Psi_{\mu}\nonumber\\
&  \;\;\;\;+\frac{c_{B}^{S}}{\Lambda}B_{\mu\nu}\times\overline{\Psi}%
\hspace{0in}^{[\mu}\Psi^{\nu]}+\frac{c_{B}^{P}}{\Lambda}B_{\mu\nu}%
\times\overline{\Psi}\hspace{0in}^{[\mu}\gamma_{5}\Psi^{\nu]}+\frac{c_{B}^{T}%
}{\Lambda}B_{\mu\nu}\times\overline{\Psi}_{\rho}\sigma^{\mu\nu}\Psi^{\rho
}\nonumber\\
&  \;\;\;\;+\frac{c_{\tilde{B}}^{S}}{\Lambda}\tilde{B}_{\mu\nu}\times
\overline{\Psi}\hspace{0in}^{[\mu}\Psi^{\nu]}+\frac{c_{\tilde{B}}^{P}}%
{\Lambda}\tilde{B}_{\mu\nu}\times\overline{\Psi}\hspace{0in}^{[\mu}\gamma
_{5}\Psi^{\nu]}+\frac{c_{\tilde{B}}^{T}}{\Lambda}\tilde{B}_{\mu\nu}%
\times\overline{\Psi}_{\rho}\sigma^{\mu\nu}\Psi^{\rho}\nonumber\\
&  \;\;\;\;+\frac{c_{H}^{V}}{\Lambda^{2}}iH^{\dagger}\overleftrightarrow
{\mathcal{D}}_{\mu}H\times\overline{\Psi}^{\rho}\gamma^{\mu}\Psi_{\rho}+\frac{c_{H}^{A}%
}{\Lambda^{2}}iH^{\dagger}\overleftrightarrow{\mathcal{D}}_{\mu}H\times\overline{\Psi
}^{\rho}\gamma^{\mu}\gamma_{5}\Psi_{\rho}\;.
\label{Hspin32Gauge}
\end{align}
After the electroweak symmetry breaking, the situation is similar as for spin 1/2 fields. The dimension-five operators in the first line generate a correction to the $\Psi$ mass (upon enforcing the $\!\not\!\Psi=0$ constraint), those in the second, third, and fourth line couple $\Psi$ to the photon and to the $Z$ boson. Note, though, that $\Psi$ does not become millicharged in the usual sense, as these effective operators do not match those derived from the minimal substitution principle (which, in any case, is not consistent for spin 3/2 particles~\cite{MSspin32}).

The last class is made of operators involving a single $\Psi$ field. As for the spin-1/2 case, Lorentz invariance requires an odd number of SM fermion fields, so these operators break either baryon or lepton number:%
\begin{align}
\mathcal{H}_{\Delta\mathcal{B},\Delta\mathcal{L}}^{\Psi}  &  =\frac{c_{1}%
^{\Delta\mathcal{L}}}{\Lambda}\mathcal{D}_{\mu}H\times\overline{\Psi}\hspace{0in}^{\mu
}L+\frac{c_{B}^{\Delta\mathcal{L}}}{\Lambda^{2}}B_{\mu\nu}H\times
\overline{\Psi}\hspace{0in}^{[\mu}\gamma^{\nu]}L+\frac{c_{W}^{\Delta
\mathcal{L}}}{\Lambda^{2}}W_{\mu\nu}^{i}H\sigma^{i}\times\overline{\Psi
}\hspace{0in}^{[\mu}\gamma^{\nu]}L\nonumber\\
&  \;\;\;\;+\frac{c_{2}^{\Delta\mathcal{L}}}{\Lambda^{2}}\bar{E}\sigma_{\mu
\nu}L\times\overline{\Psi}\hspace{0in}^{[\mu}\gamma^{\nu]}L+\frac{c_{3}%
^{\Delta\mathcal{L}}}{\Lambda^{2}}\bar{D}\sigma_{\mu\nu}Q\times\overline{\Psi
}\hspace{0in}^{[\mu}\gamma^{\nu]}L+\frac{c^{\Delta\mathcal{B}}}{\Lambda^{2}%
}\bar{D}\sigma_{\mu\nu}D^{C}\times\overline{\Psi}\hspace{0in}^{[\mu}%
\gamma^{\nu]}U^{C}+h.c.\;. \label{Spin32BL}%
\end{align}
Notably, compared to the spin-1/2 case~(\ref{Spin12BL}), no renormalizable coupling can be constructed, and thanks to the extra derivative in the dimension-five operator, there is (of course) no direct mixing between $\Psi$ and $\nu_{L}$ after the electroweak symmetry breaking. Phenomenologically, the signatures for the $\Delta\mathcal{B}$ operator would again require specific searches in $B$ decays, while those for the $\Delta\mathcal{L}$ operators are to be found in semileptonic decays. Note, however, that the interactions in Eq.~(\ref{Spin32BL}) are more difficult to access than those for spin 1/2 invisible particles, Eq.~(\ref{Spin12BL}), because the tensor matrix elements vanish, $\langle0|\bar{d}^{I}\sigma^{\mu\nu}u^{J}|P^{+}\rangle=0$. This means that $c_{3}^{\Delta\mathcal{L}}$ does not contribute to the $P^{+}\to\ell^{+}\Psi$ decays, but only enters in the $P\to P^{\prime}\ell\Psi$ decays for which the helicity-allowed SM contribution $P\to P^{\prime}\ell\nu$ is large. Purely leptonic processes are only sensitive to higher-dimensional operators involving covariant derivatives acting on the quark or lepton fields, for example $\bar{D}\mathcal{D}_{\mu}Q\times\overline{\Psi}\hspace{0in}^{\mu}L$. So, the $\Delta\mathcal{L}$ operators do not appear promising in searching for dark spin 3/2 particles, and will not be further considered here.

\subsubsection*{Reduction and phenomenology}

The non-conserved quark flavor-changing neutral currents break the gauge symmetry appearing in the Lagrangian~(\ref{RSLagr}) when $m_{\Psi}\to0$. As a result, the $1/m_{\Psi}$ terms in the spin sum~(\ref{RSPol}) are not projected out, the massless limit is singular, and we can force $\Lambda$ up to arbitrarily high values simply by decreasing $m_{\Psi}$. The situation is analogous to that encountered for the massive vector case in Sec.~\ref{Simplest}, and may resolve itself in a fully dynamical theory in a similar way. To get physically meaningful bounds on the scale $\Lambda$, there are two possible routes.

The first procedure is inspired from the supergravity setting~\cite{Nilles}, where the spin $3/2$ gravitino mass is related to the supersymmetry breaking scale as $\Lambda_{SUSY}=(\sqrt{3}m_{\Psi}M_{Planck})^{1/2}$ with $M_{Planck}=(8\pi G_{N})^{-1/2}$. In some sense, this can be understood as the fermionic equivalent of the constraint $m_{V}\sim gv_{dark}$ enforced for the vector bosons (which would here be insufficient given the harder $(m_{\Psi}^{-2})^{2}$ singularities occurring when there are two spin-3/2 particles in the final state). Indeed, when $\Lambda_{SUSY}\ll M_{Planck}$, $m_{\Psi}\ $is very small and only those terms originating from the $m_{\Psi}^{-2}$ singularity of the spin sum~(\ref{RSPol}) are relevant~\cite{Gravitino},%
\begin{equation}
\Pi(p)_{\mu\nu}\overset{m_{\Psi}\to0}{=}\frac{2p_{\mu}p_{\nu}}{3m_{\Psi}^{2}}\slash  \hspace{-0.19cm}p\;. 
\label{RSsum0}
\end{equation}
This projects out the $\pm3/2$ helicity states, leaving the $\pm1/2$ goldstino helicity states in a way similar to Eq.~(\ref{VectorScalar}) for the massive vector boson. Specifically, the spin-3/2 operators become equivalent, in the $m_{\Psi}\to0$ limit, to the spin-1/2 derivative operators obtained by replacing%
\begin{equation}
\Psi_{\mu}\overset{m_{\Psi}\to0}{=}\sqrt{\frac{2}{3}}\frac{\partial_{\mu}\psi}{m_{\Psi}}\,.
\end{equation}
Given that the effective operators are at least of dimension six, there are at least two powers of $\Lambda=M_{Planck}$ which can be eaten away by enforcing $\Lambda m_{\Psi}\to\Lambda_{SUSY}^{2}$.

The supergravity scenario is thus characterized by the rescaling $\Lambda_{SUSY}=(\sqrt{3}m_{\Psi}M_{Planck})^{1/2}$. So, even if $\Psi$ is here not necessarily identified with a light gravitino~\cite{LightGravitino}, let us assume that
\begin{equation}
\Lambda\to\bar{\Lambda}^{2}/m_{\Psi}\;, 
\label{Spin32RescaleG}
\end{equation}
where $\bar{\Lambda}$ may not be related to $\Lambda_{SUSY}$ in any way but could denote some dark sector symmetry-breaking scale. Phenomenologically, this rescaling permits to derive sensible bounds on $\bar{\Lambda}$ from the rare $P\to P^{\prime}\Psi\overline{\Psi}$ decays, even when $m_{\Psi}\ll m_{P}$.

A second route would start from a basis made entirely of gauge-invariant operators, i.e. involving the field-strength $\Psi_{\mu\nu}\equiv\partial_{\mu}\Psi_{\nu}-\partial_{\nu}\Psi_{\mu}$ and its dual $\tilde{\Psi}_{\mu\nu}\equiv\varepsilon_{\mu\nu\rho\sigma}\partial^{\rho}\Psi^{\sigma}$. Though the $m_{\Psi}\to0$ limit would always be smooth, we do not perform this construction explicitly because with two field strengths, there are too many operators for the basis to be useful phenomenologically. Instead, we simply remark that starting from such a basis of gauge-invariant operators, it must be possible to generate the $\mathcal{H}_{mat}^{\overline{\Psi}\Psi}$ operators by partial integration and use of the EOM, exactly as for the massive vector boson.

Specifically, when only the spin-3/2 EOM is used, $i\gamma_{\mu}\Psi^{\mu\nu}=m_{\Psi}\Psi^{\nu}$ and $\gamma_{\mu}\tilde{\Psi}^{\mu\nu}=-m_{\Psi}\gamma_{5}\Psi^{\nu}$, an extra factor $m_{\Psi}^{2}/\Lambda^{2}$ is generated. For example,
\begin{subequations}
\begin{align}
\frac{1}{\Lambda^{5}}H^{\dagger}\bar{D}\gamma^{\rho}\gamma^{\nu}Q\times \overline{\Psi}_{\sigma\rho}\gamma^{\sigma}\gamma^{\mu}\Psi_{\mu\nu}  & \sim\frac{m_{\Psi}^{2}}{\Lambda^{5}}H^{\dagger}\bar{D}\gamma^{\rho}\gamma^{\nu}Q\times\overline{\Psi}_{\rho}\Psi_{\nu}\;,\\\frac{1}{\Lambda^{4}}\bar{Q}\gamma_{\rho}Q\times\varepsilon^{\rho\alpha\mu\nu}\overline{\Psi}_{\gamma\mu}\gamma^{\gamma}\gamma^{\alpha}\gamma^{\beta}\Psi_{\beta\nu}  &  \sim\frac{m_{\Psi}^{2}}{\Lambda^{4}}\bar{Q}\gamma_{\alpha}Q\times\overline{\Psi}_{\mu}\gamma^{\alpha}\Psi^{\mu}\;.
\end{align}
\end{subequations}
This is analogous to the reduction~(\ref{ReducV1}) for the vector boson. By contrast, whenever the reduction involves the quark field EOM~(\ref{EOM}), the gauge invariance ends up hidden in relationships among the $c_{i}$ of $\mathcal{H}_{mat}^{\overline{\Psi}\Psi}$, exactly like in Eq.~(\ref{ReducV2}). The resulting operators are then suppressed either by the light quark masses, or by derivatives acting on the quark fields.

Enforcing some cancellations among the operators is incompatible with our procedure of turning on one operator at a time. So, we consider only the situation where the $P\to P^{\prime}\Psi\overline{\Psi}$ decay rates are finite in the $m_{\Psi}\to0$ limit thanks to the rescaling%
\begin{equation}
c_{i}\to\bar{c}_{i}\frac{m_{\Psi}^{2}}{\Lambda^{2}}\;,
\label{Spin32Rescale}%
\end{equation}
with $\bar{c}_{i}\sim\mathcal{O}(1)$. Note that this cures the singularity~(\ref{RSsum0}). Away from the strict $m_{\Psi}=0$ limit, the other terms of the spin sum~(\ref{RSPol}) also contribute and tend to suppress the rates.

To derive the bounds on the scale $\Lambda$ from the rare decays, we must impose one of the above two prescriptions~(\ref{Spin32RescaleG}) or~(\ref{Spin32Rescale}) to make sense of the $m_{\Psi}\to0$ singularities. Comparing them, these rescalings appear precisely equivalent for the dimension-six operators of $\mathcal{H}_{mat}^{\overline{\Psi}\Psi}$. For the dimension-seven operators, the gravitino-like rescaling~(\ref{Spin32RescaleG}) leads to an additional suppression by $m_{\Psi}/\Lambda$ compared to~(\ref{Spin32Rescale}), making them completely irrelevant (remember that $m_{\Psi}$ is assumed smaller than $m_{K,B}$). Thus, the bounds we quote for these dimension-seven operators are understood to hold only for the second scenario.

As for the other types of invisible particles, we rewrite the various operators in terms of currents of definite $C$ and $P$ by introducing the fourteen complex couplings (for each $s\to d$, $b\to s$, and $b\to d$ operators)%
\begin{equation}
f_{XV,XA}=\frac{\bar{c}_{R}^{X}\pm\bar{c}_{L}^{X}}{2}\;(X=V,A)\;,\;\;
f_{XS,XP}=\frac{v}{\Lambda}\frac{\bar{c}_{R}^{X}\pm\bar{c}_{L}^{X}}{2}\;(X=S,P),\;\;
f_{TX,\tilde{T}X}=\frac{v}{\Lambda}\frac{\bar{c}_{RX}^{T}\pm\bar{c}_{LX}^{T}}{2}\;(X=S,P,T)\;.
\end{equation}
The rates and differential rates are in App.~\ref{AppKspin32}~(\ref{AppBspin32}) for $K$ ($B$) decays, and the corresponding bounds are shown in Tables~\ref{TableBND2} and~\ref{TableBNDB2}. As these numbers show, the rescaling prescription pushes the dimensionality of the operators to eight or nine. For such dimensions, the accessible scales $\Lambda$ are at or even below the electroweak scale, as expected from Eq.~(\ref{SMreach}) and Table~\ref{Reach}, and the viability of the $SU(2)_{L}\otimes U(1)_{Y}$ effective operator formalism becomes questionable. Said differently, if the scale $\Lambda$ is above the electroweak scale, the presence of a dark spin 3/2 fermion should have no impact on rare FCNC decays.

\section{Conclusions}

In this paper, we presented a complete basis of $SU(3)_C\otimes SU(2)_L\otimes U(1)_Y$ invariant operators involving SM fields together with a yet undiscovered light invisible spin 0, 1/2, 1, or 3/2 state, neutral under the SM gauge group. As summarized in Table~\ref{TableRefs}, the operators are organized into three classes: couplings to SM fermions, couplings to SM gauge and/or Higgs fields, and baryon/lepton number violating couplings. We retained the operators of lowest dimensions separately for each class. As a result, most of them do not strictly qualify as portals since they are suppressed by the NP scale $\Lambda$. However, it makes sense to extend this denomination to those operators for which the experimental constraints push $\Lambda$ far above the electroweak scale. For example, the typical scale for a dimension-five FCNC operator is greater than 10000 TeV (see Table~\ref{SMreach}), while it can even be close to the Planck scale for those inducing proton decay. For this reason, in the present paper, we systematically investigated the FCNC operators, and derived bounds from the rare FCNC transitions.

\begin{table}[t]
\centering
\begin{tabular}
[c]{cc|cc|cc|cc}\hline
&  & $\mathcal{H}_{mat}$  & Dim & $\mathcal{H}_{int}$  & Dim & $\mathcal{H}_{\Delta\mathcal{B},\Delta\mathcal{L}}$  & Dim\\\hline
1/2 & \multicolumn{1}{r|}{$\begin{array}[c]{r}\psi:\\\psi\psi:\end{array}$} &
\tpair{-\;}{\text{(\ref{Hspin12})}} & \tpair{\;}{6} & \tpair{-\;\;}{\text{(\ref{Spin12Gauge})}} & \tpair{\;}{\bar{5}} & \tpair{\text{(\ref{Spin12BL})}}{\;} & \tpair{6\;(\bar{4})}{(\geq 8)}\\\hline
0 & \multicolumn{1}{r|}{$\begin{array}[c]{r}\phi:\\\phi\phi:\end{array}$} & \tpair{\text{(\ref{Axion})}}{\text{(\ref{H2scalars})}} & \tpair{\bar{5}}{\bar{6}} & \tpair{\text{(\ref{HScalarGauge1})}}{(\text{\ref{HScalarGauge2})}} & \tpair{\bar{3}}{\bar{4}} & (\ref{HSBL}) & \tpair{7\;(\bar{6})}{8\;(\bar{7})}\\\hline
1 & \multicolumn{1}{r|}{$\begin{array}[c]{r}\text{Direct,}\\\;\\\text{Gauge,}\\\;\end{array}$ $\begin{array}[c]{r}V:\\VV:\\V:\\VV:\end{array}$} & $\begin{array}[c]{c}\text{(\ref{HvI})}\\\text{(\ref{HVVI})}\\\text{(\ref{HvII1})}\\\text{(\ref{HvII2})}\end{array}$ & $\begin{array}[c]{c}4\\\bar{6}\\\bar{6}\\\bar{8}\end{array}$ & (\ref{HvGauge}) & $\begin{array}[c]{c}\bar{4}\\\bar{4}\\4\\\bar{6}\end{array}$ & $(\ref{HVBL})$ & $\begin{array}[c]{c}7\;(\bar{7})\\8\;(\bar{7})\\8\;(\bar{7})\\10\;(\bar{9})\end{array}$\\\hline
3/2 & \multicolumn{1}{r|}{\tpair{\Psi:}{\Psi\Psi:}} & \tpair{-\;\;}{\text{(\ref{Hspin3/2})}} & \tpair{\;}{6} & \tpair{-\;\;}{(\text{\ref{Hspin32Gauge}})} & \tpair{\;}{\bar{5}} & \tpair{\text{(\ref{Spin32BL})}}{\;} & \tpair{6\;(5)}{(\geq 8)} \\ \hline
\end{tabular}\caption{References in the text for the various pieces of the basis of effective operators. Dimensions are denoted with a bar when the leading operator involves a Higgs field reducible to its vacuum expectation value after the electroweak symmetry breaking. In the last column, the dimensions are indicated for $\Delta \mathcal{B}$ operators, irrespective of their $\Delta \mathcal{L}$ components, while the dimensions in parenthesis are those of the purely $\Delta \mathcal{L}$ operators. For dark vector fields, we distinguish between direct couplings, for which the $m_V\to 0$ limit is formally divergent, from those where a dark gauge invariance effectively survives (even though full invariance is not imposed, since we allow for $m_V > 0$). Finally, the dimensions of the operators involving spin 3/2 fields increase when ensuring a sensible $m_{\Psi}\to 0$ limit, see Eqs.~(\ref{Spin32RescaleG}) and~(\ref{Spin32Rescale}).}
\label{TableRefs}
\end{table}

Our results can be split into two parts: those concerning the basis of operators itself, and those related to its phenomenological impact on the rare decays. Starting with the basis, some of the main features are:

\begin{enumerate}
\item First and foremost, it must be stressed that even though we concentrated on the rare $K$ and $B$ decays involving missing energy to derive bounds on the operators, our basis is completely general and could be used equally well to investigate signals e.g. in lepton flavor violating transitions, flavor-blind quark or lepton observables, or at high energy colliders (see e.g. Ref.~\cite{HFCNC}).

\item For all spins, there is in the basis an operator involving only the SM Higgs field coupled to the dark state. Though specific models may not generate such operators, their presence would have two consequences. First, in general, the invisible state can no longer be naturally massless, though it can be very light, since a mass shift arises after the electroweak symmetry breaking. Second, the NP scale should always be greater than the electroweak scale, $\Lambda > v\approx246$ GeV, otherwise these corrections grow unchecked as powers of $H^{\dagger}H$ are inserted (this is actually true for all our effective operators). It should be noted though that for spin 1 and spin 3/2 particles, enforcing a dark gauge invariance on the couplings to SM fields explicitly forbids such mass terms.

\item The contribution to FCNC transitions of the flavor-blind operators, i.e. either those involving SM gauge and/or Higgs fields or those involving SM fermion fields of the same flavor, has been clarified. Specifically, for scales $\Lambda$ much higher than the electroweak scale, we pointed out that it is always advantageous to dress flavor-blind operators at the low scale with a $W$ exchange (see Table~\ref{TableClasses}). Indeed, such a low-scale GIM breaking does not necessitate additional Higgs tadpoles, which would each bring in a $1/\Lambda$ suppression.

\item  For spin 1 and 3/2 dark particles, special care was devoted to maintaining a sensible massless limit, or at least to interpret the seemingly divergent limit. Indeed, the flavor-changing neutral currents are not conserved in general, so that the $1/m_V^2$ ($1/m_{\Psi}^2$) term of the polarization (spin) sum is not projected out in physical observables. This is particularly relevant for massive vector states, for which renormalizable couplings to non-conserved quark currents can be constructed. Several mechanisms were discussed, and the corresponding NP scales derived from experimental bounds on the rare decay branching ratios were compared. As shown in Figs.~\ref{FigVectorK} and~\ref{FigVectorB}, these scales strongly depend on the assumed dark sector dynamics.

\item All the leading operators producing a single dark fermion, whether of spin 1/2 or 3/2, violate either baryon ($\mathcal{B}$) or lepton ($\mathcal{L}$) number, but not both simultaneously, and have dimensions smaller or equal to that of the FCNC operators. By contrast, most of the $\Delta\mathcal{B}$ and $\Delta\mathcal{L}$ violating operators involving a dark vector or scalar particle directly derive from the dimension-six $\Delta\mathcal{B}=\Delta\mathcal{L}=1$ Weinberg operators, or from the dimension-five $\Delta\mathcal{L}=2$ operator. The former are negligible when they induce proton decay, i.e. when $m_{\phi,V}< m_p$, but could induce exotic $B$ decays into an odd number of baryons plus missing energy when $m_B > m_{\phi,V}> m_p$, since then proton decay is kinematically forbidden. The latter $\Delta\mathcal{L}=2$ operators do not induce the quark flavor transitions, hence have a negligible impact on rare decays.

\item To implement the $SU(2)_L\otimes U(1)_Y$ invariance, the FCNC operators are constructed in terms of the SM chiral fermions. Though theoretically sound, and particularly convenient to implement the MFV flavor restrictions, such a basis is not convenient phenomenologically because many of these operators interfere in physical observables. So, our bounds are always derived using the alternative basis of operators obtained by projecting on currents of definite $C$ and $P$. This minimizes interference terms, since in most cases these currents produce different final states.

\item Although many examples of NP models involving new light states were mentioned, no attempt was made to precisely match them onto our basis of operators. Indeed, this would require dwelling into the detailed dynamics and parameters of each model, and would bring us too far from our main objectives, which were to construct the most general basis and constrain its operators from rare FCNC decays.
\end{enumerate}

Concerning the rare FCNC decays of the $K$ and $B$ mesons, let us remind that provided the non-standard light states are neutral and sufficiently long-lived, they would show up as missing energy. Those modes cannot be experimentally distinguished from the SM processes producing a neutrino pair in the final state. However, these SM processes are very suppressed, and thus in principle, the rare decays could permit to identify even tiny NP effects. The main outcomes of our detailed phenomenological analysis of such effects are:

\begin{enumerate}
\setcounter{enumi}{7}
\item First, we stressed the importance of including the correct kinematical dependences for probing NP operators. This is crucial because experimentally, the rare decay modes with missing energy do not allow for a complete kinematical reconstruction, and require aggressive background suppressions. In practice, most experimental analyses implicitly assume at various stages that the differential rates have the shapes predicted by the SM, and this seeps through down to the final bounds on the branching ratios. Note, importantly, that these dependences are not always accounted for by simply enforcing the various experimental kinematical cuts. For these reasons, the specific kinematical dependences of each NP effect may have to be implemented by the experimentalists (those are detailed in App.~A for $K$ decays, and App.~B for $B$ decays). This provision has to be kept in mind when interpreting our bounds.

\item In the $K$ sector, the sensitivities of the $K_L\to X$, $K\to\pi X$, $K_L\to\gamma X$, and $K\to\pi\pi X$ channels were compared, see Tables~\ref{TableBND2} and~\ref{TableBND1}, with $X$ a single or a pair of dark particles. The two-body $K\to X$ mode turns out to be the most sensitive, though for a very limited number of operators, but it is also the most difficult to deal with experimentally. At the other extreme, the $K\to\pi\pi X$ channels are sensitive to nearly all possible operators, but do not appear competitive given their phase-space and chiral suppressions. This leaves the $K\to\pi X$ and $K_L\to\gamma X$ channels, whose sensitivities to NP operators are in general comparable. Still, it should be noted that the  latter, not yet considered experimentally, has some advantages. First, its SM contribution $K_L\to\gamma \nu \bar{\nu}$ is at the $10^{-13}$ level (see App.~\ref{AppSMrates}), and thus cannot obscure even a tiny NP contribution. Second, for massive flavor-blind dark vector bosons or for millicharged fermions, as derived e.g. from a dark $U(1)$ kinetically mixed with $U(1)_Y$~\cite{Holdom}, the $K_L\to\gamma X$ mode is significantly superior to $K\to\pi X$, whose relevant matrix elements vanish at leading order in Chiral Perturbation Theory.

\item In the $B$ sector, the sensitivities of the $B_{s,d}\to X$, $B\to(K,K^{\ast})X$, and $B\to(\pi,\rho)X$ decay channels were compared, see Tables~\ref{TableBNDB2} and~\ref{TableBNDB1}. As for $K$, the fully invisible decays are both the most sensitive and the most difficult to probe experimentally. The main new feature compared to the $K$ sector, besides an extended kinematical range allowed for $m_X$, is that the modes with two light mesons in the final states can resonate, so that $B\to K^{\ast} X$, and $B\to \rho X$ are competitive. This is particularly interesting since these modes are sensitive to all the quark currents but the scalar $\bar{b}s$ and $\bar{b}d$. It should be said also that the present sensitivity of $b\to s$ and $b\to d$ decays, in terms of NP scales, is very similar. So, if the NP operators do not follow an MFV-like scaling, a small NP effect could be easier to identify in the latter.

\item The relative sensitivity of $K$ and $B$ decays was also compared. As expected, if the flavor structures of the NP operators involving X are generic, $K$ decays are far more sensitive than $B$ decays. However, it is well-known that in the visible sector, generic NP flavor structures are at odds with current experimental constraints. If MFV is imposed on both the visible and dark sector operators, the constraints from $B$ decays become often tighter than from $K$ decays (see Fig.~\ref{FigKandB}), especially for chirality flipping currents $q^I(1,\gamma_5,\sigma^{\mu\nu})q^J$, relatively suppressed by $m_s/m_b$, and for low-dimensional operators. Indeed, the impact of MFV on the scale $\Lambda$ for an operator of dimension $n$ decreases as $n$ increases, since it is approximatively given by $\Lambda_{MFV}/\Lambda \approx (V_{tI}^{\ast}V_{tJ})^{1/(n-4)}$ for the $d^{J}\to d^I$ transitions, and with the CKM coefficients given in Eq.~(\ref{MFVscaling}). Note, however, that $n$ cannot be too large, since rare decay constraints give $\Lambda\lesssim v$ when $n\gtrsim 8$, see Table~(\ref{Reach}). In other words, for $\Lambda \gtrsim v$, the impact of such operators on the rare decays is beyond reach.

\item The $\Delta \mathcal{B}$ and $\Delta \mathcal{L}$ operators have low dimensions only for $X = \psi$ or $\Psi$. The $\Delta \mathcal{B}=1$ operators can only be probed with specific searches in $B$ decays involving an odd number of baryons plus missing energy in the final state, and should certainly be included in future experimental programs. For the $\Delta \mathcal{L}=1$ effects, the low-dimensional operators are accessible only for $X = \psi$, which contribute to $P^{+}\to \ell^{+}\psi$ ($P=K,D,B$), and would thus apparently enhance the purely leptonic transitions $P^{+}\to \ell^{+}\nu$. If the flavor structure of these NP operators is non-universal, this could resolve the persistent discrepancy in $B\to \tau \nu$ while remaining consistent with the $B\to (e,\mu) \nu$ bounds, as well as, if $m_{\psi}<m_K$, with the tight lepton universality constraint derived from $K_{\ell 2}$ decays, see Eq.~(\ref{univlept}).
\end{enumerate}

In conclusion, the presence of a light invisible state weakly coupled to SM particles is not only far from excluded, but is even compelling in many NP models. To find such states, a host of experimental facilities currently available or in planning are called in, from high-intensity meson and lepton factories to high energy colliders, neutrino detectors, earth or space-based direct or indirect dark matter searches, high intensity lasers,... In this big picture, the very rare FCNC decays of the $K$ and $B$ mesons, with their unique sensitivities and kinematical ranges, could play a crucial role in the very near future thanks to the leap in luminosity expected at the next generation of experiments, namely NA62 at CERN and K0TO at J-Parc dedicated to these $K$ decays, and Super-B in Italy and Belle II at KEK aiming for the $B$ decays.

\subsection*{Acknowledgements}

J.F.K. would like to thank Bo\v{s}tjan Golob for sharing his insights regarding experimental acceptances and background suppression in rare decay measurements. This work is supported in part by the Slovenian Research Agency, by the National Science Foundation under Grant No. 1066293 and the hospitality of the Aspen Center for Physics. C. S. would like to thank the J. Stefan Institute, where part of this work was done, for its hospitality.

\vfill                    
\appendix
\pagebreak 

\section{Differential rates for $K$ decays}

\subsection{Experimental observables in rare $K$ decays\label{AppKexp}}

Let us start by reviewing the kinematics and current experimental limits for the various $K$ decays induced by neutral currents and involving missing energy.

\subsubsection{$K\to\pi+$ missing energy}

When the missing energy consists of two invisible particles, the differential rate depends only on the invariant mass of these particles, $z\equiv q^{2}/m_{K}^{2}$, or equivalently, on the pion momentum $P_{\pi}\equiv|\mathbf{p}_{\pi}|/m_{K}=\sqrt{\lambda}/2$, with $\lambda\equiv\lambda(1,z,r_{\pi}^{2})$ defined in Eq.~(\ref{LKin}), $r_{\pi}\equiv m_{\pi}/m_{K}$. The phase-space integral is then%
\begin{equation}
\mathcal{I}_{\pi XX}=\int_{4r_{X}^{2}}^{\left(  1-r_{\pi}\right)  ^{2}%
}dz\,\frac{d\Gamma}{dz}[z]=\int_{\lambda^{1/2}(1,4r_{X}^{2},r_{\pi}^{2}%
)/2}^{(1-r_{\pi}^{2})/2}\frac{2P_{\pi}dP_{\pi}}{\sqrt{r_{\pi}^{2}+P_{\pi}^{2}%
}}\frac{d\Gamma}{dz}[z(P_{\pi})]\;, \label{DiffKpiXX}%
\end{equation}
with $r_{X}=m_{X}/m_{K}$.

In the SM, the only available invisible particles are the neutrinos. The SM spectrum for $K^{+}\to\pi^{+}\nu\bar{\nu}$ and $K_{L}\to \pi^{0}\nu\bar{\nu}$ then derives entirely from the vector current matrix element $\langle\pi|\bar{s}\gamma^{\mu}d|K\rangle$, and involves the corresponding form-factor (see Eq.~(\ref{KMatEL}) in the next Section) slopes $\lambda_{+}^{\prime}$ and $\lambda_{+}^{\prime\prime}$:%
\begin{equation}
\mathcal{I}_{\pi\nu\bar{\nu}}=\int_{0}^{\left(  1-r_{\pi}\right)  ^{2}%
}dz\lambda^{3/2}\left|  \frac{f_{+}^{K\pi}\left(  z\right)  }{f_{+}^{K\pi
}\left(  0\right)  }\right|  ^{2}=\int_{0}^{\left(  1-r_{\pi}\right)  ^{2}%
}dz\lambda^{3/2}\left(  1+\lambda_{+}^{\prime}\frac{z}{r_{\pi}^{2}}%
+\lambda_{+}^{\prime\prime}\frac{z^{2}}{2r_{\pi}^{4}}\right)  ^{2}\;\;,
\end{equation}
where $i=+,0$. Translated in terms of the pion momentum, i.e. using $z(P_{\pi
})=1+r_{\pi}^{2}-2\sqrt{r_{\pi}^{2}+P_{\pi}^{2}}$, this becomes%
\begin{equation}
\mathcal{I}_{\pi}^{tot}=\int_{0}^{(1-r_{\pi}^{2})/2}dP_{\pi}\frac{16P_{\pi
}^{4}}{\sqrt{r_{\pi}^{2}+P_{\pi}^{2}}}\left(  1+\lambda_{+}^{\prime
}\frac{z(P_{\pi})}{r_{\pi}^{2}}+\lambda_{+}^{\prime\prime}\frac{z(P_{\pi}%
)^{2}}{2r_{\pi}^{4}}\right) ^{2}\;.
\end{equation}
The slopes $\lambda_{+}^{\prime}$ and $\lambda_{+}^{\prime\prime}$ are conventionally normalized by the charged pion mass, and are equal for the $K^{+}$ and $K^{0}$ decays to an excellent approximation (see Ref.~\cite{MesciaS07}). They are extracted from $K_{\ell3}$ decays as%
\begin{equation}
\lambda_{+}^{\prime}=r_{\lambda}\left(  24.82\pm1.10\right)  \cdot
10^{-3}\;,\;\lambda_{+}^{\prime\prime}=r_{\lambda}\left(  1.64\pm0.44\right)
\cdot10^{-3}\;,\;r_{\lambda}=0.990(5)\;,
\end{equation}
with $r_{\lambda}$ a rescaling factor accounting for the $K^{\ast+}-K^{\ast0}$ mass difference. These (highly correlated) errors are negligible compared to the experimental and theoretical errors on the integrated rate, and are neglected in Table~\ref{TableWindows}.

\begin{figure}[t]
\centering     \includegraphics[width=7.7cm]{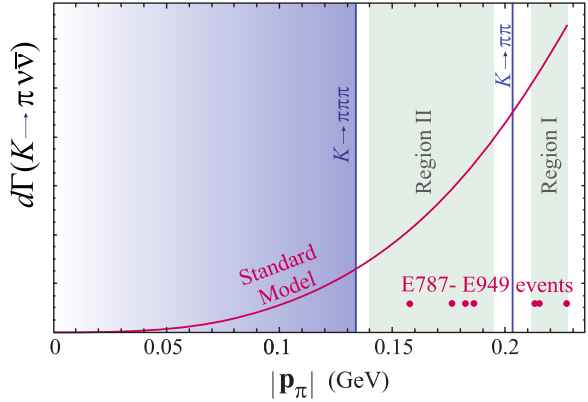}
\caption{The experimental windows in $\pi^{+}$ momentum used for controlling backgrounds in the $K^{+}\to\pi^{+}\nu\bar{\nu}$ measurements, with the seven events seen at Brookhaven. The SM spectrum corresponds to a vector coupling $\bar{s}\gamma_{\mu}d\times\bar{\nu}_{L}\gamma^{\mu}\nu_{L}$, and is implicitly implied in computing the branching ratios from the events.}%
\label{FigWindows}%
\end{figure}

Currently, only the charged decay has been observed at Brookhaven~\cite{KPpnunu}, in two momentum windows separated by the $K^{+}\to\pi^{+}\pi^{0}$ peak, and with the lower end corresponding to the $K\to\pi\pi\pi$ threshold (see Table~\ref{TableWindows} and Fig.~\ref{FigWindows}). The proposed charged $K$ experiment at J-Parc would use the region above the $K^{+}\to\pi^{+}\pi^{0}$ peak~\cite{JPARC}, while the two windows planned at NA62~\cite{NA62P326} and proposed at Fermilab~\cite{ProjectX} are similar. Note that these experiments use very different techniques (stopped vs. in flight), but in both cases, the momentum of the initial and final charged particles are in principle accessible. It is important to stress that not only the combination of the measurements done for each specific window (see Table~\ref{TableWindows}) assumes the SM spectrum, but also that within each window.

For the neutral mode, the $K_{S}\to\pi^{0}\nu\bar{\nu}$ mode is CP-conserving but difficult to access given the very short $K_{S}$ lifetime, so we concentrates on the CP-violating $K_{L}\to\pi^{0}\nu\bar{\nu}$ mode. The best limit~\cite{K0pnunu}%
\begin{equation}
\mathcal{B}(K_{L}\to\pi^{0}\nu\bar{\nu})<2.6\cdot10^{-8}\;,
\label{KLp0nn}%
\end{equation}
was obtained by the E391a experiment at KEK, and will be further improved using the same techniques at J-Parc~\cite{JPARC}. In these experiments, the $K_{L}$ momentum is not fixed. So, the high hermeticity of the detector is essential to ensure sufficient suppression of the backgrounds. Though less effective in this case, kinematical cuts are still useful. In particular, the transverse momentum $P_{T}$ of the reconstructed $\pi^{0}$ is required to be large, between $120$ and $240$ MeV. This does not cut away the background from $K_{L}\to\pi^{0}\pi^{0}$, but rather ensures that the two extra photons have high momentum, and are thus difficult to miss. Since the momentum spectrum of the $\pi^{0}$ cannot be directly measured at KEK or J-Parc, and since the SM decay spectrum is implicitly assumed in the analysis, it is far from immediate to translate the current limit~(\ref{KLp0nn}) into bounds on non-standard currents involving other types of invisible particles.

So, for both the charged and neutral modes, it is not currently possible to deconvolute the SM spectrum from the experimental numbers. To proceed and derive the bounds quoted in the text, we require that the predicted branching ratio for the production of new invisible states does not exceed $10^{-10}$ when integrated over the momentum windows of the charged mode. This is a rather loose approach, which could significantly underestimate the experimental reach in case the spectrum is very different than the SM one. To illustrate this, note that the bounds for two-body decays are already slightly tighter~\cite{KPpX},
\begin{equation}
\mathcal{B}(K^{+}\to\pi^{+}X^{0})<0.73\cdot10^{-10}\;(m_{X}=0)\;.
\label{KTBlim}
\end{equation}
To improve our naive bounds, either the specific modulations of the spectrum in the presence of NP have to be included throughout the experimental analysis\footnote{Brookhaven extracted bounds for purely vector (SM), scalar, or tensor currents, but not for a combination of the SM plus non-standard interactions. Since we cannot turn off the SM rate, this is not directly useful for our purpose.}, or the true momentum spectrum must be measured (maybe using TOF techniques for the neutral mode~\cite{ProjectX}). Finally, independently of the NP spectrum, it should be noted that the sensitivity to the $K\to\pi X(X)$ processes is ultimately bounded at around $10^{-12}$ by the theoretical error on the SM predictions for the $K\to\pi\nu\bar{\nu}$ branching ratios.

\begin{table}[t]
\centering
\begin{tabular}
[c]{ccccc}\hline
$|p_{\pi}|\;(MeV)$ & $z$ & $\mathcal{I}_{\pi}/\mathcal{I}_{\pi}^{tot}$ &
SM & Extrapolated total\\\hline
$\lbrack\;211\;,\;229\;]$ & $[\;0.000\;,\;0.062\;]$ & $27.6\%$ & $0.228(18)$ &
$1.47_{-0.89}^{+1.30}$\\
$\lbrack\;140\;,\;195\;]$ & $[\;0.116\;,\;0.289\;]$ & $39.7\%$ & $0.328(25)$ &
$7.89_{-5.10}^{+9.26}$\\\hline
Combined (Total BR) &  & -- & $0.825(64)$ & $1.73_{-1.05}^{+1.15}$\\\hline
\end{tabular}
\caption{Experimental measurement of $K^{+}\to\pi^{+}\nu\bar{\nu}$ and SM prediction within each momentum window \cite{KPpnunu}, in units of $10^{-10}$.}%
\label{TableWindows}%
\end{table}

\subsubsection{$K\to\pi\pi+$ missing energy}

The phase-space integration for the $K\to\pi(K_{1})\pi(K_{2})X(q)$ decays is%
\begin{equation}
\mathcal{I}_{\pi\pi X}=\int_{4r_{\pi}^{2}}^{(1-r_{X})^{2}}dy\;\frac{d\Gamma}{dy}\;, \label{DiffKpipiX}%
\end{equation}
with $y=(K_{1}+K_{2})^{2}/m_{K}^{2}$ the invariant mass of the pion pair. Compared to $K\to\pi X$, these modes are suppressed by the smaller hadronic matrix elements and by phase-space. Further, the kinematical range is much reduced. Currently, the best limits are (see the respective papers for different $m_{X}$ values)%
\begin{subequations}
\begin{align}
\mathcal{B}(K^{+}  &  \to\pi^{+}\pi^{0}X^{0})<4\cdot10^{-5}\;[m_{X}=50\;MeV]\;\;\text{\cite{PPXE787}},\\
\mathcal{B}(K_{L}  &  \to\pi^{0}\pi^{0}X^{0})<7\cdot10^{-7}\;[m_{X}=50\;MeV]\;\;\text{\cite{PPXE391a}},
\end{align}
and are thus very far from Eq.~(\ref{KTBlim}). Note that the hadronic matrix elements $\langle\pi^{+}\pi^{0}|\bar{s}\Gamma d|K^{+}\rangle$, $\langle\pi^{0}\pi^{0}|\bar{s}\Gamma d|K^{0}\rangle$, and $\langle\pi^{+}\pi^{-}|\bar{s}\Gamma d|K^{0}\rangle$ are related in the isospin limit, see Eq.~(\ref{KppIso}) below, so that these experimental constraints suffice to completely bound the $K\to\pi\pi X$ system.

The $K\left(  P\right)  \to\pi(K_{1})\pi(K_{2})X(p_{1})\bar{X}(p_{2})$ decays are similarly suppressed, and the experimental information is less precise. The phase-space integrals reduce to that over the invariant mass of the invisible pair $T^{2}=(p_{1}+p_{2})^{2}=zm_{K}^{2}$ and of the pion pair $K^{2}=(K_{1}+K_{2})^{2}=ym_{K}^{2}$, over the range%
\end{subequations}
\begin{equation}
\mathcal{I}_{\pi\pi XX}=\int_{4r_{\psi}^{2}}^{\left(  1-2r_{\pi}\right) ^{2}}dz\int_{4r_{\pi}^{2}}^{(1-\sqrt{z})^{2}}dy\;\frac{d^{2}\Gamma}{dydz}\;.
\label{DiffKpipiXX}%
\end{equation}
Here again, the SM spectrum critically enters and the current experimental bound%
\begin{subequations}
\begin{align}
\mathcal{B}(K^{+} & \to\pi^{+}\pi^{0}\nu\bar{\nu})<4.3\cdot 10^{-5}\;\;\text{\cite{PPXE787}\ ,}\\
\mathcal{B}(K_{L} & \to\pi^{0}\pi^{0}\nu\bar{\nu})<8.7\cdot 10^{-7}\;\;\text{\cite{PPXE391a}\ ,}%
\end{align}
cannot immediately be translated into bounds for invisible particles of a different type.

Though probably optimistic given the limited phase-space and complicated signatures, we assume bounds of $10^{-10}$ on each of these $K\to \pi\pi X$ and $K\to\pi\pi XX$ modes are achievable to derive the numbers in Tables~\ref{TableBND2} and~\ref{TableBND1}. In any case, if the bounds are different, it is a simple matter to rescale the numbers accordingly. Further, using the same $10^{-10}$ branching ratio bounds as for the $K\to\pi X(X)$ modes permits to clearly illustrate the reduced sensitivity of the $K\to\pi\pi X(X)$ channels.

\subsubsection{$K\to\gamma+$ missing energy}

Compared to $K\to\pi\pi X$ and $K\to\pi\pi XX$, the modes $K\left(  P\right)  \to\gamma(k)X(p_{1})\bar{X}(p_{2})$ and $K\left( P\right)  \to\gamma(k)X(T)$ are less suppressed and could offer simpler experimental signatures. The phase-space integral for the three-body decay is%
\end{subequations}
\begin{equation}
\mathcal{I}_{\gamma XX}=\int_{4r_{X}^{2}}^{1}dz\;\frac{d\Gamma}{dz}\;,
\label{DiffKgaXX}%
\end{equation}
with $z=(p_{1}+p_{2})^{2}/m_{K}^{2}$ the invariant mass of the invisible pair. Despite their theoretical sensitivity, there is currently no experimental limit on these modes. So, for now, we assume that the next generation of experiments will reach $\mathcal{B}(K_{L}\to\gamma X)<10^{-10}$. Note that with about $10^{12}-10^{13}$ $K_{L}$ decays, as required to measure $K_{L}\to\pi^{0}\nu\bar{\nu}$, this may be pessimistic. Further, while the $K\to\pi\nu\bar{\nu}$ processes ultimately limits the sensitivity to $K\to\pi X(X)$ at a few $10^{-12}$ given the current theoretical errors, the SM rate for $K_{L}\to\gamma\nu\bar{\nu}$ is at the $10^{-13}$ level, so bounds at or even below that level are in principle achievable.

Other modes with photon and missing energy will not be considered, as the $K\to n\pi+\gamma+X$ processes are either too suppressed and difficult to access experimentally when fully neutral (given the many photons from the $\pi^{0}$s), or superseded by the non-radiative processes $K\to n\pi+X$ when some mesons are charged (since the photon of $K\to n\pi+\gamma+X$ is essentially a bremsstrahlung radiation off the charged meson~\cite{RadLO1,RadLO2}, the amplitude for $K\to n\pi+\gamma+X$ is actually driven by that for $K\to n\pi+X$~\cite{Low}).

\subsubsection{$K\to$ missing energy}

The simplest decays are those where the $K_{L}$ or $K_{S}$ simply disappear. Though difficult to probe experimentally, the simpler matrix element together with the minimal number of final state particles strongly enhance their sensitivity to NP effects. In the case of the $\pi^{0}\to XX$ process, the best bound is~\cite{pi0nn}%
\begin{equation}
\mathcal{B}(\pi^{0}\to XX)<0.27\cdot10^{-6}\;.
\end{equation}
But this measurement is actually a by-product of the study of the $K^{+}\to\pi^{+}XX$ decay, since a bound on $K^{+}\to\pi^{+}XX$ indirectly constrains $K^{+}\to\pi^{+}\pi^{0}[\to XX]$. Doing the same for $K_{L}\to XX$ would require very tight bounds on some $B$ or $D$ decays with missing energy, well beyond current capabilities (see Sec.~\ref{AppBexp}). Alternatively, a direct bound on $K_{L}\to XX$ may be obtained from $\phi$ factories, where the other $K$ can be tagged. In deriving NP scales in the text, we will use $\mathcal{B}(K_{L}\to XX)<10^{-10}$ to simplify the comparison with the other modes, but it should be kept in mind that such a bound appears extremely challenging.

\subsection{Matrix elements for $K$ decays\label{AppKmat}}

In the $K$ sector, the quark currents are represented within Chiral Perturbation Theory (ChPT)~\cite{ChPT}. For simplicity, only the leading chiral order is kept. Specifically, the vector and axial-vector currents start at $\mathcal{O}(p)$:%
\begin{equation}
\bar{q}_{L}^{I}\gamma^{\mu}q_{L}^{J}=i\frac{F^{2}}{2}(D^{\mu}U^{\dagger}U)^{JI},\;\;\;\;\bar{q}_{R}^{I}\gamma^{\mu}q_{R}^{J}=i\frac{F^{2}}{2}(D^{\mu}UU^{\dagger})^{JI}\;,\label{VectorC}%
\end{equation}
with, at leading order, $F=F_{\pi}=92.4$ MeV. Thanks to the QED gauge invariance, there is no unknown low-energy constant in these currents. The scalar and pseudoscalar currents start at $\mathcal{O}(p^{0})$:%
\begin{equation}
\bar{q}_{L}^{I}q_{R}^{J}=-\frac{F^{2}}{2}B_{0}U^{JI},\;\;\;\bar{q}_{R}^{I}q_{L}^{J}=-\frac{F^{2}}{2}B_{0}U^{\dagger JI}\;\;,\label{ScalarC}%
\end{equation}
with the low-energy constant $B_{0}$ related to the quark masses,%
\begin{equation}
B_{s}\equiv\frac{B_{0}}{m_{K}}\approx(1-r_{\pi}^{2})\frac{m_{K}}{m_{s}}=4.6(8)\;,
\end{equation}
with $r_{\pi}=m_{\pi}/m_{K}$, and using $m_{s}(2\,$GeV$)=100\pm20$ MeV (so when deriving bounds on $c_{i}/\Lambda^{n}$, $c_{i}\equiv c_{i}(2\,$GeV$)$ is understood)~\cite{PDG}. The scalar and vector currents are related by the EOM. Specifically, the most general matrix elements for the scalar or vector $K\to\pi$ transitions have the form ($z=q^{2}/m_{K}^{2}$, $q=P-K$)%
\begin{equation}
\langle\pi\left(  K\right)  |\bar{s}d|K\left(  P\right)  \rangle
\sim\frac{m_{K}^{2}-m_{\pi}^{2}}{m_{s}-m_{d}}f_{0}\left(  z\right)
\;,\;\langle\pi\left(  K\right)  |\bar{s}\gamma^{\mu}d|K\left(  P\right)
\rangle\sim\left(  P+K\right)  ^{\mu}f_{+}\left(  z\right)  +\left(
P-K\right)  ^{\mu}f_{-}\left(  z\right)  \;,\label{KMatEL}%
\end{equation}
up to some simple Clebsch-Gordan coefficient. Taking the divergence of the vector current produces $q^{2}f_{-}\left(  z\right)  =(m_{K}^{2}-m_{\pi}^{2})\left(  f_{0}\left(  z\right)  -f_{+}\left(  z\right)  \right)  $. So, at the leading chiral order, $f_{+,0}\left(  z\right)  =1$ and $f_{-}\left(z\right)  =0$. Refinements are only needed for a precise prediction of the SM rates, but are not numerically relevant for the bounds on the production of new invisible states. Note that in practice, the scalar and vector currents do not need to be parametrized as external couplings, but can be directly introduced through the ChPT source terms. Doing this using the leading $\mathcal{O}(p^{2})$ Lagrangian reproduces Eq.~(\ref{VectorC}) and Eq.~(\ref{ScalarC}). We do not consider the next-to-leading $\mathcal{O}(p^{4})$ meson loops and local terms, except for the odd-parity contact interactions obtained by introducing the vector and axial vector sources in the $\mathcal{O}(p^{4})$ anomalous Wess-Zumino-Witten (WZW) action~\cite{ChPT}. Indeed, owing to their opposite parity, these interactions drive the leading order contributions for some amplitudes.

Finally, from Lorentz and chiral symmetry, combined with parity and charge conjugation (valid for the strong interactions), the most general parametrization for the tensor current starts at $\mathcal{O}(p^{2})$, where it is given by~\cite{RadPaper}
\begin{subequations}
\label{TensorC}%
\begin{align}
\bar{q}^{I}\sigma_{\mu\nu}P_{L}q^{J} &  =-i\frac{F^{2}}{2}a_{T}\left(  D_{\mu}U^{\dagger}D_{\nu}UU^{\dagger}-D_{\nu}U^{\dagger}D_{\mu}UU^{\dagger}-i\varepsilon_{\mu\nu\rho\sigma}D^{\rho}U^{\dagger}D^{\sigma}UU^{\dagger
}\right)  ^{JI}\nonumber\\
&  \;\;\;\;+\frac{F^{2}}{2}a_{T}^{\prime}((F_{\mu\nu}^{L}-i\tilde{F}_{\mu\nu}^{L})U^{\dagger}+U^{\dagger}(F_{\mu\nu}^{R}-i\tilde{F}_{\mu\nu}^{R}))^{JI}\;\;,\\
\bar{q}^{I}\sigma_{\mu\nu}P_{R}q^{J} &  =-i\frac{F^{2}}{2}a_{T}\left(  D_{\mu}UD_{\nu}U^{\dagger}U-D_{\nu}UD_{\mu}U^{\dagger}U+i\varepsilon_{\mu\nu\rho\sigma}D^{\rho}UD^{\sigma}U^{\dagger}U\right)  ^{JI}\nonumber\\
&  \;\;\;\;+\frac{F^{2}}{2}a_{T}^{\prime}(U(F_{\mu\nu}^{L}+i\tilde{F}_{\mu\nu}^{L})+(F_{\mu\nu}^{R}+i\tilde{F}_{\mu\nu}^{R})U)^{JI}\;.
\end{align}
Two new low-energy constants $a_{T}$ and $a_{T}^{\prime}$ occur, for which we use the Lattice estimates (see the discussion in Ref.~\cite{RadPaper})%
\end{subequations}
\begin{equation}
B_{T}(2\;GeV)=2m_{K}a_{T}=1.21(12)\text{~\cite{Mescia00}}\;,\;\;B_{T}^{\prime}(2\;GeV)=2F_{\pi}a_{T}^{\prime}=0.6(2)\text{~\cite{MS}}\;.
\end{equation}
Note that a more recent lattice estimate $B_{T}=0.65(2)$~\cite{NewBT} is two times smaller, and thus suppresses the sensitivity of $K$ decays to the tensor currents. Still, in terms of the NP scales $\Lambda$ given in the Tables~\ref{TableBND2} and~\ref{TableBND1}, the precise value of $B_{T}$ is not that relevant at present since these numbers are to be understood as order of magnitude estimates.

So, altogether, and defining $\Gamma\equiv c_{S}+c_{P}\gamma_{5}+c_{V}\gamma^{\mu}+c_{A}\gamma^{\mu}\gamma_{5}+c_{T}\sigma_{\mu\nu}+c_{\tilde{T}}\sigma_{\mu\nu}\gamma_{5}$, the matrix elements in the isospin limit and to the leading chiral order are\footnote{A number of sign conventions are implicitly defined by these equations. We closely follow the ChPT conventions of~\cite{ChPT,RadPaper}}
\begin{subequations}
\label{KMat1}%
\begin{align}
\sqrt{2}\langle0|\bar{s}\Gamma d|K^{0}(P)\rangle &  =2iF_{\pi}\left(
-B_{0}c_{P}+P^{\mu}c_{A}\right)  \;,\\
\sqrt{2}\langle\gamma(k,\alpha)|\bar{s}\Gamma d|K^{0}(P)\rangle &
=-\frac{4}{3}eF_{\pi}(a_{T}^{\prime}c_{\tilde{T}}\left(  k^{\mu}g^{\alpha\nu
}-k^{\nu}g^{\alpha\mu}\right)  -ia_{T}^{\prime}c_{T}\epsilon^{\alpha\mu\nu
\rho}k_{\rho}+\frac{c_{V}N_{C}}{8\pi^{2}F_{\pi}^{2}}\varepsilon^{\alpha\mu
\rho\sigma}k_{\rho}P_{\sigma})\;\;,\\
\sqrt{2}\langle\pi^{0}(K)|\bar{s}\Gamma d|K^{0}(P)\rangle &  =-\langle\pi
^{+}(K)|\bar{s}\Gamma d|K^{+}(P)\rangle\nonumber\\
&  =-B_{0}c_{S}+(P^{\mu}+K^{\mu})c_{V}+2a_{T}\left(  ic_{T}K^{[\mu}P^{\nu
]}-c_{\tilde{T}}\varepsilon^{\mu\nu\rho\sigma}K_{\rho}P_{\sigma}\right)  \;\;,
\end{align}
and
\end{subequations}
\begin{subequations}
\label{KMat2}%
\begin{gather}
-\langle\pi^{+}(K_{1})\pi^{0}(K_{2})|\bar{s}\Gamma d|K^{+}(P)\rangle
=\mathcal{M}_{-}\left(  K\to\pi\pi\right)  ,\;\sqrt{2}\langle\pi
^{0}(K_{1})\pi^{0}(K_{2})|\bar{s}\Gamma d|K^{0}(P)\rangle=\mathcal{M}%
_{+}\left(  K\to\pi\pi\right)  \;\;,\nonumber\\
\sqrt{2}\langle\pi^{+}(K_{1})\pi^{-}(K_{2})|\bar{s}\Gamma d|K^{0}%
(P)\rangle=\mathcal{M}_{+}\left(  K\to\pi\pi\right)  +\mathcal{M}%
_{-}\left(  K\to\pi\pi\right)  \;\;,
\end{gather}
with
\end{subequations}
\begin{subequations}
\label{KMat3}%
\begin{align}
\mathcal{M}_{+}(K\to\pi\pi)  &  =\frac{iB_{0}c_{P}}%
{F_{\pi}}\frac{K_{+}\cdot T_{-}}{m_{K}^{2}-T_{-}^{2}}-\frac{ic_{A}}{F_{\pi}%
}\left(  K_{+}^{\mu}+\frac{T_{-}^{\mu}K_{+}\cdot T_{-}}{m_{K}^{2}-T_{-}^{2}%
}\right)  +\frac{a_{T}}{F_{\pi}}\left(  c_{\tilde{T}}K_{+}^{[\mu}T_{-}^{\nu
]}-ic_{T}\varepsilon^{\mu\nu\rho\sigma}T_{-\rho}K_{+\sigma}\right)  ,\\
\mathcal{M}_{-}(K\to\pi\pi)  &  =-\frac{iB_{0}c_{P}%
}{F_{\pi}}\frac{K_{-}\cdot T_{-}}{m_{K}^{2}-T_{-}^{2}}+\frac{ic_{A}}{F_{\pi}%
}\left(  K_{-}^{\mu}+\frac{T_{-}^{\mu}K_{-}\cdot T_{-}}{m_{K}^{2}-T_{-}^{2}%
}\right)  +\frac{a_{T}}{F_{\pi}}\left(  c_{\tilde{T}}K_{-}^{[\nu}T_{+}^{\mu
]}+ic_{T}\varepsilon^{\mu\nu\rho\sigma}T_{+\rho}K_{-\sigma}\right) \nonumber\\
&  \;\;\;\;-\frac{c_{V}N_{C}\varepsilon^{\mu\nu\rho\sigma}K_{+\nu}K_{-\rho
}T_{-\sigma}}{12\pi^{2}F_{\pi}^{3}}\;,
\end{align}
where $K_{\pm}=K_{2}\pm K_{1}$, $T_{\pm}=P\pm K_{+}$, and $X^{[\mu}Y^{\nu]}=X^{\mu}Y^{\nu}-X^{\nu}Y^{\mu}$. Terms proportional to the number of QCD colors, $N_{C}=3$, come from the WZW action. The $m_{K}^{2}-T_{-}^{2}$ denominators arise from the kaon pole topologies, $K\to\pi\pi K^{0}$ followed by $K^{0}\to0$ (from Eq.~(\ref{KMat1})). Note that $\mathcal{M}_{+}(K\to\pi\pi)$ is even under $K_{1}\leftrightarrow K_{2}$, while $\mathcal{M}_{-}(K\to\pi\pi)$ is odd, hence these amplitudes describe two-pion states with even and odd orbital angular momentum, respectively. The $\pi^{0}\pi^{0}$ state is purely even due to Bose statistics, while the $\pi^{+}\pi^{0}$ state has total isospin one, hence is purely odd.

Only the tensor currents contribute to the $\mathcal{M}(K\to\gamma)$ amplitude at tree-level. So, it may seem that together with the $\mathcal{O}(p^{4})$ WZW amplitude, we should include also the even-parity meson loops along with their counterterms. However, for neutral current sources, i.e., in terms of Gell-Mann matrices, for $v_{\mu},a_{\mu},s,p \sim\lambda_{6}\pm i\lambda_{7}$, the only allowed even-parity $K\to\gamma$ matrix elements vanish when the photon is on-shell
\end{subequations}
\begin{subequations}
\label{MatElgK}%
\begin{align}
\langle\gamma(q,\nu)|\bar{s}\gamma_{5}d|K^{0}(P)\rangle &  =\frac{ieB_{0}%
}{72\sqrt{2}\pi^{2}F_{\pi}}\frac{q^{2}T^{\nu}-q^{\nu}q\cdot T}{T^{2}-m_{K}%
^{2}}\frac{\Phi(q^{2},m_{K}^{2})-\Phi(q^{2},m_{\pi}^{2})}{q^{2}}\;,\\
\langle\gamma(q,\nu)|\bar{s}\gamma^{\mu}\gamma_{5}d|K^{0}(P)\rangle &
=\frac{ie}{72\sqrt{2}\pi^{2}F_{\pi}}\left[  q^{\mu}q^{\nu}-q^{2}g^{\mu\nu
}+T^{\mu}\frac{q^{2}T^{\nu}-q^{\nu}q\cdot T}{T^{2}-m_{K}^{2}}\right]
\frac{\Phi(q^{2},m_{K}^{2})-\Phi(q^{2},m_{\pi}^{2})}{q^{2}}\;,
\end{align}
where $T=P-q$, and $\Phi(q^{2},m^{2})$ the loop functions occurring for $K\to\pi\gamma^{\ast}$ (see e.g. Ref.~\cite{RadLO1}), defined in terms of the standard scalar one-loop integrals as
\end{subequations}
\begin{equation}
\Phi(q^{2},m^{2})=3(q^{2}-4m^{2})B_{0}(q^{2},m^{2},m^{2})+12m^{2}B_{0}(0,m^{2},m^{2})-2q^{2}\;.
\end{equation}
Contrary to $K\to\pi\gamma^{\ast}$, the FCNC matrix elements are finite since the UV divergences cancel in the difference between the $K^{\pm}$ and $\pi^{\pm}$ loop contributions, and there are no counterterms. Though in principle, the $K\to\ell^{+}\ell^{-}X$ modes could thus offer precise probes, their rates are far too suppressed by $\alpha$, the loop factors, and the cancellation between the $K^{\pm}$ and $\pi^{\pm}$ loops. So, only $K\to\gamma X$ will be considered here, which is thus induced exclusively by tensor currents and $\mathcal{O}(p^{4})$ anomalous interactions.

Finally, the operator basis also includes vector and scalar currents with a covariant derivative. Though these operators are never retained in deriving bounds on the new physics scale in the main text, for the sake of completeness, let us nevertheless write down the relevant chiral realizations. For the vector currents, extending the analysis of Ref.~\cite{BuchallaI98}, we write%
\begin{subequations}
\begin{align}
\bar{q}^{I}\overleftarrow{D}_{\alpha}\gamma_{\mu}P_{L}q^{J}  &  =i\frac{F^{2}%
}{4}(\partial_{\mu}U^{\dagger}\partial_{\alpha}U+\partial_{\alpha}%
\partial_{\mu}U^{\dagger}U-\frac{1}{8}g_{\alpha\mu}(\chi^{\dagger}%
U+U^{\dagger}\chi)+a_{V}i\varepsilon_{\alpha\mu\beta\nu}\partial^{\beta
}U^{\dagger}\partial^{\nu}U)^{JI}\;,\\
\bar{q}^{I}\overleftarrow{D}_{\alpha}\gamma_{\mu}P_{R}q^{J}  &  =i\frac{F^{2}%
}{4}(\partial_{\mu}U\partial_{\alpha}U^{\dagger}+\partial_{\alpha}%
\partial_{\mu}UU^{\dagger}-\frac{1}{8}g_{\alpha\mu}(\chi U^{\dagger}%
+U\chi^{\dagger})-a_{V}i\varepsilon_{\alpha\mu\beta\nu}\partial^{\beta
}U\partial^{\nu}U^{\dagger})^{JI}\;,\\
\bar{q}^{I}\gamma^{\mu}P_{L}\overrightarrow{D}_{\alpha}q^{J}  &
=-i\frac{F^{2}}{4}(\partial_{\alpha}U^{\dagger}\partial_{\mu}U+U^{\dagger
}\partial_{\alpha}\partial_{\mu}U-\frac{1}{8}g_{\alpha\mu}(U^{\dagger}%
\chi+\chi^{\dagger}U)+a_{V}i\varepsilon_{\alpha\mu\beta\nu}\partial^{\beta
}U^{\dagger}\partial^{\nu}U)^{JI}\;,\\
\bar{q}^{I}\gamma_{\mu}P_{R}\overrightarrow{D}_{\alpha}q^{J}  &
=-i\frac{F^{2}}{4}(\partial_{\alpha}U\partial_{\mu}U^{\dagger}+U\partial
_{\alpha}\partial_{\mu}U^{\dagger}-\frac{1}{8}g_{\alpha\mu}(U\chi^{\dagger
}+\chi U^{\dagger})-a_{V}i\varepsilon_{\alpha\mu\beta\nu}\partial^{\beta
}U\partial^{\nu}U^{\dagger})^{JI}\;.
\end{align}
Most of the terms are fixed by taking divergences and imposing the EOM, but for the constant $a_{V}$, a priori of $\mathcal{O}(1)$. For the scalar currents, the chiral realizations start at $\mathcal{O}(p)$,%
\end{subequations}
\begin{subequations}
\begin{align}
\bar{q}^{I}P_{R}\overrightarrow{D}_{\alpha}q^{J}  &  =\bar{q}^{I}%
P_{R}\overleftarrow{D}_{\alpha}q^{J}=-\frac{F^{2}}{4}B_{0}(D_{\alpha}%
U)^{JI},\\
\bar{q}^{I}P_{L}\overrightarrow{D}_{\alpha}q^{J}  &  =\bar{q}^{I}%
P_{L}\overleftarrow{D}_{\alpha}q^{J}=-\frac{F^{2}}{4}B_{0}(D_{\alpha
}U^{\dagger})^{JI}\;.
\end{align}
These currents are completely fixed by imposing parity and charge conjugation, together with
\end{subequations}
\begin{equation}
\bar{q}^{J}P_{L,R}\overrightarrow{D}_{\alpha}q^{I}+\bar{q}^{J}\overleftarrow
{D}_{\alpha}P_{L,R}q^{I}=\partial_{\alpha}(\bar{q}^{J}P_{L,R}q^{I})\;.
\end{equation}
Note, however, that the above chiral representations lead to $\bar{q}^{I}(\overrightarrow{D}_{\alpha}-\overleftarrow{D}_{\alpha})P_{L,R}q^{J}=0$ instead of $(m_{I}^{2}-m_{J}^{2})\bar{q}^{I}P_{L,R}q^{J}$. This is because while $\bar{q}^{I}(\overrightarrow{D}_{\alpha}+\overleftarrow{D}_{\alpha})P_{L,R}q^{J}$ is of $\mathcal{O}(p)$, the difference $\bar{q}^{I} (\overrightarrow{D}_{\alpha}-\overleftarrow{D}_{\alpha})P_{L,R}q^{J}$ is actually of $\mathcal{O}(p^{3})$ since quark masses are $\mathcal{O}(p^{2})$. So, terms at that order would be required to get a correct divergence.

\subsubsection{Decay rates in the isospin limit}

The strong matrix elements~(\ref{KMat1}--\ref{KMat3}) are derived in the isospin limit. As a result, all the differential decay rates can be reconstructed entirely from those of the $K_{L}\approx K_{2}$. The contributions coming from the $\varepsilon K_{1}$ piece of the $K_{L}$, suppressed by $\varepsilon\sim2\cdot10^{-3}$, are neglected here.

For the $K\to(\gamma)X$ modes, the $K_{S}\approx K_{1}$ rates are obtained from those for $K_{L}\approx K_{2}$ by interchanging the real and imaginary parts of the couplings $x=f_{i}$, $g_{i}$, $h_{i}$:%
\begin{equation}
\frac{d\Gamma\left(  K_{S}\to(\gamma)X\right)  }{dz}=\frac{d\Gamma
\left(  K_{L}\to(\gamma)X\right)  }{dz}\left[  \Im(x)\leftrightarrow
\Re(x)\right]  \;.
\end{equation}
The $K^{+}\to\pi^{+}X$ decay rates are proportional to the sum $\Gamma(K_{S}\to\pi^{0}X)+\Gamma(K_{L}\to\pi^{0}X)$, hence are
obtained from $\Gamma(K_{L}\to\pi^{0}X)$ through the substitutions ($x,y=f_{i}$, $g_{i}$, $h_{i}$)%
\begin{subequations}
\begin{align}
\frac{d\Gamma\left(  K_{S}\to\pi^{0}X\right)  }{dz}  &  =\frac{d\Gamma
\left(  K_{L}\to\pi^{0}X\right)  }{dz}\left[  \Im(x)\leftrightarrow
\Re(x)\right]  \;\;,\;\\
\frac{d\Gamma\left(  K^{+}\to\pi^{+}X\right)  }{dz}  &  =\frac{d\Gamma
\left(  K_{L}\to\pi^{0}X\right)  }{dz}\left[
\begin{array}[c]{l}%
\Im(x)^{2},\Re(x)^{2}\to|x|^{2}\;,\\
\Im(x)\Im(y),\Re(x)\Re(y)\to\Re(xy^{\ast})
\end{array}
\right]\;.
\end{align}
Finally, the whole set of $K\to\pi\pi X$ decay rates can be reconstructed from the $K_{L}\to\pi^{+}\pi^{-}X$ rate as follows. Denoting $\Gamma\left(  K_{L}\to\pi^{+}\pi^{-}X\right)  _{\pm}\sim|\mathcal{M}_{\pm}(K_{L}\to\pi\pi)|^{2}$ from Eqs.~(\ref{KMat1},~\ref{KMat2}), and noting that these two amplitudes do not interfere, we get
\end{subequations}
\begin{subequations}
\label{KppIso}%
\begin{align}
\frac{d^{2}\Gamma\left(  K_{L}\to\pi^{0}\pi^{0}X\right)  }{dzdy}  &
=\frac{1}{2}\frac{d^{2}\Gamma\left(  K_{L}\to\pi^{+}\pi^{-}X\right)
_{+}}{dzdy}\;,\\
\frac{d^{2}\Gamma\left(  K_{L}\to\pi^{+}\pi^{-}X\right)  }{dzdy}  &
=\frac{d^{2}\Gamma\left(  K_{L}\to\pi^{+}\pi^{-}X\right)  _{+}}%
{dzdy}+\frac{d^{2}\Gamma\left(  K_{L}\to\pi^{+}\pi^{-}X\right)  _{-}%
}{dzdy}\;,\\
\frac{d^{2}\Gamma\left(  K^{+}\to\pi^{+}\pi^{0}X\right)  }{dzdy}  &
=\frac{d^{2}\Gamma\left(  K_{L}\to\pi^{+}\pi^{-}X\right)  _{-}}%
{dzdy}\left[
\begin{array}
[c]{l}%
\Im(x)^{2},\Re(x)^{2}\to|x|^{2}\;,\\
\Im(x)\Im(y),\Re(x)\Re(y)\to\Re(xy^{\ast})
\end{array}
\right]\;, \\
\frac{d^{2}\Gamma\left(  K_{S}\to\pi^{0}\pi^{0}X\right)  }{dzdy}  &
=\frac{d^{2}\Gamma\left(  K_{L}\to\pi^{0}\pi^{0}X\right)  }%
{dzdy}\left[  \Im(x)\leftrightarrow\Re(x)\right]  \;,\\
\frac{d^{2}\Gamma\left(  K_{S}\to\pi^{+}\pi^{-}X\right)  }{dzdy}  &
=\frac{d^{2}\Gamma\left(  K_{L}\to\pi^{+}\pi^{-}X\right)  }%
{dzdy}\left[  \Im(x)\leftrightarrow\Re(x)\right]  \;.
\end{align}

In the following sections, the differential rates are given for the various scenarios adopting the notations of Eq.~(\ref{DiffKpiXX}) for $K\to\pi XX$ modes, Eq.~(\ref{DiffKpipiXX}) for $K\to\pi\pi XX$ modes, Eq.~(\ref{DiffKpipiX}) for $K\to\pi\pi X$ modes, and finally, Eq.~(\ref{DiffKgaXX}) for $K\to\gamma XX$ modes. For $K\to\pi X$ and $K\to\gamma X$, the total integrated rate is directly written down. Given their regular occurrence, let us also introduce specific notations for the usual kinematical functions. First,%
\end{subequations}
\begin{equation}
\lambda_{\alpha\beta}=\lambda(1,\alpha,\beta)\;,\;\;\lambda(a,b,c)=a^{2}%
+b^{2}+c^{2}-2(ab+ac+bc)\;, \label{LKin}%
\end{equation}
with $\alpha$,$\beta$ standing for the reduced variable $y$, $z$, or the
reduced mass $r_{\pi}$, $r_{X}$ (in which case we simply denote $\alpha
,\beta=\pi,X$). Similarly, we define%
\begin{equation}
\beta_{y}^{2}=1-4r_{\pi}^{2}/y\;,\;\beta_{z}^{2}=1-4r_{X}^{2}/z\;,\;\beta
_{i\neq y,z}^{2}=1-4r_{i}^{2}\;.
\end{equation}

\subsection{Spin 1/2 invisible particles in the final states\label{AppKspin12}}

The rate for the fully invisible decay is:%
\begin{gather}
\Gamma\left(  K_{L}\to\bar{\psi}\psi\right)  =\frac{m_{K}^{4}}%
{\Lambda^{4}}\Gamma_{0}\left\{  \mathcal{I}_{1}\Im(f_{PS})^{2}+\mathcal{I}%
_{1}^{\prime}\Re(f_{PP})^{2}+\mathcal{I}_{2}\Re(f_{AA})^{2}+\mathcal{I}%
_{12}\Re(f_{AA})\Re(f_{PP})\right\}  \;,\nonumber\\
\Gamma_{0}=\frac{F_{\pi}^{2}\beta_{\psi}}{2\pi m_{K}},\;\mathcal{I}%
_{1}=\mathcal{I}_{1}^{\prime}\beta_{\psi}^{2},\;\mathcal{I}_{1}^{\prime}%
=B_{s}^{2},\;\mathcal{I}_{2}=4r_{\psi}^{2},\;\mathcal{I}_{12}=-4r_{\psi}%
B_{s}\;. \label{Defsxx}%
\end{gather}
The differential rates for the decays into a pion plus invisibles are%
\begin{align}
\frac{d\Gamma}{dz}\left(  K_{L}\to\pi^{0}\bar{\psi}\psi\right)   &
=\frac{m_{K}^{4}}{\Lambda^{4}}\Gamma_{0}^{\pi}\left\{  \mathcal{J}_{1}%
\Im(f_{TT})^{2}+\mathcal{J}_{1}^{\prime}\Re(f_{\tilde{T}T})^{2}+\mathcal{J}%
_{2}\Im(f_{VV})^{2}+\mathcal{J}_{12}\Im(f_{TT})\Im(f_{VV})\right. \nonumber\\
&  \;\;\;\;\;\;\;\;\;\;\;\;\;\;\;\;\;\left.  +\mathcal{J}_{2}^{\prime}%
\Im(f_{VA})^{2}+\mathcal{J}_{3}\Re(f_{SS})^{2}+\mathcal{J}_{3}^{\prime}%
\Im(f_{SP})^{2}+\mathcal{J}_{23}\Im(f_{VA})\Im(f_{SP})\right\}  \;,
\end{align}
with the normalization and kinematical functions%
\begin{align}
\Gamma_{0}^{\pi}  &  =\frac{m_{K}\beta_{z}\lambda_{z\pi}^{1/2}}{96\pi^{3}%
},\;\;\mathcal{J}_{1}=\mathcal{B}_{tz}\left(  1+8r_{\text{$\psi$}}%
^{2}/z\right)  \;,\;\mathcal{J}_{1}^{\prime}=\mathcal{B}_{tz}\beta_{z}%
^{2}\;,\;\mathcal{J}_{12}=3r_{\text{$\psi$}}B_{T}\lambda_{z\pi}\;,\nonumber\\
\mathcal{J}_{2}  &  =\frac{\lambda_{z\pi}}{2}\left(  1+2r_{\text{$\psi$}}%
^{2}/z\right)  \;,\;\mathcal{J}_{2}^{\prime}=\mathcal{J}_{4}+\frac{3r_{\psi
}^{2}}{z}\left(  1-r_{\pi}^{2}\right)  ^{2}\;,\;\mathcal{J}_{3}=\mathcal{J}%
_{3}^{\prime}\beta_{z}^{2}\;,\;\mathcal{J}_{3}^{\prime}=\frac{z}{2}%
\mathcal{J}_{5}\;,\nonumber\\
\mathcal{J}_{23}  &  =-3r_{\text{$\psi$}}B_{s}\left(  1-r_{\pi}^{2}\right)
\;,\;\mathcal{J}_{4}=\frac{\lambda_{z\pi}}{2}\beta_{z}^{2},\;\mathcal{J}%
_{5}=\frac{3}{2}B_{s}^{2}\;,\;\mathcal{B}_{tz}=\frac{z}{4}B_{T}^{2}%
\lambda_{z\pi}\;. \label{Defspxx}%
\end{align}
In the massless limit, this expression simplifies a lot because the fermion helicity states do not mix. The interference terms drop out while the parity of the current becomes indistinguishable; $\mathcal{J}_{i}=\mathcal{J}_{i}^{\prime}$, $i=1,2,3$.

The decays $K\to\pi\pi\bar{\psi}\psi$ also receive contributions from all the currents:%
\begin{align}
\frac{d^{2}\Gamma}{dzdy}\left(  K_{L}\to\pi^{+}\pi^{-}\bar{\psi}%
\psi\right)  _{-}  &  =\frac{m_{K}^{4}}{\Lambda^{4}}\Gamma_{0}^{\pi\pi
}\left\{  \mathcal{F}_{1}\Re(f_{TT})^{2}+\mathcal{F}_{1}^{\prime}\Im
(f_{\tilde{T}T})^{2}+\mathcal{F}_{2}\Im(f_{AV})^{2}+\mathcal{F}_{12}%
\Im(f_{\tilde{T}T})\Im(f_{AV})\right. \nonumber\\
&  \;\;\;\;\;\;\;\;\;\;\;\;\;\;\;+\mathcal{F}_{2}^{\prime}\Im(f_{AA}%
)^{2}+\mathcal{F}_{3}\Re(f_{PS})^{2}+\mathcal{F}_{3}^{\prime}\Im(f_{PP}%
)^{2}+\mathcal{F}_{23}\Im(f_{PP})\Im(f_{AA})\nonumber\\
&  \;\;\;\;\;\;\;\;\;\;\;\;\;\;\;\left.  +\mathcal{F}_{7}\Re\left(
f_{VV}\right)  ^{2}+\mathcal{F}_{7}^{\prime}\Re\left(  f_{VA}\right)
^{2}+\mathcal{F}_{27}\Re\left(  f_{TT}\right)  \Re\left(  f_{VV}\right)
\right\}  \;,\\
\frac{d^{2}\Gamma}{dzdy}\left(  K_{L}\to\pi^{+}\pi^{-}\bar{\psi}%
\psi\right)  _{+}  &  =\frac{m_{K}^{4}}{\Lambda^{4}}\Gamma_{0}^{\pi\pi
}\left\{  \mathcal{F}_{4}\Im(f_{PS})^{2}+\mathcal{F}_{4}^{\prime}\Re
(f_{PP})^{2}+\mathcal{F}_{5}^{\prime}\Re(f_{AA})^{2}+\mathcal{F}_{45}%
\Re(f_{AA})\Re(f_{PP})\right. \nonumber\\
&  \;\;\;\;\;\;\;\;\;\;\;\;\;\;\;\left.  +\mathcal{F}_{5}\Re(f_{AV}%
)^{2}+\mathcal{F}_{6}\Im(f_{TT})^{2}+\mathcal{F}_{6}^{\prime}\Re(f_{\tilde
{T}T})^{2}+\mathcal{F}_{56}\Re(f_{\tilde{T}T})\Re(f_{AV})\right\}  \;,
\end{align}
with the normalization and kinematical functions%
\begin{align}
\Gamma_{0}^{\pi\pi}  &  =\frac{m_{K}^{3}\beta_{y}\lambda_{yz}^{1/2}}%
{3072\pi^{5}F_{\pi}^{2}},\;\lambda_{1}=\beta_{y}^{2}(y(1-y)^{2}+\frac{\lambda
_{yz}}{12}(4y+z)),\;\lambda_{2}=\beta_{y}^{2}(yz+\frac{\lambda_{yz}}%
{12}),\nonumber\\
\lambda_{3}  &  =\frac{\beta_{y}^{2}\lambda_{yz}}{2(1-z)^{2}},\;\lambda
_{4}=\frac{3}{2}\frac{4yz+\lambda_{yz}}{(1-z)^{2}},\;\lambda_{5}%
=\frac{\lambda_{yz}}{4},\;\lambda_{6}=y\beta_{y}^{2}\lambda_{yz}%
,\;\mathcal{F}_{1}=\frac{B_{T}^{2}}{2}(\lambda_{1}\beta_{z}^{2}+(8r_{\psi}%
^{2}/z)\lambda_{6}),\;\nonumber\\
\mathcal{F}_{1}^{\prime}  &  =\frac{B_{T}^{2}}{2}(\lambda_{1}(1+8r_{\psi}%
^{2}/z)-(8r_{\psi}^{2}/z)\lambda_{6})\;,\;\mathcal{F}_{2,5}=\lambda
_{2,5}(1+2r_{\psi}^{2}/z),\;\mathcal{F}_{2,5}^{\prime}=\mathcal{F}%
_{2,5}^{\prime\prime}+\frac{r_{\psi}^{2}}{z}\lambda_{3,4},\;\mathcal{F}%
_{2,5}^{\prime\prime}=\lambda_{2,5}\beta_{z}^{2}\;,\nonumber\\
\mathcal{F}_{3,4}  &  =\mathcal{F}_{3,4}^{\prime}\beta_{z}^{2},\;\mathcal{F}%
_{3,4}^{\prime}=\frac{z}{2}\mathcal{F}_{3,4}^{\prime\prime},\;,\mathcal{F}%
_{3,4}^{\prime\prime}=\frac{1}{2}\lambda_{3,4}B_{s}^{2}\;,\;\mathcal{F}%
_{6}=B_{T}^{2}\frac{z\lambda_{yz}}{8}\beta_{z}^{2}\;,\;\;\mathcal{F}%
_{6}^{\prime}=B_{T}^{2}\frac{z\lambda_{yz}}{8}(1+8r_{\psi}^{2}%
/z)\;,\;\nonumber\\
\mathcal{F}_{7}  &  =\mathcal{F}_{7}^{\prime\prime}\left(  1+2r_{\psi}%
^{2}/z\right)  ,\;\mathcal{F}_{7}^{\prime}=\mathcal{F}_{7}^{\prime\prime}%
\beta_{z}^{2}\;,\mathcal{F}_{7}^{\prime\prime}=\frac{z}{24}\lambda_{6}%
A_{WZW}^{2},\ \mathcal{F}_{27}=-r_{\psi}B_{T}\lambda_{6}A_{WZW}%
\;,\;\;\nonumber\\
\mathcal{F}_{12}  &  =\frac{r_{\psi}}{2}B_{T}\beta_{y}^{2}\left(  \lambda
_{yz}+12y(1-y)\right)  ,\;\mathcal{F}_{23,45}=-r_{\psi}\lambda_{3,4}%
B_{s},\;\mathcal{F}_{56}=\frac{3}{2}r_{\psi}B_{T}\lambda_{yz}\;,\;\;A_{WZW}%
=\frac{N_{C}m_{K}^{2}}{6\pi^{2}F_{\pi}^{2}} \;.\label{Defsppxx}%
\end{align}
Again, the interference terms drop out and $\mathcal{F}_{i}^{\prime}=\mathcal{F}_{i}$, $i=1,...,6$, when $m_{\psi}\to0$.

The rate $K_L\to\gamma\bar{\psi}\psi$ is driven either by the anomalous vertices at $\mathcal{O}(p^{4})$ for vector currents, or directly by the tensor currents, and has the differential rate%
\begin{gather}
\frac{d\Gamma}{dz}\left(  K_{L}\to\gamma\bar{\psi}\psi\right)
=\frac{m_{K}^{4}}{\Lambda^{4}}\Gamma_{0}^{\gamma}\left\{  \mathcal{G}_{1}%
(\Im(f_{\tilde{T}T})^{2}+\Re(f_{TT})^{2})+\mathcal{G}_{2}\Re(f_{VV})\Re
(f_{TT})+\mathcal{G}_{3}\Re(f_{VA})^{2}+\mathcal{G}_{4}\Re(f_{VV}%
)^{2}\right\}  \;,\nonumber\\
\Gamma_{0}^{\gamma}=\frac{\alpha m_{K}\beta_{z}(1-z)^{3}}{27\pi^{2}%
},\;\;\mathcal{G}_{1}=\frac{B_{T}^{\prime2}}{2}(1+2r_{\text{$\psi$}}%
^{2}/z),\;\mathcal{G}_{2}=3r_{\psi}B_{T}^{\prime}A_{\gamma\gamma
}\;,\;\nonumber\\
\mathcal{G}_{3}=\frac{z\beta_{z}^{2}}{2}A_{\gamma\gamma}^{2},\;\mathcal{G}%
_{4}=\frac{z}{2}(1+2r_{\text{$\psi$}}^{2}/z)A_{\gamma\gamma}^{2}%
,\;A_{\gamma\gamma}=\frac{N_{C}m_{K}}{8\pi^{2}F_{\pi}}\;, \label{Defsgxx}%
\end{gather}
where $N_{C}$ is the number of QCD colors. The $\mathcal{O}(p^{4})$ loops and normal-parity counterterms cancel out for the (axial-)vector and (pseudo-)scalar currents when the photon is on-shell, see Eq.~(\ref{MatElgK}).

\subsubsection{Standard Model rates\label{AppSMrates}}

The standard model rates are recovered by setting all the coefficients to zero but for%
\begin{equation}
\frac{c_{LL}^{V}}{\Lambda^{2}}=\frac{4G_{F}\alpha\left(  M_{Z}\right)  }%
{\sqrt{2}}\frac{y_{\nu}}{2\pi\sin^{2}\theta_{W}}\;,\;\;y_{\nu}=\left(
\Re(\lambda_{t})+i\Im(\lambda_{t})\right)  X_{t}+\left|  V_{us}\right|
^{4}\Re(\lambda_{c})P_{u,c}\;.
\end{equation}
Numerically, $X_{t}=1.465(16)$~\cite{BrodGS10}, $P_{c}=0.372(15)$~\cite{BurasGHN05}, $\delta P_{u,c}=0.04(2)$~\cite{IsidoriMS05} (with
$\bar{\lambda}=0.2255$), so that for each $\ell=e,\mu,\tau$,
\begin{equation}
y_{\nu}=2\pi\sin^{2}\theta_{W}\times\lbrack4.84(22)-i1.359(96)]\times
10^{-4}\;\;.
\end{equation}

The full set of differential rates is, in the isospin limit,
\begin{subequations}
\label{SMdiffrates}%
\begin{align}
\Gamma\left(  K_{L}\to\nu\bar{\nu}\right)   &  =0\;,\\
\Gamma\left(  K_{L}\to\pi^{0}\nu\bar{\nu}\right)   &  =\Gamma_{\nu
\bar{\nu}}\times\int dz\lambda_{\pi z}^{3/2}\times\Im(y_{\nu})^{2}\;,\\
\Gamma\left(  K_{L}\to\pi^{+}\pi^{-}\nu\bar{\nu}\right)  _{+}  &
=\Gamma_{\nu\bar{\nu}}\times\int dzdy\frac{\beta_{\pi}\lambda_{yz}^{3/2}%
m_{K}^{2}}{64\pi^{2}F_{\pi}^{2}}\times\Re(y_{\nu})^{2}\;,\\
\Gamma\left(  K_{L}\to\pi^{+}\pi^{-}\nu\bar{\nu}\right)  _{-}  &
=\Gamma_{\nu\bar{\nu}}\times\int dzdy\frac{\beta_{\pi}^{3}\lambda_{yz}%
^{1/2}m_{K}^{2}}{192\pi^{2}F_{\pi}^{2}}\left[  \Im(y_{\nu})^{2}\times\left(
\lambda+12yz\right)  +\Re(y_{\nu})^{2}\times\frac{yz\lambda_{yz}m_{K}^{4}%
N_{c}^{2}}{72\pi^{4}F_{\pi}^{4}}\right]  \;,\\
\Gamma\left(  K_{L}\to\gamma\nu\bar{\nu}\right)   &  =\Gamma_{\nu
\bar{\nu}}\times\int dz\frac{\alpha m_{K}^{2}z(1-z)^{3}}{2\pi^{3}F_{\pi}^{2}%
}\times\Re(y_{\nu})^{2}\;,
\end{align}
\end{subequations}
where $\Gamma_{\nu\bar{\nu}}=G_{F}^{2}\alpha\left(  M_{Z}\right)^{2}m_{K}^{5}/(256\pi^{5}\sin^{4}\theta_{W})$, and the ranges for the phase-space integrals are given in Eqs.~(\ref{DiffKpiXX},~\ref{DiffKpipiXX},~\ref{DiffKgaXX}). Note that $K_{L}\to\gamma\nu\bar{\nu}$ is purely CP-conserving because the parity even matrix element vanishes at $\mathcal{O}(p^{4})$, see Eq.~(\ref{MatElgK}).

The corresponding branching ratios are
\begin{subequations}
\label{SMrates}%
\begin{gather}
\mathcal{B}\left(  K_{L}\to\pi^{0}\nu\bar{\nu}\right)  =2.3\times
10^{-11},\;\mathcal{B}\left(  K^{+}\to\pi^{+}\nu\bar{\nu}\right)
=7.6\cdot10^{-11},\\
\mathcal{B}\left(  K_{L}\to\pi^{+}\pi^{-}\nu\bar{\nu}\right)
=1\cdot10^{-13},\;\mathcal{B}\left(  K_{L}\to\pi^{0}\pi^{0}\nu
\bar{\nu}\right)  =6\cdot10^{-14},\;\mathcal{B}\left(  K^{+}\to
\pi^{+}\pi^{0}\nu\bar{\nu}\right)  =5\cdot10^{-15}\;,\\
\mathcal{B}\left(  K_{L}\to\gamma\nu\bar{\nu}\right)  =3.4\times
10^{-13},\;\mathcal{B}\left(  K_{L}\to e^{+}e^{-}\nu\bar{\nu}\right)
=2\cdot10^{-15},\;\mathcal{B}\left(  K_{L}\to\mu^{+}\mu^{-}\nu
\bar{\nu}\right)  =7\cdot10^{-18}\;.
\end{gather}
\end{subequations}
Adopting the usual chiral counting, the typical error on these LO estimates is expected to be of about $30\%$ at the amplitude level. Note, in this respect, that the $K\to\pi\nu\bar{\nu}$ rates given above are just indicative, as higher order corrections as well as isospin-breaking effects are known and more precise estimates have been obtained~\cite{MesciaS07}. For $K\to \pi\pi\nu\bar{\nu}$, the anomalous term gives negligible percent-level contributions, so that $\Im(y_{\nu})<\Re(y_{\nu})$ implies $\mathcal{B}\left(K_{L}\to\pi^{+}\pi^{-}\nu\bar{\nu}\right)  \approx2\mathcal{B}\left(K_{L}\to\pi^{0}\pi^{0}\nu\bar{\nu}\right)  \gg\mathcal{B}\left(K^{+}\to\pi^{+}\pi^{0}\nu\bar{\nu}\right)  $, in fair agreement with Ref.~\cite{Kppnn}. For $K_{L}\to\gamma\nu\bar{\nu}$, previous estimates incorrectly rely on $K_{\ell2\gamma}$ for the matrix elements~\cite{Kgnn}, hence included a parity-even contribution in contradiction with Eq.~(\ref{MatElgK}). Numerically, this is however without consequences since these contributions are CP-violating hence subleading for the rate. Finally, the rates for $K_{L}\to\ell^{+}\ell^{-}\nu\bar{\nu}$ are dominated by the Dalitz emission from the purely anomalous $K_{L}\to\gamma\nu\bar{\nu}$. The rates from the even-parity contributions arising from the matrix elements Eq.~(\ref{MatElgK}) are in the $10^{-19}$ range. Note that we did not consider the tree-level process $K_{L}\to W^{+}W^{-}[\to\ell^{+}\ell^{-}\nu_{\ell}\bar{\nu}_{\ell}]$, which may actually be competitive given these strong suppressions. In any case, these rates are far too small to be accessible any time soon.

\subsection{Spin 0 invisible particles in the final states\label{AppKspin0}}

The rates for the production of a single invisible scalar from the $H^{\dagger}(\bar{D}Q)\phi$ operator of Eq.~(\ref{Axion}) are%

\begin{equation}%
\begin{array}
[c]{ll}%
\Gamma\left(  K_{L}\to\pi^{0}\phi\right)  =\bar{\Gamma}_{0}^{\pi}%
B_{s}^{2}\Re(g_{S})^{2}\;,\; & \dfrac{d\Gamma}{dy}\left(  K_{L}\to
\pi^{+}\pi^{-}\phi\right)  _{-}=\bar{\Gamma}_{0}^{\pi\pi}\;\mathcal{H}_{1}%
\Re(g_{P})^{2}\;,\;\;\\
\Gamma\left(  K_{L}\to\gamma\phi\right)  =0\;, & \dfrac{d\Gamma}%
{dy}\left(  K_{L}\to\pi^{+}\pi^{-}\phi\right)  _{+}=\bar{\Gamma}%
_{0}^{\pi\pi}\;\mathcal{H}_{2}\Im(g_{P})^{2}\;,
\end{array}
\end{equation}
with%
\begin{equation}
\bar{\Gamma}_{0}^{\pi}=\frac{\lambda_{\phi\pi}^{1/2}m_{K}}{16\pi}%
\;,\;\bar{\Gamma}_{0}^{\pi\pi}=\frac{\beta_{y}\lambda_{\phi y}^{1/2}m_{K}^{3}%
}{1024\pi^{3}F_{\pi}^{2}},\;\mathcal{H}_{1}=\left(  1-\frac{y}{1-r_{\phi}^{2}%
}\right)  ^{2}B_{s}^{2},\;\mathcal{H}_{2}=\frac{\beta_{y}^{2}\lambda_{\phi y}%
}{3(1-r_{\phi}^{2})^{2}}B_{s}^{2}\;.\label{DefsSx}%
\end{equation}
The matrix element $\langle\gamma|\bar{s}d|K_{L}\rangle$ occurring for $K_{L}\to\gamma\phi$ vanishes at $\mathcal{O}(p^{4})$. The rates corresponding to the derivative couplings $(\bar{s}\gamma_{\mu}d)\partial^{\mu}\phi$ and $(\bar{s}\gamma_{\mu}\gamma_{5}d)\partial^{\mu}\phi$ are obtained through the replacement~(\ref{SubsAxion}), i.e.%
\begin{equation}
g_{S,P}\to-ig_{V,A}\frac{m_{s}\mp m_{d}}{\Lambda}\text{ \ \ with
}m_{s}-m_{d}=\frac{1-r_{\pi}^{2}}{B_{s}}\frac{m_{K}}{\Lambda}\text{ and }%
m_{s}+m_{d}=\frac{1}{B_{s}}\frac{m_{K}}{\Lambda}\;.
\end{equation}
Note that if simultaneously present, the $g_{S,P}$ and $g_{V,A}$ currents obviously interfere.

The differential rates for two-scalar final states can be written in terms of the same kinematical functions as for fermionic final states, but for the obvious replacement $m_{\psi}\to m_{\phi}$ everywhere. They are significantly simpler though, because the angular momentum of the two-scalar states is purely orbital. Specifically,
\begin{align}
\Gamma\left(  K_{L}\to\bar{\phi}\phi\right)   &  =\frac{\Gamma_{0}}%
{2}\mathcal{I}_{1}^{\prime}\Im(g_{PS})^{2}\frac{m_{K}^{2}}{\Lambda^{2}%
}\;,\nonumber\\
\frac{d\Gamma}{dz}\left(  K_{L}\to\pi^{0}\bar{\phi}\phi\right)   &
=\frac{\Gamma_{0}^{\pi}}{4}\left\{  \mathcal{J}_{4}\Im(g_{VV})^{2}%
\frac{m_{K}^{4}}{\Lambda^{4}}+\mathcal{J}_{5}\Re(g_{SS})^{2}\frac{m_{K}^{2}%
}{\Lambda^{2}}\right\}  \;,\nonumber\\
\frac{d^{2}\Gamma}{dzdy}\left(  K_{L}\to\pi^{+}\pi^{-}\bar{\phi}%
\phi\right)  _{-} &  =\frac{\Gamma_{0}^{\pi\pi}}{4}\left\{  \mathcal{F}%
_{2}^{\prime\prime}\Im(g_{AV})^{2}\frac{m_{K}^{4}}{\Lambda^{4}}+\mathcal{F}%
_{7}^{\prime}\Re(g_{VV})^{2}\frac{m_{K}^{4}}{\Lambda^{4}}+\mathcal{F}%
_{3}^{\prime\prime}\Re(g_{PS})^{2}\frac{m_{K}^{2}}{\Lambda^{2}}\right\}
\;,\nonumber\\
\frac{d^{2}\Gamma}{dzdy}\left(  K_{L}\to\pi^{+}\pi^{-}\bar{\phi}%
\phi\right)  _{+} &  =\frac{\Gamma_{0}^{\pi\pi}}{4}\left\{  \mathcal{F}%
_{5}^{\prime\prime}\Re(g_{AV})^{2}\frac{m_{K}^{4}}{\Lambda^{4}}+\mathcal{F}%
_{4}^{\prime\prime}\Im(g_{PS})^{2}\frac{m_{K}^{2}}{\Lambda^{2}}\right\}
\;,\nonumber\\
\frac{d\Gamma}{dz}\left(  K_{L}\to\gamma\bar{\phi}\phi\right)   &
=\frac{\Gamma_{0}^{\gamma}}{4}\mathcal{G}_{3}\Im(g_{VV})^{2}\frac{m_{K}^{4}%
}{\Lambda^{4}}\;,
\end{align}
with the kinematical quantities defined in Eqs.~(\ref{Defsxx},~\ref{Defspxx},~\ref{Defsppxx},~\ref{Defsgxx}). Note that if $\phi=\bar{\phi}$, Bose statistics has to be enforced and these rates should be divided by two.

\subsection{Spin 1 invisible particles in the final states\label{AppKspin1}}

From the single dark vector production from the operators of $\mathcal{H}_{mat}^{\mathrm{V}}[$I$]$ and $\mathcal{H}_{mat}^{\mathrm{V}}[$II$]$, Eqs.~(\ref{HvI}) and~(\ref{HvII1}), we find%
\begin{equation}%
\begin{tabular}
[c]{lll}%
$K_{L}\to\pi^{0}V:$ & $\Gamma^{\lbrack\text{I}]}=\bar{\Gamma}_{0}%
^{\pi}\dfrac{\lambda_{V\pi}}{r_{V}^{2}}\Im(\varepsilon_{V})^{2}\;,$ &
$\Gamma^{\lbrack\text{II}]}=\bar{\Gamma}_{0}^{\pi}B_{T}^{2}\lambda_{V\pi
}\dfrac{m_{V}^{2}}{\Lambda^{2}}\Im(f_{T})^{2}\;,$\\
$K_{L}\to\gamma V:$ & $\Gamma^{\lbrack\text{I}]}=\bar{\Gamma}%
_{0}^{\gamma}\,A_{\gamma\gamma}^{2}\Re(\varepsilon_{V})^{2}\;,$ &
$\Gamma^{\lbrack\text{II}]}=\bar{\Gamma}_{0}^{\gamma}\dfrac{m_{K}^{2}}%
{\Lambda^{2}}B_{T}^{\prime2}\left\{  \Re(f_{T})^{2}+\Im(f_{\tilde{T}}%
)^{2}\right\}  \;,$\\
$K_{L}\to\pi^{+}\pi^{-}V:$ & $\dfrac{d\Gamma_{-}^{[\text{I}]}}%
{dy}=\bar{\Gamma}_{0}^{\pi\pi}(\mathcal{H}_{2}^{\prime}\Im(\varepsilon
_{A})^{2}+\mathcal{H}_{3}^{\prime}\Re(\varepsilon_{V})^{2})\;,$ &
$\dfrac{d\Gamma_{-}^{[\text{II}]}}{dy}=\bar{\Gamma}_{0}^{\pi\pi}\dfrac
{m_{K}^{2}}{\Lambda^{2}}\left\{  \mathcal{H}_{1}^{\prime\prime}\Re(f_{T}%
)^{2}+\mathcal{H}_{2}^{\prime\prime}\Im(f_{\tilde{T}})^{2}\right\}  ,$\\
& $\dfrac{d\Gamma_{+}^{[\text{I}]}}{dy}=\bar{\Gamma}_{0}^{\pi\pi}%
\mathcal{H}_{1}^{\prime}\Re(\varepsilon_{A})^{2}\;,$ & $\dfrac{d\Gamma
_{+}^{[\text{II}]}}{dy}=\bar{\Gamma}_{0}^{\pi\pi}\dfrac{m_{K}^{2}}{\Lambda
^{2}}\mathcal{H}_{3}^{\prime\prime}\Re(f_{\tilde{T}})^{2}\;,$%
\end{tabular}
\end{equation}
with the definitions in Eq.~(\ref{DefsSx}) together with%
\begin{align}
\bar{\Gamma}_{0}^{\gamma}  & =\frac{2\alpha m_{K}(1-r_{V}^{2})^{3}}%
{9},\;\mathcal{H}_{1}^{\prime}=\frac{\lambda_{Vy}}{r_{V}^{2}}\;,\;\mathcal{H}%
_{2}^{\prime}=\beta_{y}^{2}\frac{\lambda_{Vy}+12yr_{V}^{2}}{3r_{V}^{2}%
}\;,\;\mathcal{H}_{3}^{\prime}=-\lambda_{Vy}\frac{y\beta_{y}^{2}}{6}%
A_{WZW}^{2}\;,\label{DefsSx2}\\
\mathcal{H}_{1}^{\prime\prime}  & =\frac{8}{3}B_{T}^{2}y\lambda_{Vy}\beta
_{y}^{2}\;,\;\mathcal{H}_{2}^{\prime\prime}=\frac{B_{T}^{2}}{3}\left(
\beta_{y}^{2}\frac{\mathcal{H}_{3}^{\prime\prime}}{3}-\frac{\mathcal{H}%
_{1}^{\prime\prime}}{2}+12\beta_{y}^{2}y(1-y)^{2}\right)  \;,\;\mathcal{H}%
_{3}^{\prime\prime}=B_{T}^{2}r_{V}^{2}\lambda_{Vy}\;.
\end{align}
Note that the $\mathcal{H}_{3}^{\prime}$ term is finite when $r_{V}\to0$ since it is induced by the WZW anomaly, while by construction, all the $\Gamma^{\lbrack\text{II}]}$ are finite in that limit. For the two-vector modes induced by the scalar and pseudoscalar couplings of $\mathcal{H}_{mat}^{\mathrm{VV}}[$II$]$, the rates are%
\begin{align}
\Gamma^{\lbrack\text{II}]}\left(  K_{L}\to V\bar{V}\right)   &
=4\Gamma_{0}\frac{m_{K}^{6}}{\Lambda^{6}}\left\{  \mathcal{I}_{1}(\Re
(h_{PP})^{2}+\Im(h_{PS})^{2})+6r_{V}^{4}\mathcal{I}_{1}^{\prime}\Im
(h_{PS})^{2}\right\}  \;,\nonumber\\
\frac{d\Gamma^{\lbrack\text{II}]}}{dz}\left(  K_{L}\to\pi^{0}V\bar
{V}\right)   &  =4\Gamma_{0}^{\pi}\frac{m_{K}^{6}}{\Lambda^{6}}\left\{
z\mathcal{J}_{3}(\Re(h_{SS})^{2}+\Im(h_{SP})^{2})+3r_{V}^{4}\mathcal{J}_{5}%
\Re(h_{SS})^{2}\right\}  \;,\nonumber\\
\frac{d^{2}\Gamma^{\lbrack\text{II}]}}{dzdy}\left(  K_{L}\to\pi^{+}%
\pi^{-}V\bar{V}\right)  _{-} &  =4\Gamma_{0}^{\pi\pi}\frac{m_{K}^{6}}%
{\Lambda^{6}}\left\{  z\mathcal{F}_{3}(\Im(h_{PP})^{2}+\Re(h_{PS})^{2}%
)+3r_{V}^{4}\mathcal{F}_{3}^{\prime\prime}\Re(h_{PS})^{2})\right\}
,\nonumber\\
\frac{d^{2}\Gamma^{\lbrack\text{II}]}}{dzdy}\left(  K_{L}\to\pi^{+}%
\pi^{-}V\bar{V}\right)  _{+} &  =4\Gamma_{0}^{\pi\pi}\frac{m_{K}^{6}}%
{\Lambda^{6}}\left\{  z\mathcal{F}_{4}(\Re(h_{PP})^{2}+\Im(h_{PS})^{2}%
)+3r_{V}^{4}\mathcal{F}_{4}^{\prime\prime}\Im(h_{PS})^{2}\right\}  \;,
\end{align}
and the kinematical quantities defined in Eqs.~(\ref{Defsxx},\ref{Defspxx},\ref{Defsppxx},\ref{Defsgxx}) but for $m_{\psi}\to m_{V}$ everywhere. The matrix element $\langle\gamma|\bar{s}d|K_{L}\rangle$ occurring for $K_{L}\to\gamma VV$ vanishes at $\mathcal{O}(p^{4})$. Note that if the vector field is real, $V=\bar{V}$, then these rates have to be divided by two to account for Bose statistics.

\subsection{Spin 3/2 invisible particles in the final states\label{AppKspin32}}

Introducing the short-hands $\Re_{X}\equiv\Re(f_{X})$, $\Im_{X}\equiv\Im(f_{X})$, $\beta_{i}^{\prime}=(5\beta_{i}^{4}-6\beta_{i}^{2}+9)/18$, and $\beta_{i}^{\prime\prime}=(9\beta_{i}^{4}-6\beta_{i}^{2}+5)/18$, the rate for $K_{L}\to\Psi\overline{\Psi}$ is:%
\begin{gather}
\Gamma\left(  K_{L}\to\Psi\overline{\Psi}\right)  =\frac{m_{K}^{4}%
}{\Lambda^{4}}\Gamma_{0}\frac{1}{4r_{\psi}^{4}}\left(  \mathcal{I}_{S}\Im
_{PS}^{2}+\mathcal{I}_{P}\Re_{PP}^{2}+\mathcal{I}_{A}\Re_{AA}^{2}%
+\mathcal{I}_{PA}\Re_{AA}\Re_{PP}\right)  \;,\nonumber\\
\mathcal{I}_{S}=\mathcal{I}_{1}\beta_{\psi}^{\prime\prime},\;\mathcal{I}%
_{P}=\mathcal{I}_{1}^{\prime}\beta_{\psi}^{\prime},\;\mathcal{I}%
_{A}=\mathcal{I}_{2}\beta_{\psi}^{\prime}\;,\;\mathcal{I}_{PA}=\mathcal{I}%
_{12}\beta_{\psi}^{\prime}\;,
\end{gather}
where $\Gamma_{0}$, $\mathcal{I}_{1}$, $\mathcal{I}_{1}^{\prime}$, $\mathcal{I}_{2}$, and $\mathcal{I}_{12}$ are defined in Eq.~(\ref{Defsxx}). The differential rate for the decay into a pion plus invisibles is%
\begin{align}
\frac{d\Gamma}{dz}\left(  K_{L}\to\pi^{0}\Psi\overline{\Psi}\right)
&  =\frac{m_{K}^{4}}{\Lambda^{4}}\Gamma_{0}^{\pi}\frac{z^{2}}{4r_{\psi}^{4}%
}\left\{  \mathcal{J}_{S}\Im_{TS}^{2}+\mathcal{J}_{P}\Re_{TP}^{2}%
+\mathcal{J}_{T}\Im_{TT}^{2}+\mathcal{J}_{\tilde{S}}\Re_{\tilde{T}S}%
^{2}+\mathcal{J}_{\tilde{T}}\Re_{\tilde{T}T}^{2}+\mathcal{J}_{\tilde{P}}%
\Im_{\tilde{T}P}^{2}\right.  \\
&  \;\;\;\;\;\;\;\;\;-\mathcal{J}_{P\tilde{S}}\Re_{TP}\Re_{\tilde{T}%
S}+\mathcal{J}_{T\tilde{P}}\Im_{TT}\Im_{\tilde{T}P}+\mathcal{J}_{P\tilde{T}%
}\Re_{TP}\Re_{\tilde{T}T}+\mathcal{J}_{ST}\Im_{TS}\Im_{TT}\nonumber\\
&  \;\;\;\;\;\;\;\;\;-\mathcal{J}_{\tilde{S}\tilde{T}}\Re_{\tilde{T}S}%
\Re_{\tilde{T}T}+\Im_{VV}(\mathcal{J}_{VT}\Im_{TT}+\mathcal{J}_{VS}\Im
_{TS}+\mathcal{J}_{V\tilde{P}}\Im_{\tilde{T}P})\nonumber\\
&  \;\;\;\;\;\;\;\;\;\left.  +\mathcal{J}_{SS}\Re_{SS}^{2}+\mathcal{J}_{SP}%
\Im_{SP}^{2}+\mathcal{J}_{V}\Im_{VV}^{2}+\mathcal{J}_{A}\Im_{VA}%
^{2}+\mathcal{J}_{AP}\Im_{VA}\Im_{SP}\right\}  \;,
\end{align}
with the normalization and kinematical functions%
\begin{align}
\mathcal{J}_{S} &  =\mathcal{J}_{1}^{\prime}\frac{\beta_{z}^{2}}{9}\left(
1+\frac{5r_{\psi}^{2}}{z}\right)  ,\mathcal{J}_{P}=\frac{\mathcal{J}%
_{1}^{\prime}}{9}\left(  1+\frac{3r_{\psi}^{2}}{z}\right)  ,\mathcal{J}%
_{T}=\mathcal{J}_{1}\beta_{z}^{\prime}-\frac{16}{5}\mathcal{J}_{\tilde{P}%
},\mathcal{J}_{\tilde{S}}=\mathcal{J}_{1}^{\prime}\frac{r_{\psi}^{2}}%
{3z}\left(  1+\frac{10r_{\psi}^{2}}{3z}\right)  ,\mathcal{J}_{\tilde{T}%
}=\mathcal{J}_{1}^{\prime}\beta_{z}^{\prime},\nonumber\\
\mathcal{J}_{\tilde{P}} &  =\mathcal{B}_{tz}\frac{5r_{\psi}^{2}}{9z}\left(
1+\frac{2r_{\psi}^{2}}{z}\right)  ,\mathcal{J}_{P\tilde{S}}=\mathcal{J}%
_{1}^{\prime}\frac{8r_{\psi}^{2}}{9z},\mathcal{J}_{T\tilde{P}}=4\mathcal{J}%
_{\tilde{P}},\mathcal{J}_{P\tilde{T}}=\frac{4}{9}\mathcal{J}_{1}^{\prime
}\left(  1-\frac{r_{\psi}^{2}}{z}\right)  ,\mathcal{J}_{\tilde{S}\tilde{T}%
}=\mathcal{J}_{1}^{\prime}\frac{4r_{\psi}^{2}}{9z}\left(  1-\frac{10r_{\psi
}^{2}}{z}\right)  ,\nonumber\\
\mathcal{J}_{ST} &  =4\mathcal{J}_{S},\mathcal{J}_{SS}=\mathcal{J}_{3}%
\beta_{z}^{\prime\prime},\mathcal{J}_{SP}=\mathcal{J}_{3}^{\prime}\beta
_{z}^{\prime},\mathcal{J}_{V}=\mathcal{J}_{2}\beta_{z}^{\prime}-\frac{2r_{\psi
}^{2}}{9z}\lambda_{z\pi}(3-\beta_{z}^{4}),\mathcal{J}_{A}=\mathcal{J}%
_{2}^{\prime}\beta_{z}^{\prime}-\frac{8r_{\psi}^{2}}{9z}\mathcal{J}%
_{4},\mathcal{J}_{AP}=\mathcal{J}_{12}\beta_{z}^{\prime},\nonumber\\
\mathcal{J}_{VS} &  =\frac{4}{27}\mathcal{J}_{12}\beta_{\psi}^{4}%
,\mathcal{J}_{V\tilde{P}}=\frac{2}{27}\mathcal{J}_{12}\left(  1+\frac{2r_{\psi
}^{2}}{z}+\frac{6r_{\psi}^{4}}{z^{2}}\right)  ,\mathcal{J}_{VT}=\frac{4}%
{9}\mathcal{J}_{12}\left(  1-\frac{14r_{\psi}^{2}}{3z}+\frac{38r_{\psi}^{4}%
}{3z^{2}}\right) ,
\end{align}
in addition to those in Eq.~(\ref{Defspxx}). The mode with a single photon with missing energy is%
\begin{align}
\frac{d\Gamma}{dz}\left(  K_{L}\to\gamma\Psi\overline{\Psi}\right)
&  =\frac{m_{K}^{4}}{\Lambda^{4}}\Gamma_{0}^{\gamma}\frac{z^{2}}{4r_{\psi}%
^{4}}\left\{  \mathcal{G}_{S}(\Re_{TS}^{2}+\Im_{\tilde{T}S}^{2})+\mathcal{G}%
_{P}(\Im_{TP}^{2}+\Re_{\tilde{T}P}^{2})+\mathcal{G}_{T}(\Re_{TT}^{2}%
+\Im_{\tilde{T}T}^{2})\right.  \nonumber\\
&  \;\;\;\;+\mathcal{G}_{SP}(\Re_{TP}\Re_{\tilde{T}S}+\Im_{\tilde{T}S}\Im
_{TP})+\mathcal{G}_{ST}(\Re_{TP}\Re_{TT}+\Im_{\tilde{T}P}\Im_{\tilde{T}%
T})+\mathcal{G}_{PT}(\Re_{\tilde{T}P}\Re_{TT}+\Im_{TP}\Im_{\tilde{T}%
T})\nonumber\\
&  \;\;\;\;\left.  +\mathcal{G}_{V}\Re_{VV}^{2}+\mathcal{G}_{A}\Re_{VA}%
^{2}-\Re_{VV}(\mathcal{G}_{VT}\Re_{TT}+\mathcal{G}_{VP}\Re_{\tilde{T}%
P}+\mathcal{G}_{VS}\Re_{TS})\right\}\;,
\end{align}
with the definitions in Eq.~(\ref{Defsgxx}) together with%
\begin{gather}
\mathcal{G}_{S}=\frac{B_{T}^{\prime2}}{36}\left(  1-\frac{26r_{\text{$\psi$}%
}^{4}}{z^{2}}+\frac{40r_{\psi}^{6}}{z^{3}}\right)  ,\;\mathcal{G}%
_{P}=\frac{B_{T}^{\prime2}}{36}\left(  1+\frac{4r_{\text{$\psi$}}^{2}}%
{z}-\frac{2r_{\psi}^{4}}{z^{2}}\right)  ,\;\mathcal{G}_{T}=\frac{4}%
{9}\mathcal{G}_{1}\left(  1-\frac{4r_{\text{$\psi$}}^{2}}{z}+\frac{10r_{\psi
}^{4}}{z^{2}}\right)  \;,\nonumber\\
\mathcal{G}_{SP}=-\frac{B_{T}^{\prime2}}{9}\beta_{z}^{2}\frac{2r_{\psi}^{2}%
}{z},\mathcal{G}_{ST}=\frac{B_{T}^{\prime2}}{9}\beta_{z}^{2}\left(
1-\frac{10r_{\text{$\psi$}}^{4}}{z^{2}}\right)  ,\mathcal{G}_{PT}%
=\frac{B_{T}^{\prime2}}{9}\left(  1+\frac{14r_{\text{$\psi$}}^{4}}{z^{2}%
}\right)  ,\mathcal{G}_{V}=\mathcal{G}_{4}\beta_{5}+\mathcal{G}_{3}%
\frac{8r_{\psi}^{2}}{9z},\nonumber\\
\mathcal{G}_{A}=\mathcal{G}_{3}\beta_{9}-\mathcal{G}_{3}\frac{8r_{\psi}^{2}%
}{9z},\mathcal{G}_{VS}=\frac{4\mathcal{G}_{2}}{27}\beta_{z}^{4},\mathcal{G}%
_{VP}=\frac{2\mathcal{G}_{2}}{27}\left(  1+\frac{2r_{\psi}^{2}}{z}%
+\frac{6r_{\psi}^{4}}{z^{2}}\right)  ,\mathcal{G}_{VT}=\frac{4\mathcal{G}_{2}%
}{9}\left(  1-\frac{14r_{\psi}^{2}}{3z}+\frac{38r_{\psi}^{4}}{3z^{2}}\right)
.
\end{gather}

Finally, with the many possible interferences among the tensor currents and the complicated kinematical functions, the differential rates for $K_{L}\to\pi\pi\Psi\overline{\Psi}$ are too cumbersome to be given here.

\section{Differential rates for $B$ decays}

\subsection{Experimental observables in rare $B$ decays\label{AppBexp}}

Let us start by reviewing the kinematics and the current experimental limits for the various $B$ decays induced by neutral currents and involving missing energy. For many of these modes, the kinematics and phase-space integrals are similar to that of the corresponding $K$ decays, so we refer to Appendix~\ref{AppKexp} for the explicit expressions.

Specifically, the observable three-body differential distributions for the $B\to HX$ modes, $H=\pi,\rho,K,K^{\ast}$ and $X=\bar{\psi}\psi,\phi(\phi),V(V),\bar{\Psi}\Psi$, can be written in terms of the (reduced) invariant mass $z=q^{2}/m_{B}^{2}$ of the invisible particles, or equivalently in terms of the $H$ momentum in the $B$ rest-frame ($|\mathbf{p}_{H}|=m_{B}\lambda^{1/2}(1,z,r_{H}^{2})/2$) or the missing energy ($\slash \hspace{-0.19cm}E=m_{B}(1+z-r_{H}^{2})/2$), as explicitly written down in Eq.~(\ref{DiffKpiXX}) for the $K\to\pi X$ case.

\subsubsection{$b\to s+$ missing energy}

For these transitions, we can obtain bounds on the production of new invisible states by using the existing experimental data from Babar~\cite{:2008fr,delAmoSanchez:2010bk} and Belle~\cite{:2007zk} searches for $B\to K^{(\ast)}\nu\bar{\nu}$ decays. In particular, the presently most stringent bounds are
\begin{align}
\mathcal{B}(B^{+}  &  \to K^{\ast+}\nu\bar{\nu})<8\cdot 10^{-5}~\text{\cite{:2008fr}}\;,\\
\mathcal{B}(B^{+}  &  \to K^{+}\nu\bar{\nu})<1.3\cdot10^{-5}~\text{\cite{delAmoSanchez:2010bk}}\;,
\end{align}
both at $90\%$ C.L. and assuming the SM differential rates.

However, in general, the kinematical distributions (and the associated phase-space ranges) depend on the nature and couplings of the invisible particles. So, one would need to correct for the associated experimental reconstruction efficiencies and background shapes. Most notably, the SM backgrounds typically rise steeply at small final state $K^{(\ast)}$ momentum in the center of mass frame, thus reducing signal sensitivity in this region~\cite{delAmoSanchez:2010bk}. Without the detailed knowledge of the experimental analyses and detectors, we cannot faithfully reproduce the final signal sensitivity distribution. However, we consider these effects to be the least severe for massless final state invisible particles, since the kinematical distributions in this case at least cover the whole kinematical region populated by the SM signal with (almost) massless neutrinos. There we derive our tentative bounds on the individual NP operators given in Tables~\ref{TableBNDB2} and~\ref{TableBNDB1}. The impact of purely kinematical (phase-space) effects on the NP scale sensitivity away from the massless invisible particle limit is illustrated in Figure~\ref{FigBehaveB} for the case of pair production of two invisible fermions through the various considered operators.

The Super-B factories are expected to provide a sensitivity down to a fraction of the expected SM signal. Then, possible measurement of the $K^{(\ast)}$ momentum or missing energy distributions could provide additional powerful discriminants in the search for NP contributions~\cite{Altmannshofer:2009ma}. In this respect, the decay $B\to K^{\ast}X$ has the virtue that the angular distribution of the $K^{\ast}$ decay products allows to extract information about the polarization of the $K^{\ast}$. The experimental information that can be obtained from the process $B\to K^{\ast}(\to K\pi)X$ with an on-shell $K^{\ast}$ is completely described by the double differential decay distribution in terms of the two kinematical variables $s=(p_{B}-p_{K})^{2}$, corresponding to the invariant mass of the final state invisible particles, and $\theta$, the angle between the $K^{\ast}$ flight direction in the $B$ rest frame and the $K$ flight direction in the $K\pi~(K^{\ast})$ rest frame. The spectrum can be expressed in terms of $B\to K^{\ast}$ transversity rates $\Gamma_{L,T}$ corresponding to longitudinally and transversely polarized $K^{\ast}$ final states (see App.~\ref{AppBtransverse}), while the double differential spectrum can be written as
\begin{equation}
\frac{d^{2}\Gamma}{ds\,d\cos\theta}=\frac{3}{4}\frac{d\Gamma_{T}}{ds}\sin
^{2}\theta+\frac{3}{2}\frac{d\Gamma_{L}}{ds}\cos^{2}\theta\,.
\end{equation}
Thus, $d\Gamma_{L}/ds$ and $d\Gamma_{T}/ds$ can be extracted by an angular
analysis of the $K^{\ast}$ decay products. Finally, the total invisible mass
distribution is
\begin{equation}
\frac{d\Gamma}{ds}=\int_{-1}^{1}d\cos\theta\frac{d^{2}\Gamma}{ds\,d\cos\theta
}=\frac{d\Gamma_{T}}{ds}+\frac{d\Gamma_{L}}{ds}\,.
\end{equation}
As an illustration of the potential impact of such future precision measurements, in Tables~\ref{TableBNDB2} and~\ref{TableBNDB1}, we also present the accessible NP scales assuming a $20\%$ relative precision on the $B\to K^{(\ast)}X$ rates compared to their SM predictions.

\subsubsection{$b\to d+$ missing energy}

The $B^{+}\to\pi^{+}(\rho^{+})X$ processes in the SM, with $X=\nu\bar{\nu}$, receive a dominant contribution already at the tree level though the decay chain $B^{+}\to\tau^{+}\nu_{\tau}\to\pi^{+}(\rho^{+})\nu_{\tau}\bar{\nu}_{\tau}$~\cite{Kamenik:2009kc}. They have recently been measured by both Belle~\cite{Hara:2010dk} and Babar~\cite{:2010rt}. Reinterpreting their $\mathcal{B}(B^{+}\to\tau^{+}\nu_{\tau})$ values in the $\pi(\rho)$ decay channels of the $\tau$ by multiplying them with the corresponding well measured $\tau\to\pi(\rho)\nu$ branching fractions we obtain
\begin{align}
\mathcal{B}(B^{+}  &  \to\pi^{+}\nu\bar{\nu})=1.96(85)\cdot10^{-5}\;,\;\;\label{BPpirho}\\
\mathcal{B}(B^{+}  &  \to\rho^{+}\nu\bar{\nu})=0.97(50)\cdot10^{-4}\;.
\end{align}

A potential NP signal in these modes would manifest itself via differences in the measured $\mathcal{B}(B\to\tau\nu)$ values in the leptonic and hadronic decay modes of the $\tau$ -- something that future dedicated experimental searches could employ to reduce systematic uncertainties. Although the SM signal shape in this case is well determined and largely free from theoretical form factor uncertainties, the appearance of two neutrinos in the final state means that the same experimental caveats in extracting bounds on NP contributions apply as in the $B\to K^{(\ast)}X$ case~\cite{Kamenik:2009kc}. Again the polarization states of the $\rho$ in the $B\to\rho X$ mode, which are well predicted within the SM, could be reconstructed using the angular distributions of the $\rho\to2\pi$ system and aid in discriminating SM contributions from possible NP effects.

The neutral modes $B^{0}\to\pi^{0}(\rho^{0})X$ are free from long distance SM contributions, but the purely neutral final states make them more challenging experimentally. The present bounds of
\begin{align}
\mathcal{B}(B^{0}  &  \to\pi^{0}\nu\bar{\nu})<2.2\cdot 10^{-4}~\text{\cite{:2007zk}}\;,\\
\mathcal{B}(B^{0}  &  \to\rho^{0}\nu\bar{\nu})<4.4\cdot 10^{-4}~\text{\cite{:2007zk}}\;,
\end{align}
are less constraining than the charged modes analysis. Consequently, in setting our bounds on invisible particles in the massless limit in Tables~\ref{TableBNDB2} and~\ref{TableBNDB1} and away from this limit in Figure~\ref{FigBehaveB} we tentatively allow NP contributions to saturate the experimental uncertainties in the charged modes, Eq.~(\ref{BPpirho}).

\subsubsection{Other modes with missing energy}

Both Belle~\cite{BelleEPS} and Babar~\cite{Aubert:2004xy} have searched for the $B\to XX$ decay mode, which is helicity suppressed in the SM. While unresolved soft photons can partially lift this suppression~\cite{Badin:2010uh,Becirevic:2009aq}, the SM predictions for the branching ratio remain at the order of $10^{-9}$~\cite{Badin:2010uh}. Being a two body decay process in the scenarios we consider, the kinematical distributions are trivial and no model-dependent efficiency corrections are needed. The latest experimental upper limit reads%
\begin{equation}
\mathcal{B}(B\to XX)<1.3\cdot10^{-4}\,@\,90\%~C.L.~\text{\cite{BelleEPS}\ ,}%
\end{equation}
and can be employed directly to constrain the relevant interactions of invisible particles, as given in Tables~\ref{TableBNDB2} and~\ref{TableBNDB1} and Fig.~\ref{FigBehaveB}. In the future, both $B_{s,d}\to XX$ modes could potentially be probed to greater accuracy at the super flavor factories~\cite{SuperB}.

On the other hand, while $B_{s,d}\to\gamma X$ with reconstructed final state photons allow to lift the helicity suppression suffered by the NP fermionic contributions proportional to $f_{AA}$, the additional $\alpha_{\mathrm{EM}}$ suppression only makes them competitive with the $B\to\rho(K^{\ast})X$ modes in the opposite region of large invisible particle masses ($m_{B}-m_{\rho(K^{\ast})}<m_{X}<m_{B}$), where the helicity suppression is least and the $B_{s,d}\to X$ channels are effective in constraining NP effects as well. In addition, precisely in this region the SM backgrounds, most notably from misreconstructed $B_{q}\to\gamma X$ events, limit the experimental reach of $B_{s,d}\to\gamma X$~\cite{Aubert:2004xy}. For a more detailed discussion of the expected NP sensitivity, we refer to Ref.~\cite{Badin:2010uh} and do not consider these modes any further.

\subsubsection{Transversity basis in $B\to H^{\ast}$ transitions\label{AppBtransverse}}

Consider the decay $B\to H^{\ast}X$ with the $B$ meson decaying to an on-shell vector meson $H^{\ast}$ and other (invisible) particles $X$. The amplitude for this process can be written as
\begin{equation}
\mathcal{M}_{(m)}(B\to H^{\ast}X)=\epsilon_{H^{\ast}}^{\mu}(m)M_{\mu
}\,,
\end{equation}
where $\epsilon_{H^{\ast}}^{\mu}(m)$ is the polarization vector of the $H^{\ast}$. Being on-shell, the $H^{\ast}$ has only three polarization states, satisfying $\epsilon_{H^{\ast}}\cdot p_{H^{\ast}}=0$. In addition, the polarization vectors satisfy the following relations
\begin{equation}
\epsilon_{H^{\ast}}^{\ast\mu}(m)\epsilon_{H^{\ast}\mu}(m^{\prime}%
)=-\delta_{mm^{\prime}}\,,\;\;\sum_{m,m^{\prime}}\epsilon_{H^{\ast}}^{\ast\mu
}(m)\epsilon_{H^{\ast}}^{\nu}(m^{\prime})\delta_{mm^{\prime}}=-g^{\mu\nu
}+\frac{p_{H^{\ast}}^{\mu}p_{H^{\ast}}^{\nu}}{m_{H^{\ast}}^{2}}\,.
\end{equation}

In measurements where only the $H^{\ast}$ decay products are reconstructed, only two of the three $H^{\ast}$ polarization states can be disentangled. The corresponding transversity rates ($\Gamma_{L}$ and $\Gamma_{T}$) can be projected using covariant polarization projectors. First, we write the sum of squared amplitudes over $H^{\ast}$ polarizations as
\begin{equation}
\sum_{m}|\mathcal{M}_{(m)}(B\to H^{\ast}X)|^{2}=M_{\mu}M_{\nu
}^{\dagger}\mathbb{P}^{\mu\nu}\,,\;\mathbb{P}^{\mu\nu}=-g^{\mu\nu
}+\frac{p_{H^{\ast}}^{\mu}p_{H^{\ast}}^{\nu}}{m_{H^{\ast}}^{2}}\;.
\end{equation}
The longitudinal $H^{\ast}$ polarization vector should satisfy $\mathbf{p}_{H^{\ast}}\Vert\mathbf{\epsilon}_{H^{\ast}}$ in the $B$ meson rest frame and the corresponding decay rate ($\Gamma_{L}$) is obtained with the help of the projector
\begin{equation}
\mathbb{P}_{L}^{\mu\nu}=\frac{4r_{H^{\ast}}^{2}}{m_{B}^{2}\lambda_{H^{\ast}%
z}^{2}}\left(  p_{B}^{\mu}-\frac{p_{B}\cdot p_{H^{\ast}}}{m_{H^{\ast}}^{2}%
}p_{H^{\ast}}^{\mu}\right)  \left(  p_{B}^{\nu}-\frac{p_{B}\cdot p_{H^{\ast}}%
}{m_{H^{\ast}}^{2}}p_{H^{\ast}}^{\nu}\right)  \,,
\end{equation}
where $m_{B}\lambda_{H^{\ast}z}^{1/2}/2=|\mathbf{p}_{H^{\ast}}|$, $\lambda_{iz}\equiv\lambda(1,z,r_{i}^{2})$, is the absolute value of the $H^{\ast}$ momentum in the $B$ rest frame. The (unpolarized) transverse $H^{\ast}$ helicity projector entering the transverse rate $\Gamma_{T}$ is then simply obtained as $\mathbb{P}_{T}^{\mu\nu}=\mathbb{P}^{\mu\nu}-\mathbb{P}_{L}^{\mu\nu}$.

\subsection{Matrix elements for $B$ decays}

The hadronic matrix elements between a $B_{q}$ state and the vacuum are
\begin{subequations}
\begin{align}
\langle0|\bar{b}\gamma_{\mu}\gamma_{5}q|B_{q}(p_{B})\rangle &  =if_{B_{q}%
}p_{B}^{\mu}\,,\\
\langle0|\bar{b}\gamma_{5}q|B_{q}(p_{B})\rangle &  =-i\frac{m_{B_{q}}^{2}%
}{(m_{b}+m_{q})}f_{B_{q}}\,.
\end{align}
For the decay constants, we use the values of a recent lattice QCD average~\cite{Laiho:2009eu}: $f_{B_{s}}=0.2388(95)$~GeV and $f_{B_{d}}=0.1928(99)$~GeV. The matrix elements for $B\to H_{q}$ (where $H$ is a pseudoscalar meson, i.e. $K$ or $\pi$) are~\cite{Ball:2004ye}%
\end{subequations}
\begin{subequations}
\begin{align}
\langle H(p_{H})|\bar{b}\gamma^{\mu}q|B(p_{B})\rangle &  =f_{+}^{H}%
(q^{2})\left[  P^{\mu}-\frac{1-r_{H}^{2}}{z}q^{\mu}\right]  +f_{0}^{H}%
(q^{2})\frac{1-r_{H}^{2}}{z}q^{\mu}\,,\\
\langle H(p_{H})|\bar{b}q|B(p_{B})\rangle &  =\frac{m_{B}^{2}}{m_{b}-m_{q}%
}(1-r_{H}^{2})f_{0}^{H}(q^{2})\,,\\
\langle H(p_{H})|\bar{b}\sigma^{\mu\nu}q|B(p_{B})\rangle &  =i\frac{P^{\mu
}q^{\nu}-P^{\nu}q^{\mu}}{m_{B}(1+r_{H})}f_{T}^{H}(q^{2})\,,
\end{align}
where $P^{\mu}=(p_{B}+p_{H})^{\mu}$ and $q^{\mu}=(p_{B}-p_{H})^{\mu}$, while $z=q^{2}/m_{B}^{2}$ and $r_{i}=m_{i}/m_{B}$. The $\langle H(p_{H})|\bar{b}\sigma^{\mu\nu}\gamma_{5}q|B(p_{B})\rangle$ matrix element can be obtained via the Chisholm identity. We also consider $B\to H^{\ast}$ matrix elements~\cite{Ball:2004rg} (where $H^{\ast}$ is a vector meson, i.e. $K^{\ast}$ or $\rho$)
\end{subequations}
\begin{subequations}
\begin{align}
\langle H^{\ast}(p_{H^{\ast}},\epsilon_{H^{\ast}})|\bar{b}\gamma_{\mu
}q|B(p_{B})\rangle &  =2\epsilon_{\mu\nu\rho\sigma}\epsilon_{H^{\ast}}%
^{\ast\nu}p_{B}^{\rho}p_{H^{\ast}}^{\sigma}\frac{V^{H^{\ast}}(q^{2})}%
{m_{B}(1+r_{H^{\ast}})}\,,\\
\langle H^{\ast}(p_{H^{\ast}},\epsilon_{H^{\ast}})|\bar{b}\gamma^{\mu}%
\gamma_{5}q|B(p_{B})\rangle &  =i\epsilon_{H^{\ast}}^{\ast\mu}m_{B}%
(1+r_{K})A_{1}^{H^{\ast}}(q^{2})-iP^{\mu}(\epsilon_{H^{\ast}}^{\ast}\cdot
q)\frac{A_{2}^{H^{\ast}}(q^{2})}{m_{B}(1+r_{H^{\ast}})}\nonumber\\
&  \;\;\;\;-iq^{\mu}(\epsilon_{H^{\ast}}^{\ast}\cdot q)\frac{2r_{H^{\ast}}%
}{m_{B}z}[A_{3}^{H^{\ast}}(q^{2})-A_{0}^{H^{\ast}}(q^{2})]\,,\\
\langle H^{\ast}(p_{H^{\ast}},\epsilon_{H^{\ast}})|\bar{b}\gamma_{5}%
q|B(p_{B})\rangle &  =-2i\frac{m_{B}}{m_{b}+m_{q}}r_{K}(\epsilon_{H^{\ast}%
}^{\ast}\cdot q)A_{0}^{H^{\ast}}(q^{2})\,,\\
\langle H^{\ast}(p_{H^{\ast}},\epsilon_{H^{\ast}})|\bar{b}\sigma_{\mu\nu
}q|B(p_{B})\rangle &  =i\epsilon_{\mu\nu\rho\sigma}\left[  P^{\rho}%
\epsilon_{H^{\ast}}^{\ast\sigma}-\frac{1-r_{K}^{2}}{z}q^{\rho}\epsilon
_{H^{\ast}}^{\ast\sigma}+\frac{\epsilon_{H^{\ast}}^{\ast}\cdot q}{m_{B}^{2}%
z}q^{\rho}P^{\sigma}\right]  T_{1}^{H^{\ast}}(q^{2})\nonumber\\
&  \;\;\;\;-i\frac{1-r_{K}^{2}}{z}\epsilon_{\mu\nu\rho\sigma}q^{\rho}%
\epsilon_{H^{\ast}}^{\ast\sigma}T_{2}^{H^{\ast}}(q^{2})+i\frac{\epsilon
_{H^{\ast}}^{\ast}\cdot q}{m_{B}^{2}z}\epsilon_{\mu\nu\rho\sigma}P^{\rho
}q^{\sigma}\tilde{T}_{3}^{H^{\ast}}(q^{2})\,,
\end{align}
while the matrix elements $\langle H^{\ast}(p_{H^{\ast}},\epsilon_{H^{\ast}})|\bar{b}\sigma^{\mu\nu}q|B(p_{B})\rangle$ are again determined via the Chisholm identity. At $q^{2}=0$ the form factors satisfy the following relations $f_{+}^{H}(0)=f_{0}^{H}(0)$, $A_{0}^{H^{\ast}}(0)=A_{3}^{H^{\ast}}(0)$, and $T_{1}^{H^{\ast}}(0)=T_{2}^{H^{\ast}}(0)=\tilde{T}_{3}^{H^{\ast}}(0)$. For convenience we will also define the following auxiliary form factor combinations%
\end{subequations}
\begin{subequations}
\begin{align}
A_{L}^{H^{\ast}}(q^{2})  &  \equiv A_{2}^{H^{\ast}}(q^{2})-\frac{1}%
{\lambda_{H^{\ast}z}}(1+r_{H^{\ast}})^{2}(1-r_{H^{\ast}}^{2}%
-z)A_{1}^{H^{\ast}}(q^{2})\,,\\
T_{L}^{H^{\ast}}(q^{2})  &  \equiv\tilde{T}_{3}^{H^{\ast}}(q^{2}%
)-\frac{1}{\lambda_{H^{\ast}z}}(1-r_{H^{\ast}}^{2})(1-r_{H^{\ast}}%
^{2}-z)T_{2}^{H^{\ast}}(q^{2})\,,
\end{align}
where $\lambda_{iz}\equiv\lambda(1,z,r_{i}^{2})$.

In our numerical analysis, we employ the form-factor normalizations, shapes, and the associated uncertainties as determined in Refs.~\cite{Ball:2004ye,Ball:2004rg, Khodjamirian:2011ub} using light-cone QCD sum rules. In particular we employ results of Ref.~\cite{Ball:2004rg} for the $V^{K^{\ast},\rho},A_{i}^{K^{\ast},\rho},T_{i}^{K^{\ast},\rho}$ values, Ref.~\cite{Ball:2004ye} for the $f_{+,0,T}^{K}$ and $f_{T}^{\pi}$ form-factors, while we use the results of a more recent calculation~\cite{Khodjamirian:2011ub} for $f_{+,0}^{\pi}$.

\subsection{Spin 1/2 invisible particles in the final states\label{AppBspin12}}

The differential rates into a pair of invisible fermions are%
\end{subequations}
\begin{subequations}
\begin{align}
\Gamma(B_{q}\to\psi\bar{\psi}) &  =\frac{m_{B_{q}}^{4}}{\Lambda^{4}%
}\Gamma_{\psi}^{B_{q}}\left\{  \mathcal{I}_{\psi}^{PP}|f_{PP}|^{2}%
+\mathcal{I}_{\psi}^{PS}|f_{PS}|^{2}-\mathcal{I}_{\psi}^{AA,PP}\Re
(f_{AA}f_{PP}^{\ast})+\mathcal{I}_{\psi}^{AA}|f_{AA}|^{2}\right\}  \,,\\
\frac{d\Gamma}{d{z}}(B\to H\psi\bar{\psi}) &  =\frac{m_{B}^{4}%
}{\Lambda^{4}}\Gamma_{\psi}^{BH}\left\{  \mathcal{J}_{\psi}^{\tilde{T}%
}|f_{\tilde{T}T}|^{2}+\mathcal{J}_{\psi}^{T}|f_{TT}|^{2}+\mathcal{J}_{\psi
}^{VV,T}\Re(f_{VV}f_{TT}^{\ast})+\mathcal{J}_{\psi}^{VV}|f_{VV}|^{2}\right.
\nonumber\\
&  \;\;\;\;\left.  +\mathcal{J}_{\psi}^{VA}|f_{VA}|^{2}+\mathcal{J}_{\psi
}^{SP}|f_{SP}|^{2}+\mathcal{J}_{\psi}^{SS}|f_{SS}|^{2}+\mathcal{J}_{\psi
}^{SP,VA}\Re(f_{SP}f_{VA}^{\ast})\right\}  \,,\\
\frac{d\Gamma_{L}}{dz}(B\to H^{\ast}\psi\bar{\psi}) &  =\frac{m_{B}%
^{4}}{\Lambda^{4}}\Gamma_{\psi}^{BH^{\ast}}\left\{  \mathcal{J}_{\psi}%
^{\prime\tilde{T}}|f_{\tilde{T}T}|^{2}+\mathcal{J}_{\psi}^{\prime T}%
|f_{TT}|^{2}+\mathcal{J}_{\psi}^{\prime AV,T}\Re(f_{AV}f_{\tilde{T}T}^{\ast
})+\mathcal{J}_{\psi}^{\prime PP}|f_{PP}|^{2}\right.  \nonumber\\
&  \;\;\;\;\left.  +\mathcal{J}_{\psi}^{\prime PS}|f_{PS}|^{2}+\mathcal{J}%
_{\psi}^{\prime PP,AA}\Re(f_{PP}f_{AA}^{\ast})+\mathcal{J}_{\psi}^{\prime
AV}|f_{AV}|^{2}+\mathcal{J}_{\psi}^{\prime AA}|f_{AA}|^{2}\right\}  \,,\\
\frac{d\Gamma_{T}}{dz}(B\to H^{\ast}\psi\bar{\psi}) &  =\frac{m_{B}%
^{4}}{\Lambda^{4}}\Gamma_{\psi}^{BH^{\ast}}\left\{  \mathcal{J}_{\psi}%
^{\prime\prime\tilde{T}}|f_{\tilde{T}T}|^{2}+\mathcal{J}_{\psi}^{\prime\prime
T}|f_{TT}|^{2}-\mathcal{J}_{\psi}^{\prime\prime VV,T}\Re(f_{VV}f_{TT}^{\ast
})+\mathcal{J}_{\psi}^{\prime\prime VV}|f_{VV}|^{2}+\mathcal{J}_{\psi}%
^{\prime\prime VA}|f_{VA}|^{2}\right.  \nonumber\\
&  \;\;\;\;\left.  +\mathcal{J}_{\psi}^{\prime\prime AV,\tilde{T}}\Re
(f_{AV}f_{\tilde{T}T}^{\ast})+\mathcal{J}_{\psi}^{\prime\prime SP,VA}%
\Re(f_{SP}f_{VA}^{\ast})+\mathcal{J}_{\psi}^{\prime\prime AV}|f_{AV}%
|^{2}+\mathcal{J}_{\psi}^{\prime\prime AA}|f_{AA}|^{2}\right\}  \,,
\end{align}
where $\Gamma_{i}^{B_{q}}=f_{B_{q}}^{2}\beta_{i}/8\pi m_{B_{q}}$, $\Gamma
_{i}^{BH^{(\ast)}}=m_{B}\beta_{iz}\lambda_{H^{(\ast)}z}^{1/2}/96\pi^{3}$,
$\beta_{iz}=\sqrt{1-4r_{i}^{2}/z}$, $\beta_{i}=\sqrt{1-4r_{i}^{2}}$ and%
\end{subequations}
\begin{align}
&  \mathcal{I}_{\psi}^{PP}=m_{B_{q}}^{2}/(m_{b}+m_{q})^{2}\,,\mathcal{I}%
_{\psi}^{PS}=\mathcal{I}_{\psi}^{PP}\beta_{\psi}^{2}\,,\mathcal{I}_{\psi
}^{AA,PP}=4r_{\psi}m_{B_{q}}/(m_{b}+m_{q})\,,\mathcal{I}_{\psi}^{AA}=4r_{\psi
}^{2}\,,\nonumber\\
&  \mathcal{J}_{\psi}^{\tilde{T}}=f_{T}^{H}(q^{2})^{2}z\lambda_{Hz}%
\beta_{\psi z}^{2}/2(1+r_{H})^{2}\,,\mathcal{J}_{\psi}^{T}=\mathcal{J}_{\psi
}^{\tilde{T}}(1+8r_{\psi}^{2}/z)/\beta_{\psi z}^{2}\,,\mathcal{J}_{\psi
}^{VV,T}=3f_{T}^{H}(q^{2})f_{+}(q^{2})r_{\psi}\lambda_{Hz}/(1+r_{H}%
)\,,\nonumber\\
&  \mathcal{J}_{\psi}^{VV}=\lambda_{Hz}f_{+}^{H}(q^{2})^{2}(1+2r_{\psi
}^{2}/z)/2\,,\mathcal{J}_{\psi}^{VA}=\mathcal{J}_{\psi}^{VV}\beta_{\psi z}%
^{2}/(1+2r_{\psi}^{2}/z)+3f_{0}^{H}(q^{2})^{2}r_{\psi}^{2}(1-r_{H}^{2}%
)^{2}/z\,,\nonumber\\
&  \mathcal{J}_{\psi}^{SP}=3zf_{0}^{H}(q^{2})^{2}(1-r_{H}^{2})^{2}m_{B}%
^{2}/4(m_{b}-m_{q})^{2}\,,\mathcal{J}_{\psi}^{SS}=\mathcal{J}_{\psi}^{SP}%
\beta_{\psi z}^{2}\,,\mathcal{J}_{\psi}^{SP,VA}=\mathcal{J}_{\psi}%
^{SP}4r_{\psi}(m_{b}-m_{q})/m_{B}z\,,\nonumber\\
&  \mathcal{J}_{\psi}^{\prime\tilde{T}}=\lambda_{{H^{\ast}}z}^{2}%
T_{L}^{H^{\ast}}(q^{2})^{2}(1+8r_{\psi}^{2}/z)/4r_{H^{\ast}}^{2}%
z\,,\mathcal{J}_{\psi}^{\prime T}=\lambda_{{H^{\ast}}z}^{2}\beta_{\psi z}%
^{2}T_{L}^{H^{\ast}}(q^{2})^{2}/4r_{H^{\ast}}^{2}z\,,\nonumber\\
&  \mathcal{J}_{\psi}^{\prime AV,T}=3r_{\psi}\lambda_{{H^{\ast}}z}^{2}%
A_{L}^{H^{\ast}}(q^{2})T_{L}^{H^{\ast}}(q^{2})/2r_{H^{\ast}}^{2}(1+r_{H^{\ast
}})z\,,\mathcal{J}_{\psi}^{\prime PP}=3z\lambda_{Hz}A_{0}^{H^{\ast}}%
(q^{2})^{2}m_{B}^{2}/4(m_{b}+m_{q})^{2}\,,\nonumber\\
&  \mathcal{J}_{\psi}^{\prime PS}=\mathcal{J}_{\psi}^{\prime PP}\beta_{\psi
z}^{2}\,,\mathcal{J}_{\psi}^{\prime PP,AA}=3r_{\psi}\lambda_{{H^{\ast}}z}A_{0}^{H^{\ast}}(q^{2})^{2}m_{B}/2(m_{B}+m_{q})\,,\nonumber\\
&  \mathcal{J}_{\psi}^{\prime AV}=\lambda_{{H^{\ast}}z}^{2}A_{L}^{H^{\ast}%
}(z)^{2}(1+2r_{\psi}^{2}/z)/8r_{H^{\ast}}^{2}(1+r_{H^{\ast}})^{2}%
\,,\mathcal{J}_{\psi}^{\prime AA}=\mathcal{J}_{\psi}^{\prime AV}\beta_{\psi
z}^{2}/(1+2r_{\psi}^{2}/z)+3\lambda_{{H^{\ast}}z}r_{\psi}^{2}%
A_{0}^{H^{\ast}}(q^{2})^{2}/z\,,\nonumber\\
&  \mathcal{J}_{\psi}^{\prime\prime\tilde{T}}=2\lambda_{{H^{\ast}}z}%
\beta_{\psi z}^{2}T_{1}^{H^{\ast}}(q^{2})^{2}+(1-r_{H^{\ast}}^{2})^{2}%
T_{2}^{H^{\ast}}(q^{2})^{2}(1+8r_{\psi}^{2}/z)\,,\mathcal{J}_{\psi}%
^{\prime\prime VV}=\lambda_{{H^{\ast}}z}V^{H^{\ast}}(q^{2})(z+2r_{\psi
}^{2})/(1+r_{H^{\ast}})^{2}\,,\nonumber\\
&  \mathcal{J}_{\psi}^{\prime\prime T}=2\lambda_{{H^{\ast}}z}%
T_{1}^{H^{\ast}}(q^{2})^{2}(1+8r_{\psi}^{2}/z)+(1-r_{H^{\ast}}^{2})^{2}%
\beta_{\psi z}^{2}T_{2}^{H^{\ast}}(q^{2})^{2}\,,\mathcal{J}_{\psi}%
^{\prime\prime VV,T}=12r_{\psi}\lambda_{{H^{\ast}}z}T_{1}^{H^{\ast}}%
(q^{2})V(q^{2})/(1+r_{H^{\ast}})\,,\nonumber\\
&  \mathcal{J}_{\psi}^{\prime\prime VA}=\lambda_{{H^{\ast}}z}z\beta_{\psi
z}^{2}V^{H^{\ast}}(q^{2})/(1+r_{H^{\ast}})^{2}\,,\mathcal{J}_{\psi}%
^{\prime\prime AV,\tilde{T}}=12(1-r_{H^{\ast}}^{2})(1+r_{H^{\ast}})r_{\psi
}A_{1}^{H^{\ast}}(q^{2})T_{2}^{H^{\ast}}(q^{2})\,,\nonumber\\
&  \mathcal{J}_{\psi}^{\prime\prime AV}=(1+r_{H^{\ast}})^{2}zA_{1}^{H^{\ast}%
}(q^{2})^{2}(1+2r_{\psi}^{2}/z)\,,\mathcal{J}_{\psi}^{\prime\prime
AA}=\mathcal{J}_{\psi}^{\prime\prime AV}\beta_{\psi z}^{2}/(1+2r_{\psi}%
^{2}/z)\,.
\end{align}

\subsubsection{Standard Model rates\label{AppBSMrates}}

The $B^{+}\to\pi^{+}(\rho^{+})\nu\bar{\nu}$ modes are dominated by the
tree-level contributions mediated by an intermediate on-shell tau lepton. To
an excellent (small tau-width) approximation, they are given by
\begin{subequations}
\begin{align}
\frac{d\Gamma}{d{z}}(B^{+}\to\pi^{+}\nu\bar{\nu}) &  =\Gamma
_{\mathrm{SM}}^{B\pi}\left[  (1-r_{\tau}^{2})(1-r_{\pi}^{2}/r_{\tau}%
^{2})-z\right]  \,,\\
\frac{d\Gamma_{T}}{d{z}}(B^{+}\to\rho^{+}\nu\bar{\nu}) &
=\Gamma_{\mathrm{SM}}^{B\rho}\frac{z}{\lambda_{\rho z}}\left[
\lambda_{\rho z}-2(1-r_{\tau}^{2})(1-r_{\rho}^{2}/r_{\tau}^{2})+2z\right]
\,,\\
\frac{d\Gamma_{L}}{d{z}}(B^{+}\to\rho^{+}\nu\bar{\nu}) &
=\Gamma_{\mathrm{SM}}^{B\rho}\frac{(1-z-r_{\rho}^{2})^{2}}{\lambda_{\rho
z}}\left[  (1-r_{\tau}^{2})(1-r_{\rho}^{2}/r_{\tau}^{2})-z\right]  \,,
\end{align}
where $\Gamma_{\mathrm{SM}}^{B\pi(\rho)}=m_{B}^{6}r_{\tau}^{5}f_{B}^{2}f_{\pi(\rho)}^{2}|V_{ub}V_{ud}^{\ast}|^{2}G_{F}^{4}/64\pi^{2}\Gamma_{\tau}$. To obtain the total rates one needs to integrate over the available phase-space ($z$), which for the on-shell tau contributions is given by $z\in\lbrack0,(1-r_{\tau}^{2})(1-r_{\pi(\rho)}^{2}/r_{\tau}^{2})]$. Alternatively one can normalize these distributions to the experimentally determined $B\to\tau\nu$ (from purely leptonic tau reconstruction) and $\tau\to\pi(\rho)\nu$ (from prompt tau decays) branching ratios since $\mathcal{B}(B^{+}\to\pi^{+}(\rho^{+})\nu\bar{\nu})\simeq \mathcal{B}(B\to\tau\nu)\times\mathcal{B}(\tau\to\pi(\rho)\nu)$, eliminating the theoretical uncertainties in the normalization factor $\Gamma_{0}$. The neutral modes are dominated by short distance loop contributions, similar to $B\to K^{(\ast)}\nu\bar{\nu}$, albeit further CKM suppressed, leading to branching ratios almost two orders of magnitude smaller compared to the charged modes~\cite{Kamenik:2009kc}.

The dominant (short distance loop) contributions to the kaon modes ($B\to K^{(\ast)}\nu\bar{\nu}$) are given by
\end{subequations}
\begin{subequations}
\begin{align}
\frac{d\Gamma}{d{z}}(B\to K\nu\bar{\nu})  &  =\Gamma_{\mathrm{SM}%
}^{BK}\lambda_{Kz}^{3/2}f_{+}(q^{2})^{2}\,,\\
\frac{d\Gamma_{T}}{d{z}}(B\to K^{\ast}\nu\bar{\nu})  &  =\Gamma
_{\mathrm{SM}}^{BK}2z\lambda_{K^{\ast}z}^{1/2}\left[  \frac{\lambda_{{K^{\ast}}%
z}}{(1+r_{K^{\ast}})^{2}}V^{K^{\ast}}(q^{2})^{2}+(1+r_{K^{\ast}})^{2}%
A_{1}^{K^{\ast}}(q^{2})^{2}\right]  \,,\\
\frac{d\Gamma_{L}}{d{z}}(B\to K^{\ast}\nu\bar{\nu})  &  =\Gamma
_{\mathrm{SM}}^{BK}\frac{\lambda_{{K^{\ast}}z}^{5/2}A_{L}^{K^{\ast}}(q^{2})^{2}%
}{4r_{K^{\ast}}^{2}(1+r_{K^{\ast}})^{2}}\,,
\end{align}
where now $\Gamma_{\mathrm{SM}}^{BK}=m_{B}^{5}(G_{F}\alpha|V_{tb}V_{ts}^{\ast}|C_{\nu\bar{\nu}}^{\mathrm{SM}})^{2}/256\pi^{5}$ and $|C_{\nu\bar{\nu}}^{\mathrm{SM}}|=6.33\pm0.06$~\cite{Kamenik:2010na,BrodGS10}. The charged modes also receive tree-level contributions mediated by intermediate on-shell $\tau$ leptons similar to $B^{+}\to\pi^{+}(\rho^{+})\nu\bar{\nu}$. These are however always subleading in the kaon case; of about $15\%$ of the above short-distance contributions~\cite{Kamenik:2009kc}.

\subsection{Spin 0 invisible particles in the final states\label{AppBspin0}}

The rates for the production of a single invisible scalar from the $H^{\ast}\left(  \bar{D}Q\right)  \phi$ operator of Eq.~(\ref{Axion}) are%
\end{subequations}
\begin{subequations}
\begin{align}
\Gamma(B\to H\phi) &  =\Gamma_{1\phi}^{BH}|g_{S}|^{2}\frac{(1-r_{H}%
^{2})^{2}}{(m_{b}-m_{q})^{2}}f_{0}^{H}(m_{\phi}^{2})^{2}\,,\\
\Gamma_{L}(B\to H^{\ast}\phi) &  =\Gamma_{1\phi}^{BH^{\ast}}%
|g_{P}|^{2}\frac{\lambda_{H^{\ast}\phi}}{(m_{b}+m_{q})^{2}}A_{0}^{H^{\ast
}}(m_{\phi}^{2})^{2}\,,
\end{align}
where $\Gamma_{1i}^{BH^{(\ast)}}=m_{B}\lambda_{H^{(\ast)}i}^{1/2}/16\pi$, $\lambda_{ij\neq z}\equiv\lambda(1,r_{i}^{2},r_{j}^{2})$. The vector operator contributions are related to these via quark EOM, see Eq.~(\ref{SubsAxion}). The rates for the production of two scalars are%
\end{subequations}
\begin{subequations}
\begin{align}
\Gamma(B_{q}\to\bar{\phi}\phi) &  =\frac{m_{B_{q}}^{2}}{\Lambda^{2}%
}\Gamma_{\phi}^{B_{q}}\frac{|g_{PS}|^{2}m_{B_{q}}^{2}}{2(m_{b}+m_{q})^{2}%
}\,,\\
\frac{d\Gamma}{d{z}}(B\to H\bar{\phi}\phi) &  =\frac{m_{B}^{2}%
}{\Lambda^{2}}\Gamma_{\phi}^{BH}\left\{  \mathcal{J}_{\phi}^{SS}|g_{SS}%
|^{2}+\mathcal{J}_{\phi}^{VV}\frac{m_{B}^{2}}{\Lambda^{2}}|g_{VV}%
|^{2}\right\}  \,,\\
\frac{d\Gamma_{L}}{dz}(B\to H^{\ast}\bar{\phi}\phi) &  =\frac{m_{B}%
^{2}}{\Lambda^{2}}\Gamma_{\phi}^{BH^{\ast}}\left\{  \mathcal{J}_{\phi}^{\prime
PS}|g_{PS}|^{2}+\mathcal{J}_{\phi}^{\prime AV}\frac{m_{B}^{2}}{\Lambda^{2}%
}|g_{AV}|^{2}-\mathcal{J}_{\phi}^{\prime AV,PS}\frac{m_{B}}{\Lambda}\Re
(g_{AV}g_{PS}^{\ast})\right\}  \,,
\end{align}
where%
\end{subequations}
\begin{equation}
\mathcal{J}_{\phi}^{SS}=\mathcal{J}_{\psi}^{SP}/2z\,,\mathcal{J}_{\phi}%
^{VV}=\lambda_{Hz}\beta_{\phi z}^{2}f_{+}^{H}(q^{2})^{2}/8\,,\mathcal{J}%
_{\phi}^{\prime PS}=\mathcal{J}_{\psi}^{\prime PP}/2z\,,\mathcal{J}_{\phi
}^{\prime AV}=\mathcal{J}_{\psi}^{\prime AV}\beta_{\phi z}^{2}/4(1+2r\phi
^{2}/z)\,.
\end{equation}
Note that if $\phi=\bar{\phi}$, Bose statistics has to be enforced and these rates should be divided by two.

\subsection{Spin 1 invisible particles in the final states\label{AppBspin1}}

The production of a single vector using the simple FCNC operators of Eq.~(\ref{HvI}) or the gauge-invariant operators of Eq.~(\ref{HvII1}) are%
\begin{equation}%
\begin{tabular}
[c]{lll}%
$B\to HV:$ & $\Gamma^{\lbrack\text{I}]}=\Gamma_{1V}^{BH}%
\mathcal{J}_{1V}^{V}|\epsilon_{V}|^{2}\;,$ & $\Gamma^{\lbrack\text{II}%
]}=\dfrac{m_{B}^{2}}{\Lambda^{2}}\Gamma_{1V}^{BH}\mathcal{J}_{1V}^{T}%
|h_{T}|^{2}\;,$\\
$B\to H^{\ast}V:$ & $\Gamma_{L}^{[\text{I}]}=\Gamma_{1V}^{BH^{\ast}%
}\mathcal{J}_{1V}^{\prime A}|\epsilon_{A}|^{2}\;,$ & $\Gamma_{L}^{[\text{II}%
]}=\dfrac{m_{B}^{2}}{\Lambda^{2}}\Gamma_{1V}^{BH^{\ast}}\mathcal{J}%
_{1V}^{\prime\tilde{T}}|h_{\tilde{T}}|^{2}\;,$\\
& $\Gamma_{T}^{[\text{I}]}=\Gamma_{1V}^{BH^{\ast}}\left\{  \mathcal{J}%
_{1V}^{\prime\prime V}|\epsilon_{V}|^{2}+\mathcal{J}_{1V}^{\prime\prime
A}|\epsilon_{A}|^{2}\right\}  \;,$ & $\Gamma_{T}^{[\text{II}]}=\dfrac
{m_{B}^{2}}{\Lambda^{2}}\Gamma_{1V}^{BH^{\ast}}\left\{  \mathcal{J}%
_{1V}^{\prime\prime T}|h_{T}|^{2}+\mathcal{J}_{1V}^{\prime\prime\tilde{T}}|h_{\tilde{T}%
}|^{2}\right\}  \;,$%
\end{tabular}
\end{equation}
where
\begin{subequations}
\begin{align}
\mathcal{J}_{1V}^{V} &  =\frac{\lambda_{HV}}{r_{V}^{2}}f_{+}^{H}(m_{V}%
^{2})^{2}\,,\mathcal{J}_{1V}^{\prime A}=\frac{\lambda_{H^{\ast}V}^{2}}%
{r_{V}^{2}}\frac{A_{L}^{H^{\ast}}(m_{V}^{2})^{2}}{4r_{H^{\ast}}^{2}%
(1+r_{H^{\ast}})^{2}}\,,\mathcal{J}_{1V}^{\prime\prime V}=2\lambda_{H^{\ast}%
V}\frac{V^{H^{\ast}}(m_{V}^{2})^{2}}{(1+r_{H^{\ast}})^{2}}\,,\\
\mathcal{J}_{1V}^{\prime\prime A} &  =2(1+r_{H^{\ast}})^{2}A_{1}^{H^{\ast}%
}(m_{V}^{2})^{2}\,,\mathcal{J}_{1V}^{T}=4r_{V}^{2}\lambda_{HV}f_{T}%
^{H}(m_{V}^{2})^{2}/(1+r_{V})^{2}\,,\;\\
\mathcal{J}_{1V}^{\prime\tilde{T}} &  =\lambda_{H^{\ast}V}^{2}T_{L}^{H^{\ast}%
}(m_{V}^{2})^{2}/r_{V}^{2}r_{H^{\ast}}^{2}\,,\;\mathcal{J}_{1V}^{\prime\prime
T}=8\lambda_{{H^{\ast}}V}T_{1}^{H^{\ast}}(m_{V}^{2})^{2}\,,\mathcal{J}%
_{1V}^{\prime\prime\tilde{T}}=8(1-r_{H^{\ast}}^{2})^{2}T_{2}^{H^{\ast}}%
(m_{V}^{2})^{2}\,.
\end{align}
Note that in the $m_{V}\to0$ limit, $\Gamma_{L}^{[\text{II}]}(B\to H^{\ast}V)$ is well-defined while the tensor operator contributions actually vanish thanks to the form factor relation $T_{L}^{H}(0)=0$.

The rate and differential rates for the production of two vectors are%
\end{subequations}
\begin{subequations}
\begin{align}
\Gamma^{\lbrack\text{II}]}(B_{q}\to V\bar{V}) &  =\frac{m_{B_{q}}^{4}%
}{\Lambda^{4}}\Gamma_{V}^{B_{q}}\left\{  \mathcal{I}_{V}^{PP}|h_{PP}%
|^{2}+\mathcal{I}_{V}^{PS}|h_{PS}|^{2}\right\}  \,,\\
\frac{d\Gamma^{\lbrack\text{II}]}}{d{z}}(B\to HV\bar{V}) &
=\frac{m_{B}^{4}}{\Lambda^{4}}\Gamma_{V}^{BH}\left\{  \mathcal{J}_{V}%
^{SP}|h_{SP}|^{2}+\mathcal{J}_{V}^{SS}|h_{SS}|^{2}\right\}  \,,\\
\frac{d\Gamma_{L}^{[\text{II}]}}{dz}(B\to H^{\ast}V\bar{V}) &
=\frac{m_{B}^{4}}{\Lambda^{4}}\Gamma_{V}^{BH^{\ast}}\left\{  \mathcal{J}%
_{V}^{\prime PS}|h_{PS}|^{2}+\mathcal{J}_{V}^{\prime PP}|h_{PP}|^{2}\right\}
\,,
\end{align}
where
\end{subequations}
\begin{align}
\mathcal{I}_{V}^{PP} &  =\beta_{V}^{2}\mathcal{I}_{\psi}^{PP}\,,\mathcal{I}%
_{V}^{PS}=\mathcal{I}_{\psi}^{PP}(\beta_{V}^{2}+6r_{V}^{4})\,,\nonumber\\
\mathcal{J}_{V}^{SP} &  =z\beta_{Vz}^{2}\mathcal{J}_{\psi}^{SP}\,,\mathcal{J}%
_{V}^{SS}=z\mathcal{J}_{\psi}^{SP}(\beta_{Vz}^{2}+6r_{V}^{4}/z^{2}%
)\,,\nonumber\\
\mathcal{J}_{V}^{\prime PS} &  =z\beta_{Vz}^{2}\mathcal{J}_{\psi}^{\prime
PP}\,,\mathcal{J}_{V}^{\prime PP}=z\mathcal{J}_{\psi}^{\prime PP}(\beta
_{Vz}^{2}+6r_{V}^{4}/z^{2})\,.
\end{align}

\subsection{Spin 3/2 invisible particles in the final states\label{AppBspin32}}

The rate for the $\Psi\overline{\Psi}$ modes are%
\begin{subequations}
\begin{align}
\Gamma(B_{q}\to\Psi\overline{\Psi}) &  =\frac{m_{B_{q}}^{8}}%
{\Lambda^{8}}\Gamma_{\psi}^{B_{q}}\left\{  \mathcal{I}_{\psi^{\prime}}%
^{PP}|f_{PP}|^{2}+\mathcal{I}_{\psi^{\prime}}^{PS}|f_{PS}|^{2}-\mathcal{I}%
_{\psi^{\prime}}^{AA,PP}\Re(f_{AA}f_{PP}^{\ast})+\mathcal{I}_{\psi^{\prime}%
}^{AA}|f_{AA}|^{2}\right\}  \,,\\
\frac{d\Gamma}{d{z}}(B\to H\Psi\overline{\Psi}) &  =\frac{m_{B}^{8}%
}{\Lambda^{8}}\Gamma_{\psi}^{BH}\left\{  \mathcal{J}_{\psi^{\prime}}%
^{\tilde{T}}|f_{\tilde{T}T}|^{2}+\mathcal{J}_{\psi^{\prime}}^{T}|f_{TT}%
|^{2}+\mathcal{J}_{\psi^{\prime}}^{TS}|f_{TS}|^{2}+\mathcal{J}_{\psi^{\prime}%
}^{\tilde{T}S}|f_{\tilde{T}S}|^{2}\right.  \nonumber\\
&  \left.  +\mathcal{J}_{\psi^{\prime}}^{TP}|f_{TP}|^{2}+\mathcal{J}%
_{\psi^{\prime}}^{\tilde{T}P}|f_{\tilde{T}P}|^{2}+\mathcal{J}_{\psi^{\prime}%
}^{VV}|f_{VV}|^{2}+\mathcal{J}_{\psi^{\prime}}^{VA}|f_{VA}|^{2}+\mathcal{J}%
_{\psi^{\prime}}^{SP}|f_{SP}|^{2}+\mathcal{J}_{\psi^{\prime}}^{SS}|f_{SS}%
|^{2}\right.  \nonumber\\
&  -\mathcal{J}_{\psi^{\prime}}^{TP,\tilde{T}S}\Re(f_{TP}f_{\tilde{T}S}^{\ast
})+\mathcal{J}_{\psi^{\prime}}^{\tilde{T},\tilde{T}S}\Re(f_{\tilde{T}%
T}f_{\tilde{T}S}^{\ast})-\mathcal{J}_{\psi^{\prime}}^{T,TS}\Re(f_{TT}%
f_{TS}^{\ast})-\mathcal{J}_{\psi^{\prime}}^{T,\tilde{T}P}\Re(f_{TT}%
f_{\tilde{T}P}^{\ast})\nonumber\\
&  -\mathcal{J}_{\psi^{\prime}}^{\tilde{T},TP}\Re(f_{\tilde{T}T}f_{TP}^{\ast
})-\mathcal{J}_{\psi^{\prime}}^{TS,VV}\Re(f_{TS}f_{VV}^{\ast})\nonumber\\
&  \left.  -\mathcal{J}_{\psi^{\prime}}^{\tilde{T}P,VV}\Re(f_{\tilde{T}%
P}f_{VV}^{\ast})+\mathcal{J}_{\psi^{\prime}}^{VV,T}\Re(f_{VV}f_{TT}^{\ast
})+\mathcal{J}_{\psi^{\prime}}^{VA,P}\Re(f_{VA}f_{SP}^{\ast})\right\}  \,,
\end{align}
where $\beta_{i}^{\prime}=(9-6\beta_{i}^{2}+5\beta_{i}^{4})/18$, $\beta
_{i}^{\prime\prime}=(5-6\beta_{i}^{2}+9\beta_{i}^{4})/18$ and%
\end{subequations}
\begin{align}
&  \mathcal{I}_{\psi^{\prime}}^{PP}=\beta_{\psi}^{\prime}{}\mathcal{I}_{\psi
}^{PP}/4\,,\mathcal{I}_{\psi^{\prime}}^{PS}=\mathcal{I}_{\psi}^{PS}\beta
_{\psi}^{\prime\prime}/4\,,\mathcal{I}_{\psi^{\prime}}^{AA,PP}=\beta_{\psi
}^{\prime}{}\mathcal{I}_{\psi}^{AA,PP}/4\,,\mathcal{I}_{\psi^{\prime}}%
^{AA}=\beta_{\psi}^{\prime}r_{\psi}^{2}\,,\nonumber\\
&  \mathcal{J}_{\psi^{\prime}}^{\tilde{T}}=z^{2}{}\mathcal{J}_{\psi}%
^{\tilde{T}}\beta_{\psi z}^{\prime}/4\,,_{\psi^{\prime}}^{T}=z^{2}%
{}\mathcal{J}_{\psi}^{\tilde{T}}(1+2r_{\psi}^{2}/z-14r_{\psi}^{4}%
/z^{2}+80r_{\psi}^{6}/z^{3})/9\beta_{\psi z}^{2}\,,\mathcal{J}_{\psi^{\prime}%
}^{TS}=z^{2}{}\mathcal{J}_{\psi}^{\tilde{T}}(1+5r_{\psi}^{2}/z)\beta_{\psi
z}^{2}/36\,,\nonumber\\
&  \mathcal{J}_{\psi^{\prime}}^{\tilde{T}S}=z\mathcal{J}_{\psi}^{\tilde{T}%
}r_{\psi}^{2}(3+10r_{\psi}^{2}/z)/36\,,\mathcal{J}_{\psi^{\prime}}^{TP}%
=z^{2}{}\mathcal{J}_{\psi}^{\tilde{T}}(1+3r_{\psi}^{2}/z)/36\,,\mathcal{J}%
_{\psi^{\prime}}^{\tilde{T}P}=5z\mathcal{J}_{\psi}^{\tilde{T}}r_{\psi}%
^{2}(1+2r_{\psi}^{2}/z)/36\beta_{\psi z}^{2}\,,\nonumber\\
&  \mathcal{J}_{\psi^{\prime}}^{VV}=\lambda_{Hz}z^{2}f_{+}^{H}%
(q^{2})(36r_{\psi}^{6}/z^{3}-2r_{\psi}^{4}/z^{2}-2r_{\psi}^{2}%
/z+1)/18\,,\nonumber\\
&  \mathcal{J}_{\psi^{\prime}}^{VA}=z[54(r_{H}^{2}-1)^{2}r_{\psi}^{2}f_{0}%
^{H}(q)^{2}\beta_{\psi z}^{\prime}+\lambda_{Hz}zf_{+}^{H}(q^{2})^{2}%
\beta_{\psi z}^{2}(9\beta_{\psi z}^{\prime}-8r_{\psi}^{2}/z)]/72\,,\mathcal{J}%
_{\psi^{\prime}}^{SP}=z^{2}{}\mathcal{J}_{\psi}^{SP}\beta_{\psi}^{\prime
}/4\,,\nonumber\\
&  \mathcal{J}_{\psi^{\prime}}^{SS}=z_{\psi}^{2}\mathcal{J}^{SP}\beta_{\psi
z}^{2}\beta_{\psi z}^{\prime\prime}/4\,,_{\psi^{\prime}}^{TP,\tilde{T}%
S}=2z_{\psi}^{\tilde{T}}r_{\psi}^{2}/9\,,\mathcal{J}_{\psi^{\prime}}%
^{\tilde{T},\tilde{T}S}=z_{\psi}^{\tilde{T}}r_{\psi}^{2}(1-10r_{\psi}%
^{2}/z)/9\,,\mathcal{J}_{\psi^{\prime}}^{T,TS}=4_{\psi^{\prime}}%
^{TS}\,,\mathcal{J}_{\psi^{\prime}}^{T,\tilde{T}P}=4\mathcal{J}_{\psi^{\prime
}}^{\tilde{T}P}\,,\nonumber\\
&  \mathcal{J}_{\psi^{\prime}}^{\tilde{T},TP}=z^{2}\mathcal{J}_{\psi}%
^{\tilde{T}}(1+r_{\psi}^{2}/z)/9\,,\mathcal{J}_{\psi^{\prime}}^{TS,VV}%
=\mathcal{J}_{\psi}^{VV,\tilde{T}}z^{2}\beta_{\psi z}^{4}/27\,,\mathcal{J}%
_{\psi^{\prime}}^{\tilde{T}P,VV}=\mathcal{J}_{\psi}^{VV,\tilde{T}}%
z^{2}(1+2r_{\psi}^{2}/z+6r_{\psi}^{4}/z^{2})/54\,,\nonumber\\
&  \mathcal{J}_{\psi^{\prime}}^{VV,T}=\mathcal{J}_{\psi}^{VV,\tilde{T}}%
z^{2}(3-14r_{\psi}^{2}/z+38r_{\psi}^{4}/z^{2})/27\,,\mathcal{J}_{\psi^{\prime
}}^{VA,P}=\mathcal{J}_{\psi}^{SP,VA}z^{2}\beta_{\psi z}^{\prime}/4\,.
\end{align}
With the many possible interferences among the tensor currents and the
complicated kinematical functions, the differential rates for $B\to
H^{\ast}\Psi\overline{\Psi}$ are too cumbersome to be given here.

\pagebreak 


\end{document}